\documentclass{aa}  

\usepackage{graphicx}
\usepackage[usenames,dvipsnames]{xcolor}
\usepackage{natbib}
\usepackage{multirow}
\usepackage{booktabs,caption}
\usepackage[flushleft]{threeparttable}
\usepackage{appendix}
\usepackage{txfonts}
\usepackage[backref=page, hyperfootnotes=true, hidelinks, colorlinks, citecolor=blue, linkcolor=blue, linktocpage, bookmarks=true]{hyperref}
\usepackage{lscape}
\usepackage{footnote}
\usepackage{float}
\usepackage[bottom, flushmargin]{footmisc}
\usepackage{dcolumn}
\newcolumntype{d}[1]{D{.}{.}{#1}}

\makeatletter
\newcommand*{\rom}[1]{\expandafter\@slowromancap\romannumeral #1@}
\makeatother

\begin{document}

  \title{The mass discrepancy in intermediate- and high-mass eclipsing binaries: the need for higher convective core masses}
  \titlerunning{The mass discrepancy in intermediate- and high-mass eclipsing binaries}

   \author{A. Tkachenko
          \inst{1}
          \and
          K. Pavlovski\inst{2}
 %         \fnmsep\thanks{Just to show the usage
 %         of the elements in the author field}
          \and
          C. Johnston\inst{1} 
          \and
          M. G. Pedersen\inst{1}
          \and
          M. Michielsen\inst{1}
          \and
          D. M. Bowman\inst{1}
          \and
          J. Southworth\inst{3}
          \and
          V. Tsymbal\inst{4}
          \and
          C. Aerts\inst{1,5,6}
          }
\institute{Institute of Astronomy, KU Leuven, Celestijnenlaan 200D, B-3001, Leuven, Belgium; \email{andrew.tkachenko@kuleuven.be}
         \and
             Department of Physics, Faculty of Science, University of Zagreb, Bijenicka cesta 32, 10000 Zagreb, Croatia
         \and
         	Astrophysics Group, Keele University, Staffordshire, ST5 5BG, UK
         \and
             Institute of Astronomy, Russian Academy of Sciences, 119017, Pyatnitskaya str., 48, Moscow, Russia
         \and
         Department of Astrophysics/IMAPP, Radboud University Nijmegen, 6500 GL, Nijmegen, The Netherlands
         \and
         Max Planck Institute for Astronomy, Koenigstuhl 17, 69117 Heidelberg, Germany
             }

   \date{Received ; accepted }

% \abstract{}{}{}{}{} 
% 5 {} token are mandatory
 
  \abstract 
   {Eclipsing, spectroscopic double-lined binary star systems (SB2) are
     excellent laboratories for calibrating theories of stellar interior
     structure and evolution. Their precise and accurate masses and radii
     measured from binary dynamics offer model-independent constraints and
     challenge current theories of stellar evolution.}
   {We aim to investigate the mass discrepancy in binary stars. This is the
     significant difference between stellar components' masses measured from
     binary dynamics and those inferred from models of stellar evolution via
     positions of the components in the $T_{\rm eff}$--$\log\,g$ Kiel
     diagram. We study the effect of near-core mixing on the mass of the
     convective core of the stars and interpret the results in the context of
     the mass discrepancy.}
   {We fit stellar isochrones computed from a grid of {\sc mesa} stellar
     evolution models to a homogeneous sample of eleven high-mass binary
     systems. Two scenarios are considered, where individual stellar components
     of a binary system are treated independent of each other and where they are
     forced to have the same age and initial chemical composition. We also study
     the effect of the microturbulent velocity and turbulent pressure on the
     atmosphere model structure and stellar spectral lines, and its link with
     the mass discrepancy.}
   {We find that the mass discrepancy is present in our sample and that it is
     anti-correlated with the surface gravity of the star. No correlations are
     found with other fundamental and atmospheric parameters, including the
     stellar mass. The mass discrepancy can be partially accounted for by
     increasing the amount of near-core mixing in stellar evolution models. We
     also find that ignoring the microturbulent velocity and turbulent pressure in
     stellar atmosphere models of hot evolved stars results in overestimation
     of their effective temperature by up to 8\%. Together with enhanced
     near-core mixing, this can almost entirely account for the $\sim$30\% mass
     discrepancy found for the evolved primary component of V380\,Cyg.}
   {We find a strong link between the mass discrepancy and the convective
     core mass. The mass discrepancy can be solved by considering the
     combined effect of extra near-core
     boundary mixing and consistent treatment in
     the spectrum analysis of hot evolved stars. Our binary modelling
     results in convective core masses between 17 and 35\% of the stellar
       mass, in excellent agreement with results from gravity-mode
       asteroseismology of single stars. This implies larger helium core
       masses near the end of the main sequence than anticipated so far.}


   \maketitle
%
%-------------------------------------------------------------------

\section{Introduction}

The theory of stellar interior structure and evolution (SSE) plays a crucial
role in contemporary astrophysics. Many research fields rely on predictions of
this theory and are hence extremely dependent on how well physical conditions
can be described throughout the star both instantaneously and as a function of
time. Historically, the theory of SSE was calibrated on surface atmospheric
properties of stars meaning that the whole process was largely driven by
observational constraints available at the outer (atmospheric) boundary
only. The situation significantly improved in the case of the Sun with the
advent of helioseismology
\citep[e.g.,][]{Leighton1962,Evans1962a,Evans1962b,Evans1962c,Christensen-Dalsgaard1976,Harvey1996,Gough1996,Christensen-Dalsgaard2002}
which allowed more than 70\% of the outermost part of the Sun to be probed
through detection and interpretation of its acoustic waves stochastically driven
by convection \citep[][]{Claverie1979,Duvall1983}. 

Helioseismology has provided an enormous improvement in the calibration of the models for low-mass stars with a radiative core \citep{Christensen-Dalsgaard2002}. This was achieved from adapting the input physics and parametrized transport profiles in models of the Sun to solve discrepancies in the solar oscillation frequencies and those predicted by non-rotating non-magnetic 1D models. This seismically calibrated solar model is nowadays applied to the thousands of low-mass and evolved intermediate-mass stars observed with the {\it Kepler} \citep{Borucki2010} and TESS \citep{Ricker2015} space missions. The frequencies of their pressure (p-) modes allow us to determine the radii, masses and and ages of solar-like stars from scaling them with respect to the frequencies of the helioseismic solar model, leading to relative precisions for the stellar radii, masses, and ages of $\sim\!2\!$, $\sim\!\!5$, and $\sim\!\!15$\%, respectively  \citep{Chaplin2014,HekkerJCD2017}.  Higher precisions require improving the input physics, starting from the non-rotating non-magnetic solar-scaled model and adopting its ingredients to include more realistic descriptions for the transport processes due to convection, atomic diffusion, rotation, and magnetism so as to  meet the asteroseismic measurements \citep[e.g.,][]{SilvaAguirre2017,Verma2019,Eggenberger2019a,Eggenberger2019b,Fuller2019}.

Asteroseismic evaluations of intermediate- and high-mass stars in the core-hydrogen burning phase are based on a similar principle to helio- and asteroseismology of low-mass stars. However, evaluations of intermediate- and high-mass stars cannot rely on the scaling of solar-like models, since they have a convective core and a radiative envelope. For an extensive discussion of the methodology and a summary of achievements so far, we refer to \citet{Aerts2020}. For the current work, we distill the two most pertinent results obtained from some 40 intermediate-mass stars covering $M\in [1.3,8]\,$M$_\odot$ deduced from non-radial gravity (g) modes, which probe the region just outside the convective core: 1) stars with a convective core are near-rigid rotators throughout the entire core-hydrogen burning phase across the modelled range of $v_{\rm eq}/v_{\rm crit}\in  [0,70]\,$\% \citep{Aerts2019a}; 2) asteroseismic derivations of the mass of the convective core, $M_{\rm cc}$, via estimation of near-core mixing levels, lead to the broad range of $M_{\rm cc}/M\in [3,20]\%$.  These studies pinpoint the need for more massive convective cores during the main-sequence phase compared with those predicted by standard stellar evolution models, even when limiting to the slowest rotators in the sample.
  
Much in the spirit of helioseismology, the asteroseismic studies of intermediate- and high-mass stars purposefully do not specify the physical reason for the extra mixing near the convective core but rather interpret the deviations between g-mode frequencies predicted for 1D non-rotating non-magnetic stellar models and those detected in the space-based data. It is noteworthy to mention that the mixing profiles predicted from various 1D rotating stellar models differ by a lot, but they also do not lead to appropriate values of g-mode frequencies when compared with space asteroseismic data of such quantities \citep{Aerts2018}. In particular, the predicted levels of chemical mixing at the bottom of the deep envelopes of the models due to rotational instabilities are orders of magnitude too high to be in agreement with the asteroseismic data.

Eclipsing doubled-lined spectroscopic binaries offer a completely independent way to calibrate the interiors of intermediate- and high-mass stars. The ability to infer model-independent masses and radii of stars with very high
precision and accuracy makes binary stars ideal candidates for benchmarking SSE
models as well as asteroseismic analyses. \citet{Valle2018} performed a
theoretical study as to the ability to recover stellar ages and the overshooting
efficiency based on simulated binary star data. The authors assumed typical
observational uncertainties on stellar mass and effective temperature of
$\leq$1\% and $\pm$150~K, respectively. Moreover, they also found that recovered
ages and efficiencies of near core mixing are biased towards lower values in all
considered scenarios. \citet{Valle2018} also concluded that the above
observational uncertainties typically allow one to distinguish between models
without convective core overshooting and those with an intermediate amount of
convective core overshooting.

\citet{Pols1997} already showed that SSE models require systematically enhanced
near-core mixing at post-main-sequence stage of evolution to accommodate stellar
masses measured from binary dynamics. \citet{Lastennet2002} report an overall
agreement between model predictions and mass and radius measurements for 60
detached binary systems, emphasizing a large degeneracy between the age and
metallicity in the models due to lack of observational constraints for the 
metallicity. More recently, \citet{Higl2017} presented a study of detached
eclipsing binaries in a wide range of stellar masses. These authors report the
need to introduce extra near-core mixing in the form of overshooting for stars
with a convective core.

Binary and multiple stellar systems are also found to be extremely synergistic
with intrinsically variable pulsating stars. An example of such a synergy is the
class of `heart-beat stars` \citep[][]{Thompson2012} -- highly eccentric binaries
with variable component(s), where the pulsational variability is being
triggered by tidal forces
\citep[e.g.,][]{Welsh2011,Hambleton2013a,Hambleton2013b,Guo2019}. In addition to
providing a stellar pulsation excitation mechanism, tides are also known to be
important \citep{Fuller2017,Guo2017} and can affect self-excited oscillations of
stars, as evidenced recently by the TESS mission \citep[e.g.][]{Bowman2019c}.

\section{The mass discrepancy}
The SSE models of stars born with a convective core are not well calibrated. A
striking example is the mass discrepancy observed in massive stars which has
remained unsolved for almost three decades. The original formulation of the
problem comes from \citet{Herrero1992}, who presented a spectroscopic analysis of
25 luminous Galactic stars and made a comparison between spectroscopically
inferred masses and those derived from models of stellar evolution. The
spectroscopic masses are derived by exploiting the spectroscopically determined
surface gravities and radii that come from the absolute visual magnitude ${M_V}$
of the star and the integral of stellar flux $V$ \citep[Eq.\,1
in][]{Herrero1992}. The evolutionary masses are in turn obtained by fitting
evolutionary tracks to the position of the star in the Hertzsprung-Russell (HR)
diagram, where stellar luminosity is derived from the radius and the
spectroscopically inferred effective temperature of the star. From a comparison
of the component masses within the sample, the authors concluded that SSE models
overpredict the masses of the stars and that there is a tendency of the effect
to get more pronounced as the surface gravity $\log\,g$ of the star decreases
\citep[see Fig.~16 in][]{Herrero1992}.

The conclusions of \citet{Herrero1992} are confirmed by independent studies of
eclipsing, spectroscopic double-lined (SB2) binaries. In this particular case,
it is the dynamical versus evolutionary mass discrepancy that is being
reported. While the former mass measurement is model-independent and comes from
binary dynamics, the latter mass inference is based on fitting positions of
individual stellar components in the HR diagram with evolutionary tracks.

\subsection{Case study: V380\,Cyg}

The binary V380\,Cyg comprises an evolved early B-type primary star and a
main-sequence B-type secondary. It is amongst the most prominent cases of
massive binaries exhibiting the mass discrepancy. The system was originally
studied by \citet{Guinan2000} based on time-series of multi-color ground-based
photometric data and of high-resolution optical \'echelle spectroscopy. The
authors emphasized the difficulty to explain the position of the more evolved
primary in the HR diagram with evolutionary tracks that correspond to the
dynamical mass, spectroscopically measured metallicity, and no extra mixing in
the near-core regions of the primary. Large amounts of extra near-core mixing in
the form of convective penetration corresponding to $\alpha_{\rm ov}=0.6$ was proposed by \citet{Guinan2000} as a
possible solution to the mass discrepancy in the V380\,Cyg system. The need
  for a large amount of extra mixing in the near-core region 
has been confirmed quasi-independently by
the authors from measurements of the apsidal motion of the star and
subsequent inference of the internal structure constant $k$.

\begin{table}
\begin{threeparttable}
\tiny
\tabcolsep 0.7mm \caption{Observed fundamental and atmospheric parameters of the
  sample targets. Error bars are given in parentheses in terms of the last
  digit(s). The superscript in the first column identifies the study the
  parameters have been taken from. For each object, the first/second line
  corresponds to the primary/secondary component.
}
\label{Table:ObservedParameters}      % is used to refer this table in the text
\centering                          % used for centering table
\begin{tabular}{l c c c c c c}        % centered columns (4 columns)
\toprule
Object/ & $M$ & $R$ & $\log T_{\rm eff}$ & $\log\,g$ & $v\sin\,i$ & $v_{\rm eq}/v_{\rm crit}$\\
Parameter & (M$_{\odot}$) & (R$_{\odot}$) & (dex) & (dex) & (km\,s$^{-1}$) & (\%)\\   
\midrule                        % inserts single horizontal line
  \multirow{2}{*}{V578\,Mon$^1$}\rule{0pt}{9pt} & 14.54(8) & 5.41(4) & 4.477(7) & 4.133(18) & 117(4) & 21.0(9)\\
  & 10.29(6) & 4.29(5) & 4.411(7) & 4.185(21) & 94(2) & 17.9(7)\vspace{1mm} \\
  
  \multirow{2}{*}{V453\,Cyg$^2$} & 13.90(23) & 8.62(9) & 4.459(8) & 3.710(9) & 107.2(2.8) & 23.6(1.0)\\
  & 11.06(18) & 5.45(8) & 4.442(10) & 4.010(12)  & 98.3(3.7) & 19.3(1.1)\vspace{1mm} \\
  
  \multirow{2}{*}{V478\,Cyg$^2$} & 15.40(38) & 7.26(9) & 4.507(7) & 3.904(9) & 129.1(3.6) & 25.4(1.3)\\
  & 15.02(35) & 7.15(9) & 4.502(9) & 3.907(10) & 127.0(3.5) & 25.1(1.3)\vspace{1mm} \\
  
  \multirow{2}{*}{AH\,Cep$^2$} & 16.14(26) & 6.51(10) & 4.487(8) & 4.019(12) & 172.1(2.1) & 32.6(9)\\
  & 13.69(21) & 5.64(11) & 4.459(10) & 4.073(18) & 160.6(2.3) & 30.9(9)\vspace{1mm} \\
  
  \multirow{2}{*}{V346\,Cen$^3$} & 11.78(13) & 8.26(16) & 4.417(5) & 3.675(17) & 165.2(2.8) & 39.0(1.3)\\
  & 8.40(10) & 4.19(8) & 4.352(6) & 4.118(16) & 89.1(2.3) & 17.7(7)\vspace{1mm} \\
  
  \multirow{2}{*}{V573\,Car$^3$} & 15.14(39) & 5.41(5) & 4.504(5) & 4.151(7) & 184.6(2.7) & 31.4(1.1)\\
  & 12.38(20) & 4.48(5) & 4.458(5) & 4.229(7) & 155.4(3.1) & 26.5(9)\vspace{1mm} \\
  
  \multirow{2}{*}{V1034\,Sco$^3$} & 17.07(12) & 7.49(7) & 4.508(7) & 3.921(8) & 169.8(2.6) & 31.9(8)\\
  & 9.60(5) & 4.20(4) & 4.412(5) & 4.173(9) & 94.5(3.3) & 17.8(7)\vspace{1mm} \\
  
  \multirow{2}{*}{V380\,Cyg$^4$} & 11.43(19) & 15.71(13) & 4.336(6) & 3.104(6) & 98(2) & 32.6(1.1)\\
  & 7.00(14) & 3.82(5) & 4.356(22) & 4.120(11) & 38(2) & 8.0(6)\vspace{1mm} \\
  
  \multirow{2}{*}{CW\,Cep$^5$} & 13.00(7) & 5.45(5) & 4.452(7) & 4.079(10) & 105.2(2.1) & 19.3(5)\\
  & 11.94(7) & 5.09(5) & 4.440(7) & 4.102(10) & 96.2(1.9) & 17.8(6)\vspace{1mm} \\
  
  \multirow{2}{*}{U\,Oph$^5$} & 5.09(5) & 3.44(1) & 4.220(4) & 4.073(4) & 110(6) & 25.4(1.6)\\
  & 4.58(5) & 3.05(1) & 4.183(3) & 4.131(4) & 108(6) & 24.7(1.5)\vspace{1mm} \\
  
  V621\,Per$^6$ & 9.44(46) & 8.92(14) & 4.354(5) & 3.513(11) & 32.3(2.7) & 8.7(1.2)\\
\bottomrule
\end{tabular}
\begin{tablenotes}
      \small
      \item $^1$\citet{Garcia2014}; $^2$\citet{Pavlovski2018}; $^3$Pavlovski et al. (2020, in prep.); $^4$\citet{Tkachenko2014}; $^5$\citet{Johnston2019a}; $^6$Southworth et al. (2020, in prep.)
    \end{tablenotes}
\end{threeparttable}
\end{table}

The system was revisited by \citet{Pavlovski2009b} based on photometric data of
\citet{Guinan2000} and newly obtained multi-instrument extended time-series of
optical \'echelle high-resolution spectroscopy. The authors used the method
of spectral disentangling \citep[{\sc spd};][]{Simon1994} as implemented in the
{\sc FDBinary} software package \citep{Ilijic2004} to disentangle composite
spectra of the binary system into individual spectral contributions of the two
components. Atmospheric characteristics inferred from the disentangled spectra
were used with a grid of Geneva evolutionary models to derive evolutionary
masses of the two stars. In line with the findings by \citet{Guinan2000}, the
authors report a mass discrepancy in excess of 10\% for the more evolved primary
component, by comparing its dynamical mass to a set of evolutionary masses
inferred from the SSE model grids of \citet{Schaller1992}, \citet{Claret1995},
and \citet{Ekstroem2008}. In particular, \citet{Pavlovski2009b} conclude that
the current numerical implementation of rotation in SSE models does not solve
the mass discrepancy observed in V380\,Cyg.

V380\,Cyg was one of the brightest and most massive stars observed by the {\it
  Kepler\/} space mission in its Guest Observer Programme (PI:
A. Tkachenko). Hence, it was reanalyzed once again by \citet{Tkachenko2014}
based on about three months of nearly continuous {\it Kepler\/} space photometry
and newly obtained time-series of optical \'echelle spectroscopy from the {\sc
  hermes} spectrograph mounted on the Flemish Mercator telescope on La Palma
\citep{Raskin2011}. Similar to the previous studies, the authors report a mass
discrepancy in excess of 30\% for the primary component. The measurements could
only be explained by assuming a stellar mass at 3$\sigma$ of its dynamical
value, by significantly increasing the initial rotation rate compared to the
current one, and by adopting a high value of $\alpha_{\rm ov}=0.6$ in their grid
of {\sc mesa} evolutionary models computed according to \citep{Paxton2013}. In
addition, \citet{Tkachenko2014} also report the detection of stochastic
oscillations intrinsic to the primary component, as well as rotational
modulation in the high-resolution spectroscopy.

\begin{figure}
   \centering
   \includegraphics[width=8.7cm]{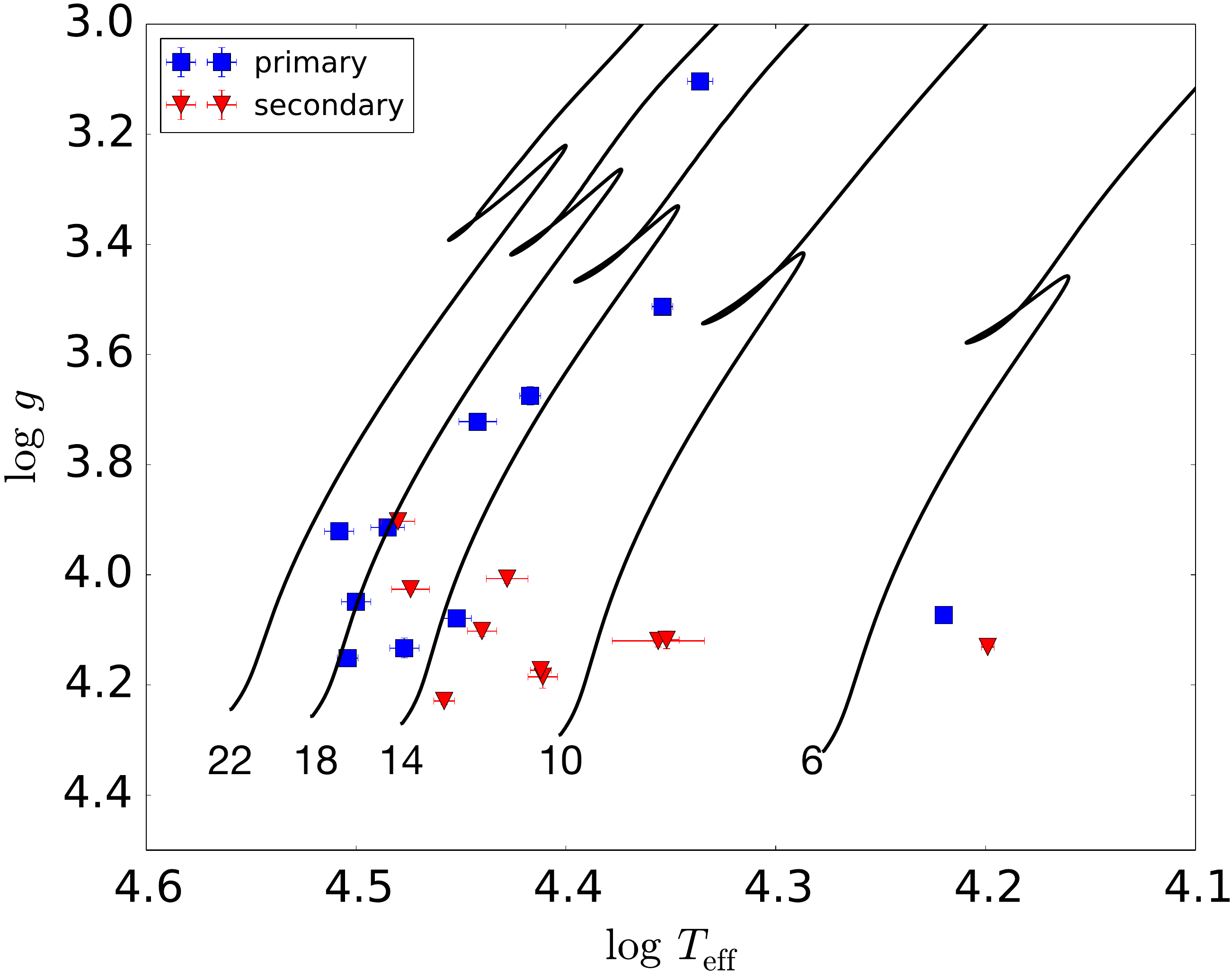}
      \caption{Positions of individual binary components from our sample in
        Table~\ref{Table:ObservedParameters} in the $T_{\rm eff}$--$\log\,g$
        Kiel diagram. Blue squares and red triangles refer to the primary and
        secondary components, respectively. Evolutionary tracks are computed
        with solar metallicity $Z=0.014$ and for an exponentially decaying CBM
        profile with $f_{\rm
          ov}=0.02$~$H_{\rm p}$. Stellar mass is indicated in M$_{\odot}$
        units.} 
         \label{Fig:StellarSample}
   \end{figure}
   
\subsection{Ensemble study results}
\citet{Schneider2014} presented a homogeneous Bayesian analysis of a sample of 18 eclipsing binaries from the \citet{Torres2010} sample and relied on Bonn stellar evolution models with rotation. The authors stressed that rotation implies a larger radius of a few percent and hence can not be ignored. Furthermore, \citet{Schneider2014} focused on the consequences of rotational mixing in stellar aging, while assuming that stars with fractional main-sequence ages less than 35\% are unaffected by convective core overshooting. Under these conditions, they were able to find a good isochrone fit for component stars cooler than 25\,000~K. Here, we take a similar approach but lift the assumption that stars do not experience near-core mixing in the first stage of the main sequence, since asteroseismology has shown young massive stars experience near-core boundary mixing \citep[see ][for a summary of measured near-core and envelope mixing levels]{Aerts2020}. Moreover, we consider models with near-core mixing without specifying its origin in terms of physical process, again following recent results from asteroseismology. We evaluate the models for a sample of eclipsing binaries with high-resolution spectroscopy analyzed in a homogeneous way so as to eliminate systematic bias in the data that is used as input for the modelling. In this respect, our study can be seen as a follow-up study of the one by \citet{Schneider2014}, but where we utilize non-rotating models with interior mixing prescriptions guided by asteroseismology.

The idea of probing the amount of near-core mixing in intermediate- and
high-mass binaries in the form of convective core overshooting has been further
elaborated upon by \citet{Claret2016,Claret2017,Claret2018,Claret2019} 
in their series of papers on a sample of some 50 eclipsing SB2 systems. Their
sample has been compiled from the catalog of
\citet{Torres2010} and the catalog of the Optical Gravitational Microlensing
Experiment (OGLE)\footnote{http://ogle.astrouw.edu.pl/}. The selection
requirements concerned precise and accurate (3\% and better) masses, radii, and
effective temperatures of stars, with additional selection criteria being
evolutionary stage and availability of the surface chemical composition
measurements. Grids of non-rotating
evolutionary models were computed with the Granada
\citep{Claret2004,Claret2012} and {\sc mesa}
\citep{Paxton2011,Paxton2013,Paxton2015} codes for variable
initial mass, metallicity, and overshooting parameter $\alpha_{\rm ov} (f_{\rm
  ov}$). The authors allowed for variable metallicity and an age tolerance of up to
5\% when fitting isochrones to the sample of the individual binary components,
deducing the best fit overshooting parameter dictated by the stars' positions in
the HR diagram. Irrespective of the assumed functional form of the overshooting,
the authors find a clear dependence of the latter on stellar mass with an almost
linear transition from no overshooting to approximately $\alpha_{\rm ov} (f_{\rm
  ov}$) = 0.2 (0.02)~$H_{\rm p}$ in the mass range from
$\sim$1.2--2.0~M$_{\odot}$. The distribution flattens beyond
$\sim$2.0~M$_{\odot}$ and no further increase of the overshooting parameter with
increasing stellar mass is deduced. In addition, \citet{Claret2016} report
systematically smaller metallicities as inferred from their evolutionary models
compared to spectroscopic measurements in the literature, without offering an
explanation for this.

The sample of detached eclipsing binaries presented by
\citet{Claret2016,Claret2017,Claret2018} has been revisited by \citet{Costa2019}
based on a new grid of rotating PARSEC models and using a Bayesian method of
analysis. The authors report a large spread in the derived values of the
overshooting parameter for stars with masses above some
1.9~M$_{\odot}$. Furthermore, \citet{Costa2019} demonstrate that the above
spread can be well explained by stars having a uniform distribution of initial
rotation in the range between 0 and 0.8 of the break-up value and a fixed mild
amount of core overshooting. \citet{Daszynska-Daszkiewicz2019} present a study
of 38 detached systems compiled from the literature where they focus on age
determination by simultaneously matching the stars' positions in the radius-age and
the $T_{\rm eff}$--$\log\,g$ Kiel diagrams. The authors rely on visual
inspection and come to the conclusion that it is necessary to adjust
values of the initial metallicity and convective core overshooting in order to
reproduce the observed properties and common ages for 33 out of 38 binary
systems.

\begin{table}
\begin{threeparttable}
\caption{Summary of the grid of {\sc mesa} stellar evolution models employed for the analysis of our sample stars. The grid comprises over 5\,500 models.}
\label{Tab:MesaGrid}      % is used to refer this table in the text
\centering                          % used for centering table
\tabcolsep=3pt
\small
\begin{tabular}{l c c c }        % centered columns (4 columns)
\toprule
\multirow{2}{*}{Parameter} & \multicolumn{3}{c}{Range}\\
& Min & Max & Step\\   
\midrule
\multirow{5}{*}{Mass (M$_{\odot}$)}\rule{0pt}{9pt} & 1.2 & 2.0 & 0.05\\
& 2.0 & 5.0 & 0.10\\
& 5.0 & 10.0 & 0.25\\
& 10.0 & 15.0 & 0.50\\
& 15.0 & 25.0 & 1.00\vspace{2mm}\\
Diffusive CBM with parameter $f_{\rm ov}$ ($H_{\rm p}$) \& $\nabla_{\rm rad}$ & 0.005 & 0.04 & 0.005\\
Core extension with parameter $\alpha_{\rm ov}$ ($H_{\rm p}$) \& $\nabla_{\rm ad}$ & 0.05 & 0.40 & 0.05\vspace{2mm}\\
Metallicity $Z$ (mass fract.) & 0.006 & 0.018 & 0.004\\
\bottomrule
\end{tabular}
\begin{tablenotes}
      \small
      \item Note that we fix the metallicity parameter $Z$ in the
        analysis, 
as argued in the text.
    \end{tablenotes}
\end{threeparttable}
\end{table}

In this paper, we study the mass discrepancy in eclipsing binaries by focusing
on a sample of (mostly) high-mass stars. The stellar sample itself is presented
in Section~\ref{Sect:Stellar_sample}, while the adopted methodology is
summarized in Section~\ref{Sect:Methodology}. 
Our method is based on isochrone fitting. 
Our analyses of the individual binary components assuming that they are effectively single stars are presented
in Section~\ref{Sect:SingleStarCase}. In these analyses, we 
investigate the influence of the assumed functional form of extra 
mixing and of the adopted
temperature gradient in the near-core region, 
without pinpointing its physical cause
(Sections~\ref{Sect: DiffusiveOvershooting} and \ref{Sect:
  ConvectivePenetration}). In practice, we consider both an exponentially
decaying and a constant near-core mixing profile for the core-boundary mixing (CBM). For each of these cases, and for each star, we 
report the most important consequence of
the adopted CBM profile and the associated mass of the convective core. Further analyses enforcing 
an equal age condition for the two stellar components of each binary
are summarized in Section~\ref{Sect:BinaryEqualAgeCase}. We
take a closer look at the V380\,Cyg system in Section~\ref{Sect:V380Cyg} before
closing the paper with the discussion and conclusions in
Section~\ref{Sect:Conclusions}.

\section{Stellar sample}
\label{Sect:Stellar_sample}

\begin{figure*}
   \centering
   \includegraphics[width=18.3cm]{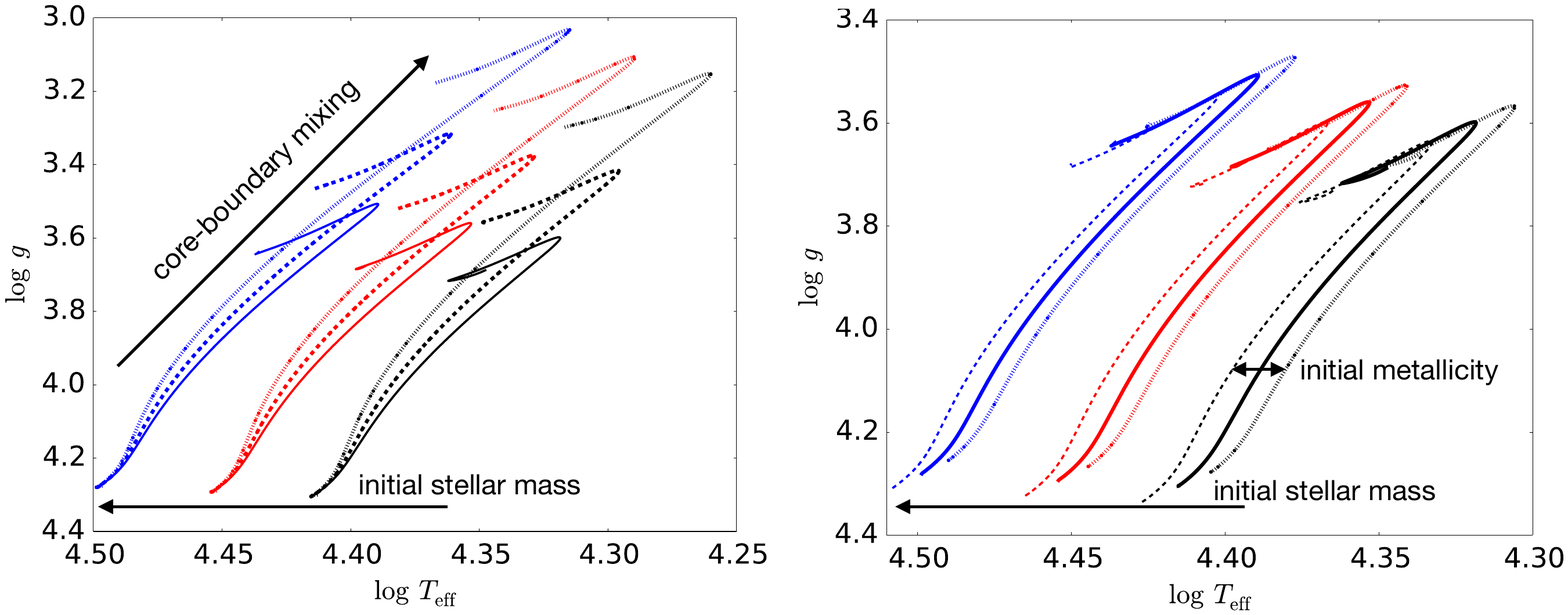}
   \caption{{\bf Left:} Effect of the initial stellar mass $M$ and of an
       exponentially decaying core-boundary mixing profile with parameter
$f_{\rm ov}$ on evolutionary tracks. Mass ($M$=10
     (black), 12 (red), and 15~M$_{\odot}$ (blue)) and mixing
     ($f_{\rm ov}$=0.005 (solid), 0.020 (dashed), and 0.040~$H_{\rm p}$
     (dotted)) sequences are shown with color and line style,
     respectively. Solar metallicity $Z$ = 0.014 is assumed. {\bf Right:} Effect
     of the initial stellar mass $M$ and the metallicity $Z$ on
     evolutionary tracks. The mass sequence is the same as in the left panel;
     line styles show the metallicity sequence $Z$ = 0.010 (dashed), 0.014
     (solid), and 0.018 (dotted) in mass fraction units. A fixed parameter
     $f_{\rm ov}$=0.005~$H_{\rm p}$ is assumed. Note the difference in X,Y-axes
     in the two panels.}
         \label{Fig:MesaTracks}
   \end{figure*}

Our stellar sample comprises eleven intermediate- and high-mass eclipsing SB2
binary systems and is presented in Table~\ref{Table:ObservedParameters}. A
distinct property of this sample and of our approach compared to previous
studies in the literature is that the fundamental and atmospheric parameters of
all stars from the sample were obtained with the same methodology
\citep{Pavlovski2018}. This concerns the use of the most recent versions of the
Wilson-Devinney \citep[{\sc WD},][]{Wilson1971}, {\sc
  phoebe}\footnote{http://phoebe-project.org/1.0} \citep{Prsa2005,Prsa2018}, and
{\sc jktebop}\footnote{https://www.astro.keele.ac.uk/~jkt/codes/jktebop.html}
\citep{Southworth2004} codes for the analysis and interpretation of (eclipsing)
binary light curves. The method of spectral disentangling as implemented in the
{\sc FDBinary} code \citep{Ilijic2004} has been employed to obtain spectroscopic
orbital elements as well as disentangled spectra of individual stellar
components for each binary system from the sample. The suite of {\sc
  surface}/{\sc detail} \citep{Giddings1981,Butler1984} codes with model
atoms listed in \citet{Pavlovski2009a} were used for NLTE analyses of the
obtained disentangled spectra to infer atmospheric characteristics of individual
binary components. In cases when the LTE assumption was adequate, the {\sc gssp}
software
package\footnote{https://fys.kuleuven.be/ster/meetings/binary-2015/gssp-software-package}
\citep{Tkachenko2015} has been employed for the analysis of stellar spectra. The
use of consistent methodology allowed us to compile a homogeneous sample of 21
stellar components. Only the primary component of the V621\,Per system is
included as the low light contribution of the secondary component prevents the
determination of precise atmospheric parameters for this star from the
corresponding disentangled spectrum. Our approach is not subject to any
systematic uncertainties that are typically expected when employing different
analysis methods.

In addition to the $\log T_{\rm eff}$ and $\log\,g$,
  Table\,\ref{Table:ObservedParameters} also contains the measured projected
  surface ($v\sin\,i$) and equatorial ($v_{\rm eq}$) rotational
velocities, which assumes the rotation axes to be
  perpendicular to the orbital axis, for the stars included in our sample. From the $M$ and $R$ estimates, we computed
  the ratio of the equatorial rotational velocity to the critical Keplerian
  rotational velocity adopting the Roche model. In general, two critical rotation
  rates occur from solving for an effective gravity equal to zero at the equator. For 
our targets, however, the gravity is not affected by a strong
  radiation-driven wind, leading to one unique value for the critical rotation
rate. This
  is given by the expression $v_{\rm crit}=\sqrt{2GM/3R_{\rm p}}$, with
  $R_{\rm p}$ the polar radius \citep{Maeder2009}. It can be seen from
  Table\,\ref{Table:ObservedParameters} that our sample stars rotate modestly,
  with $v_{\rm eq}/v_{\rm crit}$ between 8\% and 39\%. This justifies the
  use of 1D stellar evolution models in our analysis.

Our sample is represented graphically in Fig.~\ref{Fig:StellarSample} in which
the positions of all 21 individual stellar components are shown in the
$T_{\rm eff}$--$\log\,g$ Kiel diagram. Evolutionary tracks for 1D models
covering a stellar mass range from 6--22~M$_{\odot}$ are also shown.  One can
see that, in addition to its homogeneity, our sample covers a wide range of
stellar masses and evolutionary stages, which makes it suitable to study the
mass discrepancy problem.

\section{Methodology}
\label{Sect:Methodology}

In this section, we provide an overview of the adopted methodology, which
includes a grid of SSE models. We also discuss the multi-faceted
analysis approach in a concise way.

\subsection{1D Stellar evolution models}
\label{Sect:1DModels}

Even with the computational power currently available, stellar evolution models
  necessarily remain 1D simplifications of 3D gaseous spheres
  \citep[e.g.,][]{Cristini2016}.  The first steps of a solid calibration of
  stellar interiors from the bridging of 3D simulations and 1D stellar models
  are being taken from gravity-mode asteroseismology for stars in the mass range
  of our work \citep{Arnett2017}.  The level of sophistication adopted in 1D
  numerical models of stars with a convective core is diverse, even for the
  simplest phase of core-hydrogen burning upon which we focus here.
  
  While the
  simplest of these 1D models rely on mass conservation and on only the pressure force
  and gravity in the momentum equation, they already suffer from the simplified
  treatment of convection as a time-independent phenomenon described by at least
  one free parameter.  More complex main-sequence models include any of the
  Coriolis, Lorentz, and tidal forces, as well as mass loss from a
  radiation-driven wind \citep[see the monographs
  by][]{Maeder2009,Kippenhahn2012}.  Moreover, the transport equations to
  describe the change of the mass fractions of individual chemical elements as a
  function of time cannot be derived from first principles. These equations
  include parametrized profiles for the various phenomena of chemical mixing,
  involving many free parameters, several of which are connected with rotational
  or magnetic instabilities \citep[e.g.,][]{Heger2000,Palacios2013}. Finally, a
  variety of choices in numerical implementations to solve the SSE equations and
  the chosen set of boundary conditions occurs. As a result of these
  complexities, major differences of orders of magnitude occur in the mixing
  profiles of stellar models with rotation, as illustrated by comparing Fig.\,5
  in \citet{Chieffi2013} with Fig.\,29 in \citet{Paxton2013} and Fig.\,3 in
  \citet{Georgy2013}, to list a few.
  
  Consequently, SSE tracks of rotating models in the HR diagram differ a lot. For this reason, we present a complementary approach to the one by \citet{Schneider2014}, by investigating the mass-discrepancy problem from non-rotating models guided by recent asteroseismic results. Our sample is ideally suited to perform a calibration of stellar interiors independently from asteroseismology, because it covers well the considered rotation rate of the asteroseismology sample. Just as for asteroseismology of intermediate-mass stars, a good procedure is to assess deviations between the observed diagnostic properties of the binaries in our sample and the theoretical predictions for those diagnostics from 1D models. Given the slow rotation rates of our sample stars and the comparative asteroseismology sample, we consider  non-rotating non-magnetic stellar models with extra CBM without pinpointing its physical cause.

\subsection{MESA model grid}
\label{Sect: ModelsGrid}

We rely on a recent grid of {\sc mesa} models
\citep[][]{Paxton2011,Paxton2013,Paxton2015,Paxton2018,Paxton2019} but extend it
towards higher masses compared to the original one presented in
\citet{Johnston2019b}. The corresponding {\sc mesa} inlist is optimized for
intermediate- and high-mass stars as dictated by the properties of their gravity
(g-) mode oscillations. For example, the input physics includes radiative
envelope mixing as determined by \citet{Rogers2017} from 2D hydrodynamical
simulations of internal gravity waves \citep[IGWs,][]{Rogers2013} in the form of
$D_{\rm env}(r)\propto\rho^{-1/2}$. It has been implemented in {\sc mesa} in a
diffusive approximation by \citet{Pedersen2018} and was shown to have a
significant and detectable effect on C, N, and O surface abundances in stars in
the mass range under study. Such an envelope mixing profile replaces the
default {\sc mesa} implementation which is assumed to be radially constant and
is set by the parameter $\log D_{\rm mix}$. Though our study does not use any
asteroseismic information, it relies on input physics calibrated by most recent
asteroseismic findings summarized in \citet{Aerts2020}. 
We use the Ledoux criterion for convection, set the
mixing-length parameter $\alpha_{\rm mlt}$ to the solar calibrated value of 1.8,
assume a chemical mixture following the cosmic standard for abundances by
\citet{Przybilla2008} and \citet{Nieva2012}, and set the initial helium and
hydrogen fractions to $Y=0.276$ and $X=0.71$, respectively. The level of envelope mixing near the convective core amounts to $\log D_{\rm mix}=1\,{\rm cm}^2\,{\rm s}^{-1}$ in our analysis. We refer the reader
to \citet{Pedersen2018} and \citet{Johnston2019b} for details on the chosen
input physics as well as for the full {\sc mesa} inlists and {\sc
  run\_star\_extras} routines.

Global characteristics of the employed grid of {\sc mesa} models are summarized
in Table~\ref{Tab:MesaGrid}. We adopt $f_{\rm ov}=0.04$ and
$\alpha_{\rm ov}=0.40$ as upper limit for the CBM parameter following
asteroseismic findings for single stars in the considered mass range
\citep[e.g.,][]{Aerts2013,Papics2014,Moravveji2016,Buysschaert2018,Aerts2020}. Even though the
grid covers a range in metallicity values, we choose to fix the parameter $Z$ to
its solar value (corresponding to 0.014 in mass fraction), consistent with the
surface abundances measured from high-resolution spectroscopy for all stars in
our sample. By fixing the metallicity and the functional form of the
envelope mixing, we restrict ourselves to three free parameters when fitting the
position of a star in the Kiel diagram: initial stellar mass $M$, amount of CBM
($f_{\rm ov}$ or $\alpha_{\rm ov}$), and stellar age. 
Fixing the metallicity of the star also eliminates a number of model
degeneracies, in particular the metallicity-initial mass and metallicity-CBM
degeneracies. As demonstrated in Fig.~\ref{Fig:MesaTracks},
lowering the metallicity of the star has a similar effect on the evolutionary
track to increasing its initial mass and, to a certain extent, to increasing the
CBM level. In all of these cases, evolutionary tracks
experience a shift towards higher effective temperatures in the Kiel diagram,
while the CBM also leads to an extension of stellar lifetime on
the main sequence (left panel in Fig.~\ref{Fig:MesaTracks}).

\subsection{Multi-faceted analysis outline}
\label{Sect: step-by-step}

Here, we summarize our multi-faceted analysis approach. The four solutions
outlined below are first computed for the single-star scenario assuming no
relation between individual binary components, followed by the scenario where
both components of a given binary system are forced to have the same age. In
either case, we make use of individual stellar masses determined from binary
dynamics as a strong observational constraint to report and quantify the mass
discrepancy where applicable. Our four solutions are:
\begin{itemize}
\item {\bf `Reference model' (RM) solution:} we fix the initial metallicity $Z$
  to the solar value of 0.014 which is consistent with the spectroscopic
  metallicity measurements for all our targets. At this stage, we ensure the
  amount of extra near-core mixing is minimal corresponding to $f_{\rm ov} =
  0.005$ for the exponentially decaying CBM profile. The effective
  temperature $T_{\rm eff}$, surface gravity $\log\,g$, and (dynamical) stellar
  mass $M$ are the three parameters that determine the $\chi^2$-merit function
  in this particular case. Given the high precision of the dynamical mass
  measurements, the solution is largely determined by the mass with less weight
  given to the $T_{\rm eff}$ and $\log\,g$ parameters. This allows us to form a
  baseline solution (i.e., reference model) with respect to which the mass
  discrepancy is quantified in the solution below.
\item {\bf Initial mass (IM) solution:} same as the RM solution except that we
  exclude the (dynamical) stellar mass from the $\chi^2$-merit function, i.e.,
  we fit the
  position of the star in the Kiel diagram as a constraint. This particular solution allows
  us to quantify the discrepancy between the measured dynamical mass and the one
  obtained from fitting the position of the star in the Kiel diagram with
  stellar evolution models (evolutionary mass, hereafter). 
\item {\bf Core boundary mixing (CBM) solution:} same as the RM solution but
  relaxing the CBM parameter $f_{\rm ov}$, yet
  including $T_{\rm eff}$, $\log\,g$, and the dynamical stellar mass in the
  $\chi^2$-merit function. Similarly to the RM solution case scenario, $\chi^2$
  is dominated by the dynamical stellar mass due to the high precision of this
  parameter. However, the fit includes an extra free parameter, which
  is the amount of near-core mixing.
\item {\bf Initial mass-core boundary mixing (IM-CBM) solution:} same as the IM
  solution except that the (best fit) CBM parameter
  $f_{\rm ov}$ is adopted (instead of it being fixed it to
  the minimal value) and the stellar mass is relaxed for those systems where no
  satisfactory fit could be obtained in the previous solution.
\end{itemize}

\section{Results}

In this section, we discuss our results regarding the mass discrepancy across
the stellar sample as well as the connection with the near-core mixing. The
latter is assumed to have two possible functional forms and associated
temperature gradients. The default implementation in
 {\sc mesa} is the exponentially decaying efficiency of mixing in
   the near-core region with radiative temperature gradient, following the
   prescription of \citet{Freytag1996} and \citet{Herwig1997}. An alternative
   implementation of CBM concerns a step-like functional form with the adiabatic
   temperature gradient --- a prescription that has most often been used in the
   binary community when addressing the mass discrepancy problem
   \citep[e.g.,][]{Guinan2000}. In this particular case, the near-core region is
   mixed instantaneously, mimicking the effect of convective penetration and
   hence implying a global increase in the size and mass of the convective
   core. The implementation of this penetration formalism
in {\sc mesa} is
   detailed in \citet{Michielsen2019} to which we refer for details. The extent
   of the CBM region is parametrized by the $f_{\rm ov}$ ($\alpha_{\rm ov}$)
   parameter in the case of 
the exponentially decaying and step-like profile,
   respectively. In the following, the two functional forms of the CBM and their
   effect on the mass discrepancy are discussed separately.
   
   \subsection{Single-star case scenario}
\label{Sect:SingleStarCase}

As outlined in Sect.~\ref{Sect: step-by-step}, the mass discrepancy is explored
and quantified relative to the reference model, which is detailed in
Table~\ref{Table:SingleStarCaseParameters} (the `RM solution' column). In this
model, ages and
convective core masses are determined under the assumption of the
minimum amount of near-core mixing, using stellar masses measured from binary
dynamics. We fail to reproduce the stellar positions in the Kiel diagram for
almost the entire sample, with the V578\,Mon and U\,Oph binary systems and the
lower mass companion star of V346\,Cen being the exceptions. Furthermore, we
enforce the dynamical mass values for the primary components of the V380\,Cyg
and V621\,Per systems because of large inconsistencies between their dynamical
mass measurements and positions of the stars in the Kiel diagram. Since we fail
to reproduce dynamical masses for the majority of the targets with our reference
model there is a strong indication of the mass discrepancy and the need for
increased core masses in these stars.

\begin{figure*}[t]
   \centering
   \includegraphics[width=8.7cm]{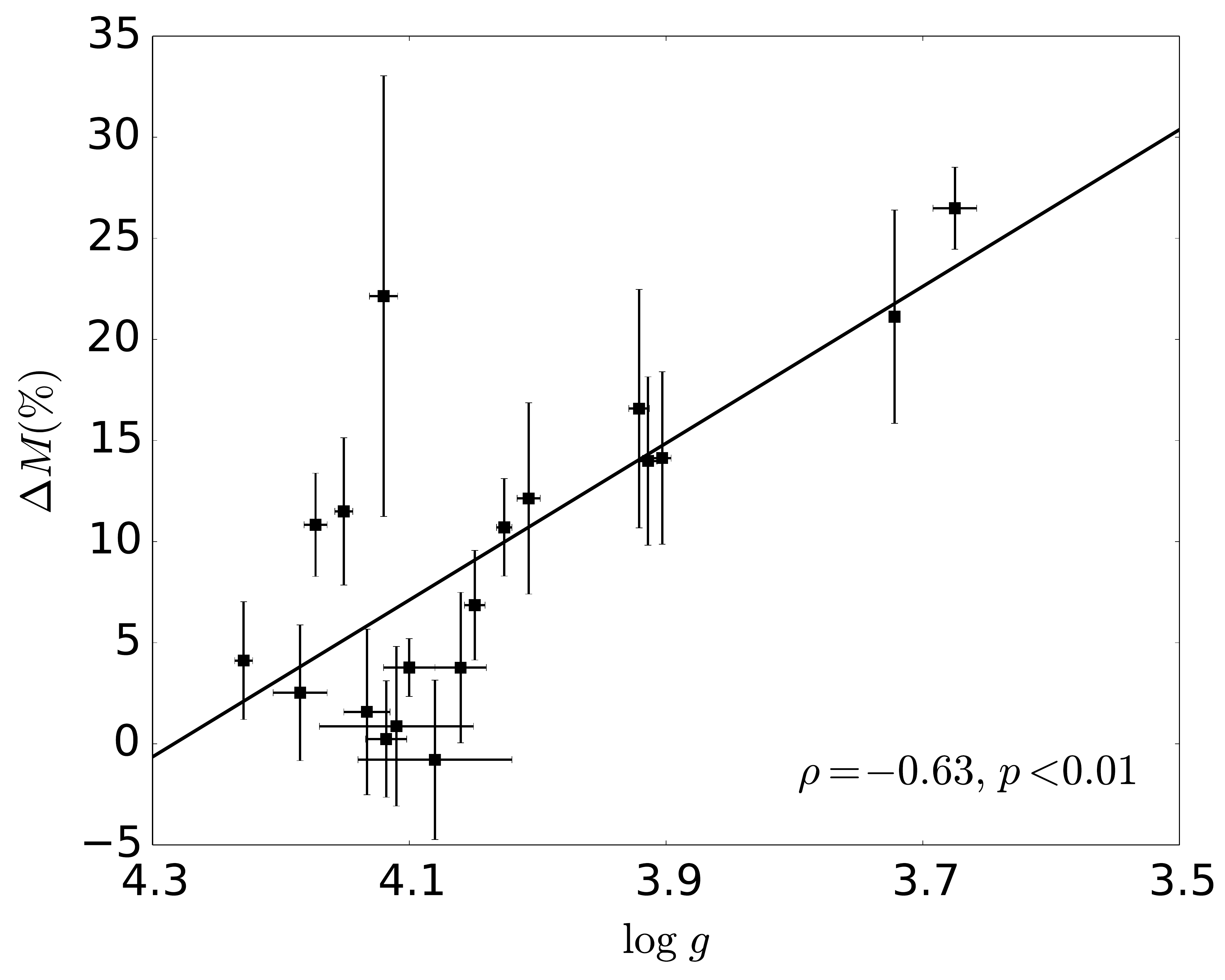}\hspace{5mm}
   \includegraphics[width=8.7cm]{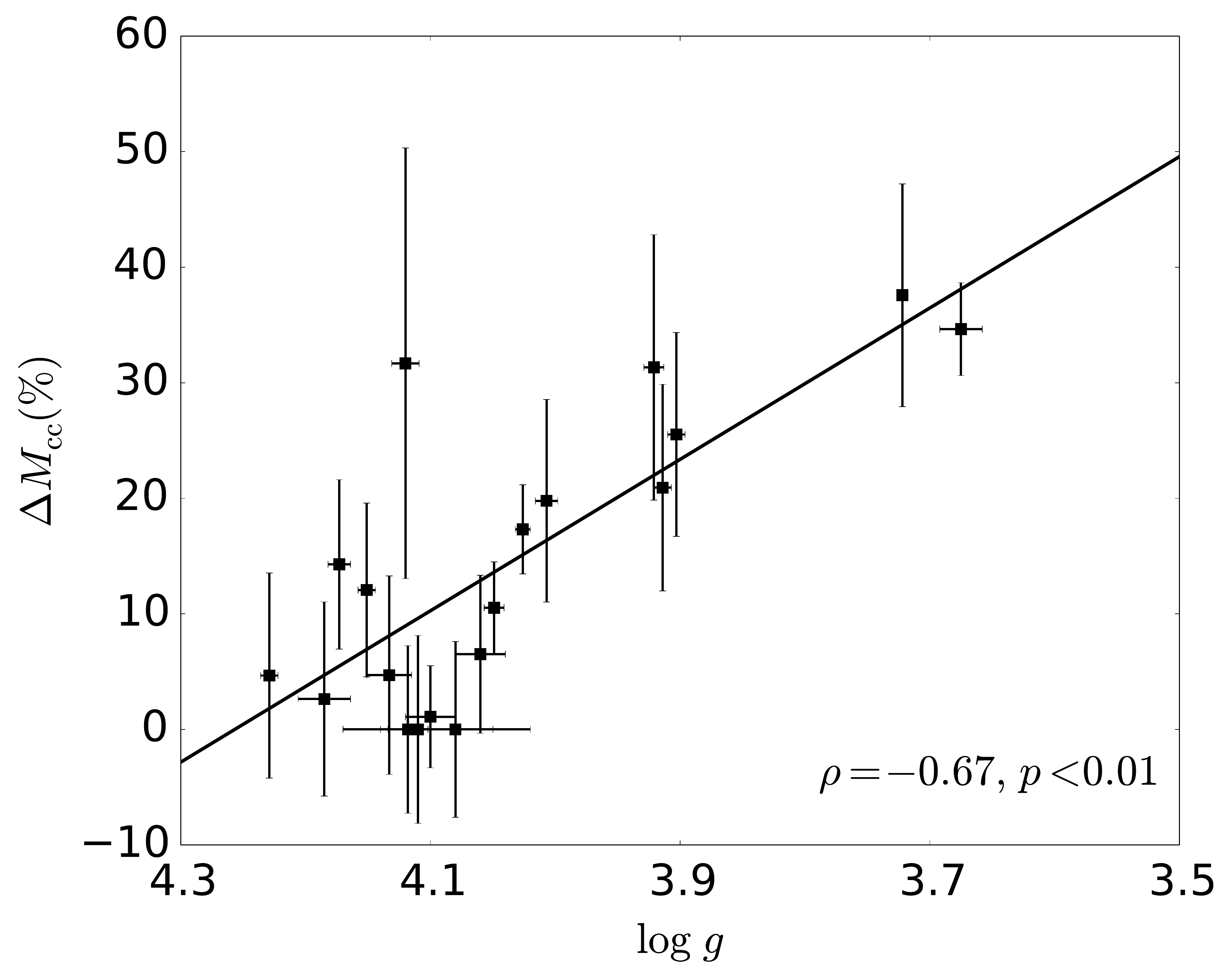}
   \includegraphics[width=8.7cm]{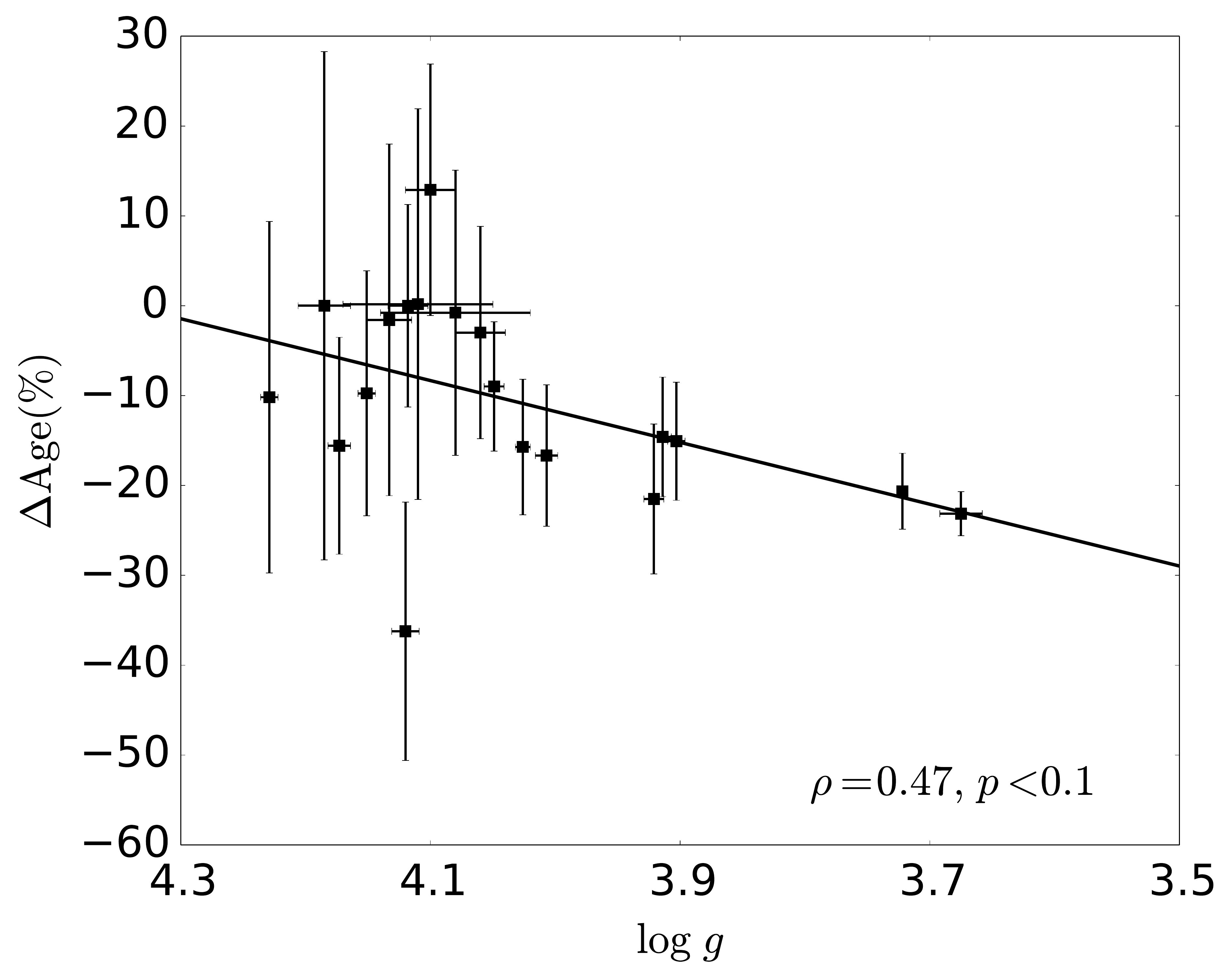}\hspace{5mm}
   \includegraphics[width=8.7cm]{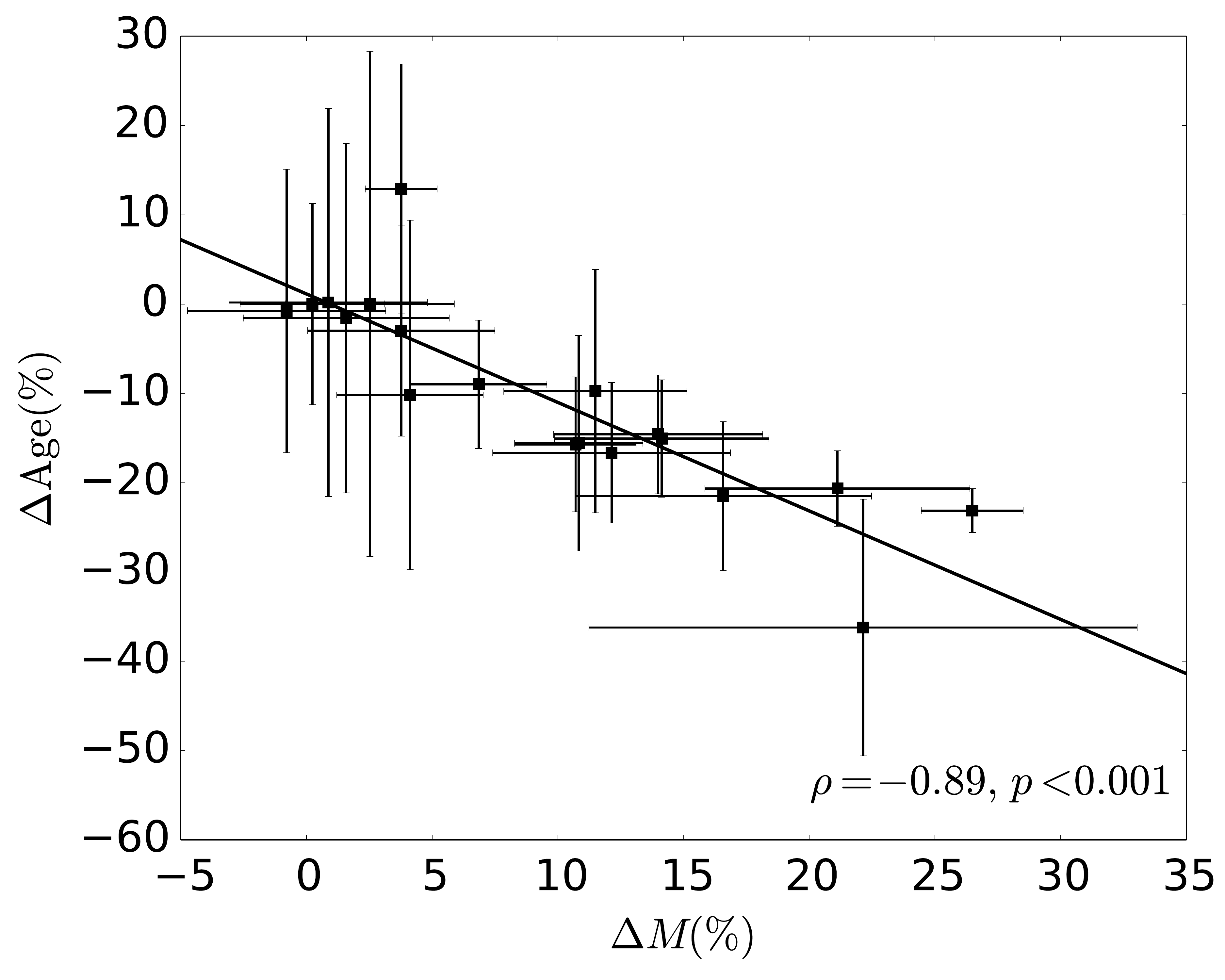}
      \caption{Stellar mass (top left), convective core mass (top right), and
        stellar age (bottom left) difference between the IM solution
        (cf. Sect~\ref{Sect: step-by-step} and
        Table~\ref{Table:SingleStarCaseParameters}, column IM solution) and the
        RM solution (cf. Sect~\ref{Sect: step-by-step} and
        Table~\ref{Table:SingleStarCaseParameters}, column RM solution) as a
        function of stellar surface gravity. The age difference as a function of
        the stellar mass difference between these two solutions is shown in the
        bottom right panel. All differences are expressed in percent relative to
        the RM solution values; the solid line depicts a linear fit in each of
        the four cases. The Spearman's rank
        correlation coefficient ($\rho$) and the $p-$value of 
null hypothesis that there is no correlation between the two sets of variables are
listed in each panel.}
         \label{Fig:Solution2}
   \end{figure*}

The IM solution (cf. Sect.~\ref{Sect: step-by-step}) quantifies the above
discrepancy, with results being detailed in
Table~\ref{Table:SingleStarCaseParameters} and visualized in
Fig.~\ref{Fig:Solution2}. We keep the amount of extra near-core mixing at the
minimal level and only require the ability to reproduce the position of the star
in the Kiel diagram. Satisfactory fits are obtained for all but two stars in our
sample, i.e., the more massive (primary) components of V380\,Cyg and
V621\,Per. These two stars are by far the most evolved targets in the entire
sample with the two lowest surface gravity values
(cf. Table~\ref{Table:ObservedParameters}). The two evident conclusions that can
be drawn from Fig.~\ref{Fig:Solution2} (top row) are: 1) the evolutionary mass
of the star is systematically overestimated compared to its dynamical mass and
the difference is cumulative with age; and 2) the mass of the convective core of
the star follows a similar trend as the stellar mass itself.

The conclusions above are also evidenced by the high absolute values of the
Spearman's rank correlation coefficients $\rho$ and the associated low
$p$-values of the null hypothesis that there is no correlation between the two
variables under consideration. Given that the mass of the convective core
of non-standard models is regulated by the amount and efficiency of near-core
mixing, we investigate this aspect in Sects.~\ref{Sect: DiffusiveOvershooting}
and \ref{Sect: ConvectivePenetration} in more detail.

It is worth mentioning that the increase of the initial (evolutionary) mass of
the star in the IM solution makes the star appear younger (bottom left panel of
Fig.~\ref{Fig:Solution2}). We find strong statistical evidence of
the effect getting more pronounced the larger the mass discrepancy is (bottom
right panel of Fig.~\ref{Fig:Solution2}). Finally, we also note that the above
mentioned (more massive) primary components of the V380\,Cyg and V621\,Per
systems are not shown in Fig.~\ref{Fig:Solution2} as their positions in the Kiel
diagram could not be reproduced in the IM solution. These findings are in line
with the conclusions by \citet{Tkachenko2014} who find that increasing the
initial stellar mass in evolutionary models alone is not sufficient to remedy
the mass discrepancy for the more evolved primary component of V380\,Cyg. Here,
we find the same result for the primary component of V621\,Per as well.

\subsubsection{CBM in the diffusive exponentially decaying approximation}
\label{Sect: DiffusiveOvershooting}

Given the strong link between the mass discrepancy and the mass of the
convective core of the star, we explore further a probable connection of the
effect with the amount and efficiency of the near-core mixing. The CBM solution
(cf. Sect.~\ref{Sect: step-by-step}) aims to reproduce stellar positions in the
Kiel diagram under the assumption of their dynamical masses while allowing for a
variable parameter $f_{\rm ov}$ in an exponentially decaying CBM profile.  The
CBM solution is different from the RM solution in that we relax this parameter,
letting the mass of the convective core increase due to the enhanced mixing in
the near-core regions, while keeping a tight constraint on the total mass.

\begin{figure*}
   \centering
   \includegraphics[width=8.7cm]{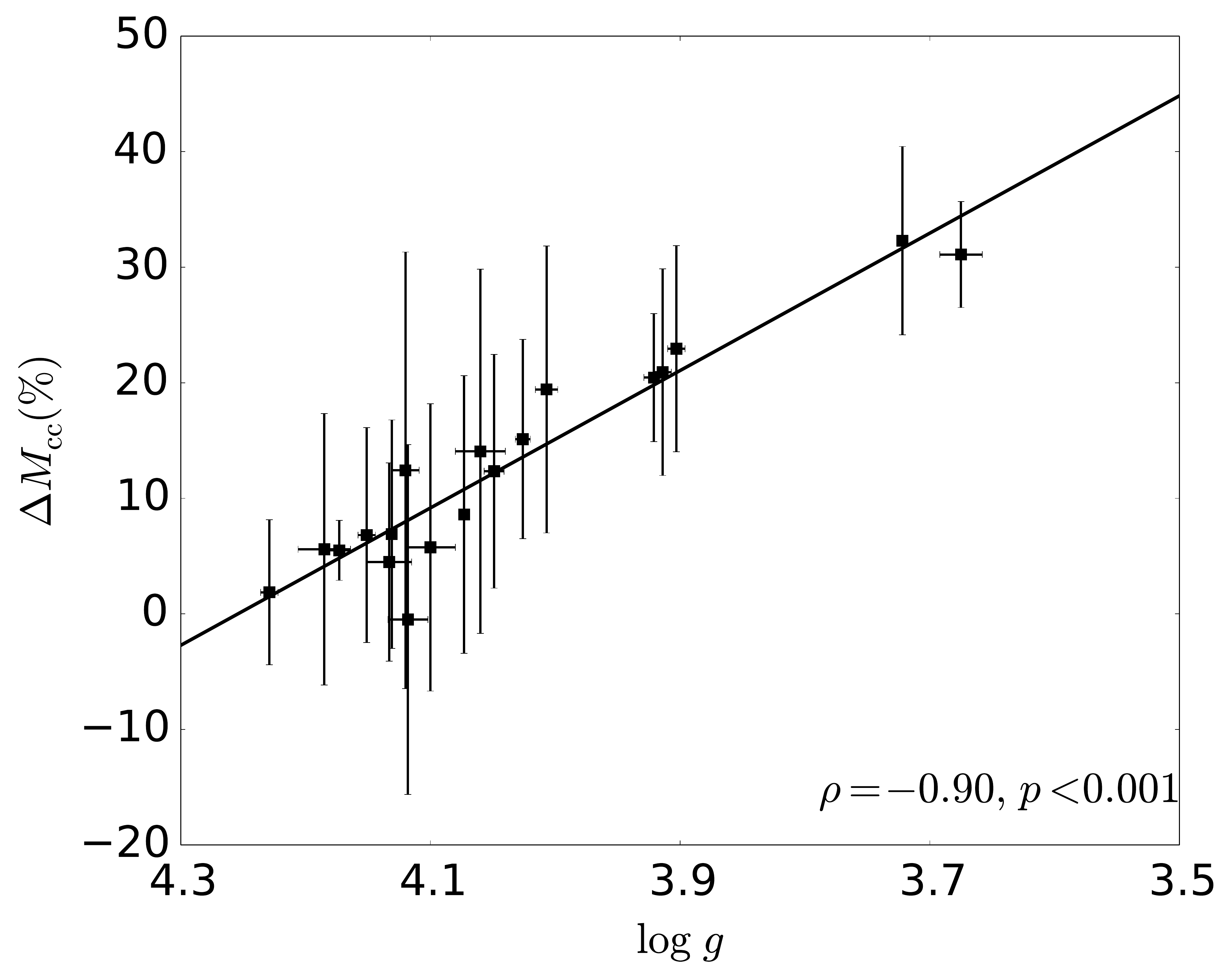}\hspace{5mm}
   \includegraphics[width=8.7cm]{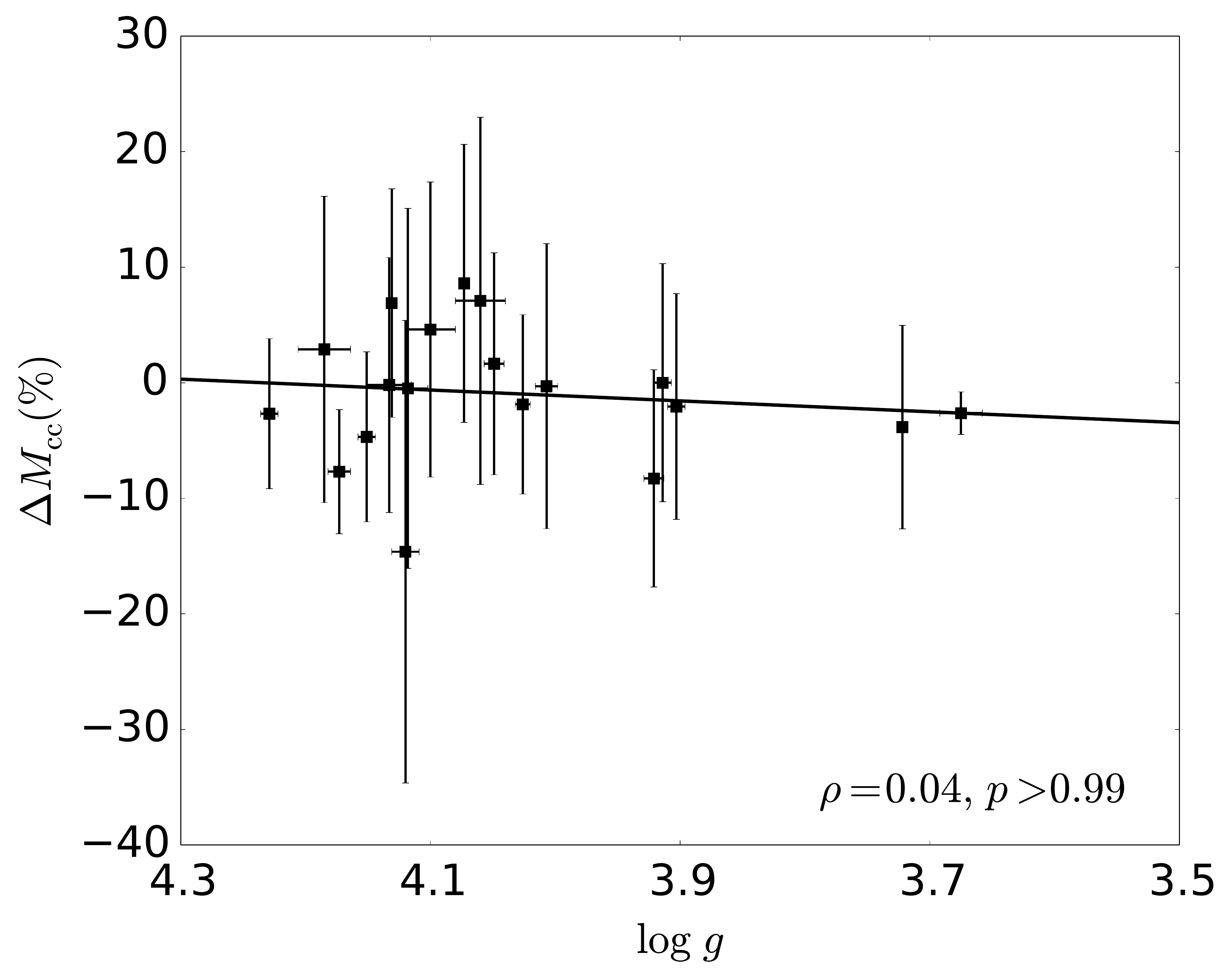}
   \includegraphics[width=8.7cm]{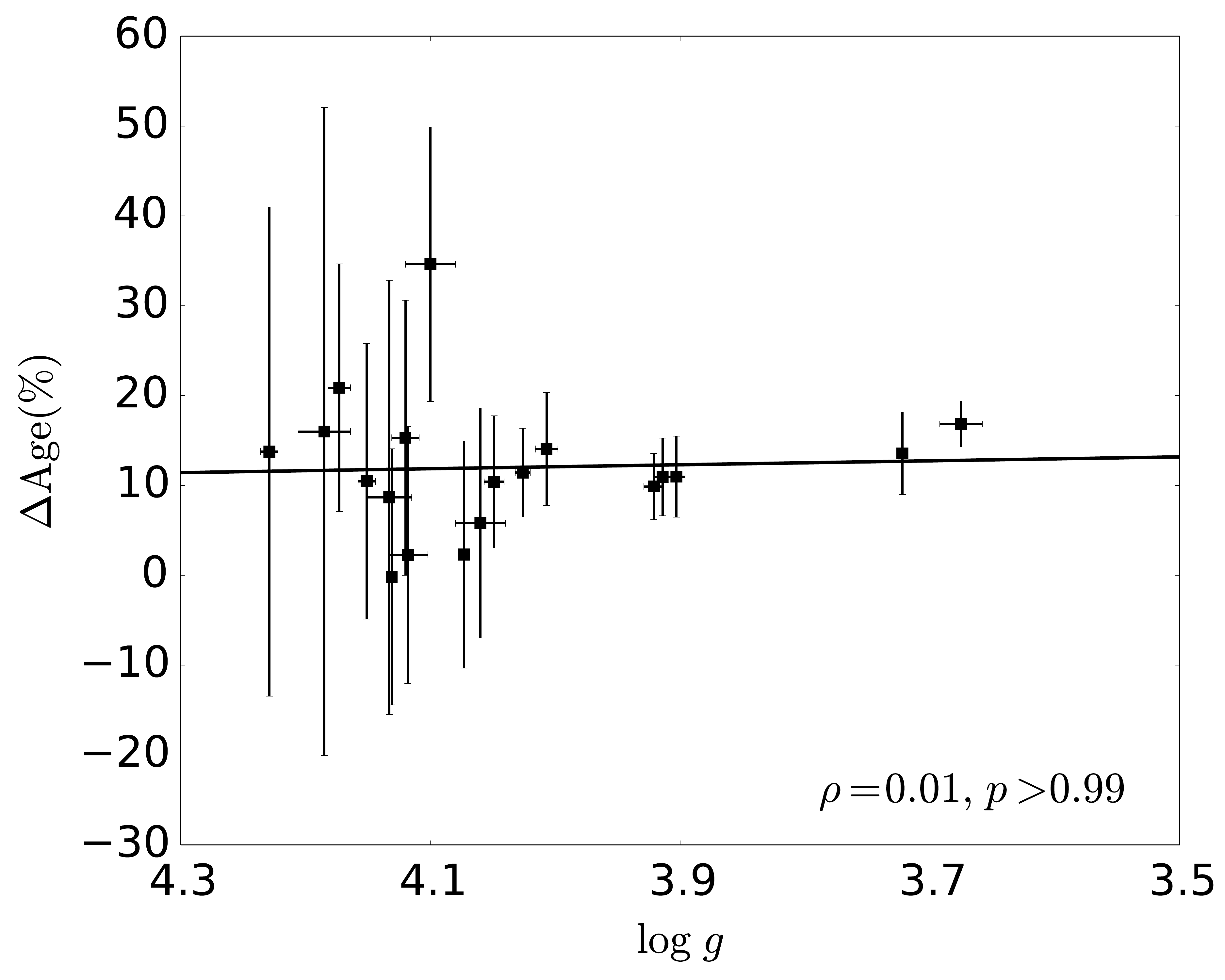}\hspace{5mm}
   \includegraphics[width=8.7cm]{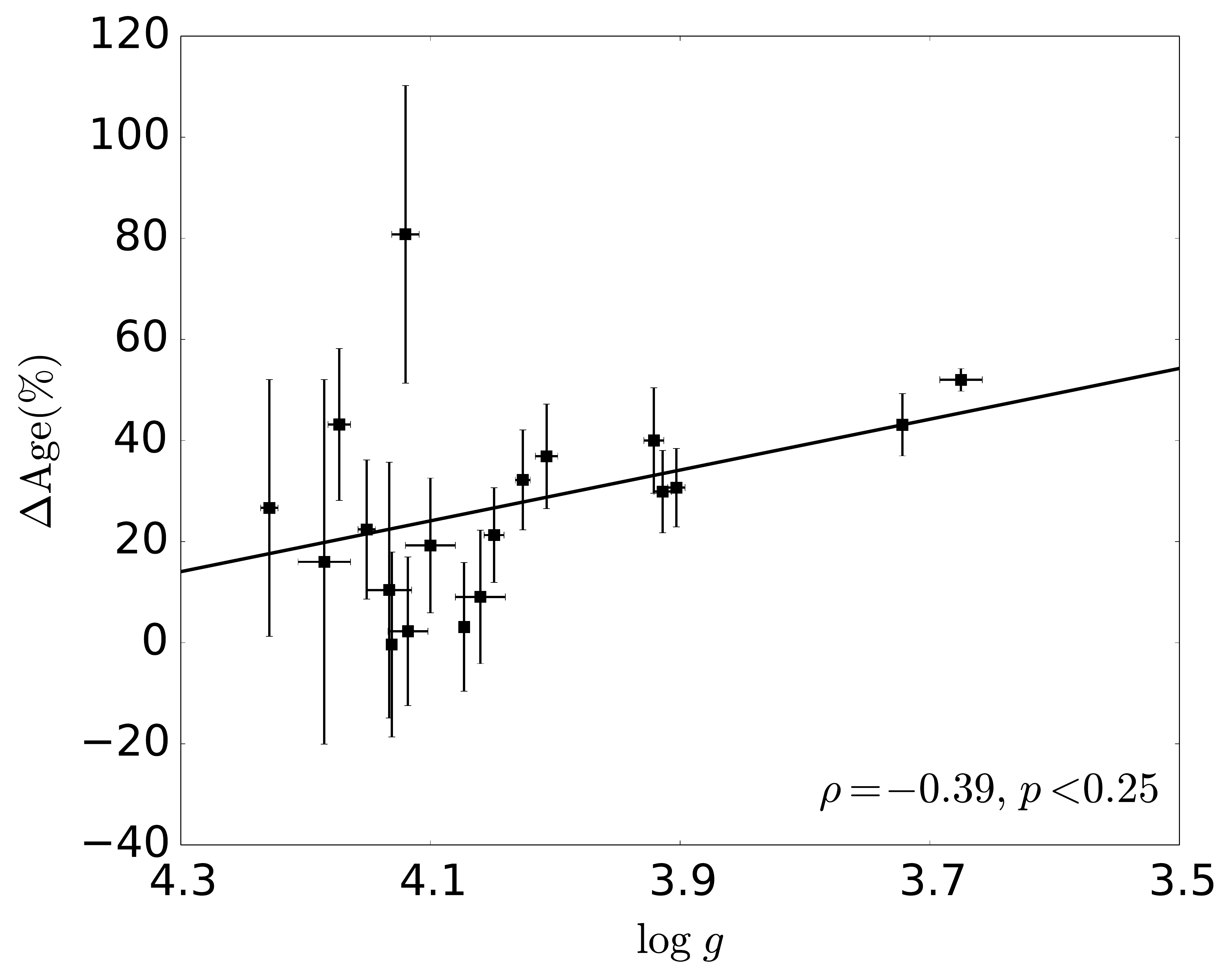}
      \caption{Convective core mass (top row) and age (bottom row) changes due
        to the inclusion of extra near-core mixing in the models
        (cf. Table~\ref{Table:SingleStarCaseParameters}, column CBM
        solution). Left and right columns represent comparisons against the RM
        solution (cf. Sect.~\ref{Sect: step-by-step} and
        Table~\ref{Table:SingleStarCaseParameters}, column RM solution) and the
        IM solution (cf. Sect.~\ref{Sect: step-by-step} and
        Table~\ref{Table:SingleStarCaseParameters}, column IM solution),
        respectively.} 
         \label{Fig:Solution3}
   \end{figure*}

The CBM solution is detailed in Table~\ref{Table:SingleStarCaseParameters} (the
`CBM solution' column) where the convergence towards the upper grid limit of the
$f_{\rm ov}$ parameter is immediately evident for almost the entire stellar
sample. This is not an unexpected outcome given the extra supply of fresh
hydrogen to the convective core and the associated increase in core mass. In
that sense, the CBM (to a certain extent) mimics the effect of varying
initial stellar mass, which is central to the stellar mass-CBM
degeneracy discussed in Sect.~\ref{Sect: ModelsGrid}
(cf. Fig.~\ref{Fig:MesaTracks}). Indeed, comparison with the RM solution
(reference model) values in the left column of Fig.~\ref{Fig:Solution3}
demonstrates that: 1) the mass of the convective core of the star increases with
the inclusion of extra CBM in stellar evolution models, and the increase is a
clear function of the surface gravity of the star (top left panel with $p$ below
0.001); and 2) all stars get systematically older with the inclusion of
CBM (bottom left panel), which is again expected given that more
near-core mixing implies an increase of the stellar lifetime on the
main sequence. This was already represented in the isochrone clouds in
\citet{Johnston2019b}.

The right column of Fig.~\ref{Fig:Solution3} compares the best CBM solution with the
one that assumes a minimum amount of the near-core mixing but allows for a
variable initial stellar mass (the IM solution; cf. Sect.~\ref{Sect:
  step-by-step} and the `IM solution' column in
Table~\ref{Table:SingleStarCaseParameters}). One can clearly see from the top
right panel in Fig.~\ref{Fig:Solution3} that there is no significant difference
between the obtained convective core masses ($p$ exceeding 0.99), strengthening
our claim that allowing the CBM parameter to vary is equivalent to
allowing the initial mass to vary. The $f_{\rm ov}$ parameter has a
non-negligible effect on stellar age as we find our sample to be systematically
older. There is also a weak indication of the age difference increasing
with the surface gravity of the star (bottom right panel with $p<$~0.25).

An important observation from Table~\ref{Table:SingleStarCaseParameters} (the
`CBM solution' column) is that the positions of stars in the Kiel diagram of
only about half of our sample stars are well reproduced with a variable CBM parameter.
The remaining half of the sample either requires more mass in the
convective core at a given age of the star or a younger age. In other words, the
mass discrepancy problem appears to be a combined effect of the age and the mass
of the convective core of the star in stellar evolution models. The above is demonstrated in our IM-CBM solution (cf. Sect.~\ref{Sect: step-by-step}) where we adopt the best fit value of the $f_{\rm ov}$ parameter from the previous
solution and relax stellar mass for systems that could not be fitted with
variable CBM alone. Increasing the initial mass of the star makes its core
slightly more massive and it makes the star younger at its (fixed) position in
the Kiel diagram. This particular combination allows us to reproduce the
atmospheric parameters for the remainder of the sample, including the primary
components of V621\,Per and V380\,Cyg --- the two most evolved stars in our
sample.
\begin{landscape}
\begin{table}
\begin{threeparttable}
\tiny
\tabcolsep 0.4mm \caption{Parameters for the single-star case scenario (cf. Sect.~\ref{Sect:SingleStarCase}) of the sample targets with error bars indicated in parentheses in terms of the last digit. The convective core mass is provided both in absolute (solar mass units) and relative (to the stellar mass) values. For each object, the first/second line corresponds to the primary/secondary component. Stars that do not have parameters for the IM-CBM solution could already be fitted in the CBM solution and did not require variable initial mass on top of the core-boundary mixing parameter. Detailed description of the individual solutions is provided in Sect.~\ref{Sect: step-by-step}.}
\begin{tabular}{lcccccccccccccccccccc}
\hline
 & \multicolumn{5}{c}{{\bf RM solution}}\rule{0pt}{9pt} & \multicolumn{5}{c}{{\bf IM solution}} & \multicolumn{5}{c}{{\bf CBM solution}} & \multicolumn{5}{c}{{\bf IM-CBM solution}}\\
Object/ & $M$ & $f_{\rm ov}$ & age & \multicolumn{2}{c}{$M_{\rm cc}$} & $M$ & $f_{\rm ov}$ & age & \multicolumn{2}{c}{$M_{\rm cc}$} & $M$ & $f_{\rm ov}$ & age & \multicolumn{2}{c}{$M_{\rm cc}$} & $M$ & $f_{\rm ov}$ & age & \multicolumn{2}{c}{$M_{\rm cc}$}\\
Parameter & (M$_{\odot}$) & ($H_{\rm p}$) & (Myr) & (M$_{\odot}$) & (\%) & (M$_{\odot}$) & ($H_{\rm p}$) & (Myr) & (M$_{\odot}$) & (\%) & (M$_{\odot}$) & ($H_{\rm p}$) & (Myr) & (M$_{\odot}$) & (\%) & (M$_{\odot}$) & ($H_{\rm p}$) & (Myr) & (M$_{\odot}$) & (\%)\\
\hline
\multirow{2}{*}{V578\,Mon}\rule{0pt}{9pt} & 14.55(9)$^{A}$ & \multirow{2}{*}{0.005} & 3.80(50) & 4.90(13) & 33.7(1.0) & 14.77(59)$^{A}$ & \multirow{2}{*}{0.005} & 3.74(55) & 5.13(40) & 34.7(4.3) & 14.55(9)$^{A}$ & 0.020(20) & 4.13(77) & 5.12(40) & 35.2(3.0) & \multicolumn{5}{c}{-----}\\
  & 10.30(6)$^{A}$ & & 5.0(1.0) & 3.04(7) & 29.5(8) & 10.55(34)$^{A}$ & & 5.0(1.0) &  3.12(22) & 29.6(3.1) & 10.30(6)$^{A}$ & 0.040(-40) & 5.8(1.5) &  3.21(35) & 31.2(3.6) & \multicolumn{5}{c}{-----}\vspace{3mm} \\

\multirow{2}{*}{V453\,Cyg} & 14.10(40)$^{B}$ & \multirow{2}{*}{0.005} & 10.9(3) & 3.22(8) & 22.8(1.3) & 16.80(70)$^{A}$ & \multirow{2}{*}{0.005} & 8.65(35) & 4.43(30) & 26.4(3.0) & 14.00(20)$^{B}$ & 0.040(-5) & 12.38(40) & 4.26(25) & 30.4(2.3) & 14.95(35)$^{A}$ & \multirow{2}{*}{0.040} & 11.00(40) & 4.82(20) & 32.2(2.2)\\
  & 11.15(15)$^{B}$ & & 10.8(4) & 2.78(6) & 24.9(9) & 12.38(50)$^{A}$ & & 9.00(75) & 3.33(23) & 26.9(3.1) & 11.11(15)$^{B}$ & 0.040(-15) & 12.32(55) & 3.32(34) & 29.9(3.5) & 11.91(57)$^{A}$ &  & 11.00(98) & 3.66(30) & 30.7(4.2)\vspace{3mm} \\

\multirow{2}{*}{V478\,Cyg} & 15.60(20)$^{B}$ & \multirow{2}{*}{0.005} & 7.40(20) & 4.54(7) & 29.1(8)& 17.60(60)$^{A}$ & \multirow{2}{*}{0.005} & 6.32(45) & 5.49(40) & 31.2(3.5) & 15.58(22)$^{B}$ & 0.040(-10) & 8.21(25) & 5.49(40) & 35.2(3.4) & 16.74(73)$^{A}$ & \multirow{2}{*}{0.040} & 7.47(52) & 6.05(50) & 36.1(4.8)\\
  & 15.10(10)$^{B}$ & & 7.83(25) & 4.27(9) & 28.3(8)& 17.20(60)$^{A}$ & & 6.65(45) & 5.36(37) & 31.2(3.3) & 15.22(22)$^{B}$ & 0.040(-10) & 8.69(25) & 5.25(37) & 34.5(3.0) & 16.39(71)$^{A}$ & & 7.87(55) & 5.97(50) & 36.4(4.9)\vspace{3mm}\\

\multirow{2}{*}{AH\,Cep} & 16.28(15)$^{B}$ & \multirow{2}{*}{0.005} & 4.90(20) & 5.51(9) & 33.8(9) & 17.30(40)$^{A}$ & \multirow{2}{*}{0.005} & 4.46(29) & 6.09(20) & 35.2(2.0) & 16.26(18)$^{A}$ & 0.040(-20) & 5.41(30) & 6.19(55) & 38.1(3.8) & \multicolumn{5}{c}{-----}\\
  & 13.80(10)$^{B}$ & & 7.00(17) & 4.10(5) & 29.7(6) & 15.20(30)$^{A}$ & & 5.90(50) & 4.81(13) & 31.6(1.6) & 13.79(14)$^{B}$ & 0.040(-15) & 7.80(30) & 4.72(35) & 34.2(2.9) & 14.86(70)$^{A}$ & 0.040 & 6.98(58) & 5.22(45) & 35.1(4.9)\vspace{3mm} \\

\multirow{2}{*}{V346\,Cen} & 12.19(15)$^{B}$ & \multirow{2}{*}{0.005} & 13.66(30) & 2.54(10) & 20.8(1.1) & 14.90(20)$^{A}$ & \multirow{2}{*}{0.005} & 10.50(15) & 3.42(2) & 23.0(4) & 12.02(14)$^{B}$ & 0.040(-2) & 15.96(18) & 3.33(6) & 27.7(8) & 13.76(40)$^{A}$ & 0.040 & 13.11(55) & 4.14(20) & 30.1(2.4)\\
  & 8.41(9)$^{A}$ & & 13.2(1.0) & 2.05(8) & 24.4(1.2)& 8.42(22)$^{A}$ & & 13.2(1.1) & 2.05(11) & 24.3(2.0) & 8.40(10)$^{A}$ & 0.005(+35) & 13.5(1.6) & 2.04(30) & 24.3(4.1) &\multicolumn{5}{c}{-----}\vspace{3mm}\\

\multirow{2}{*}{V573\,Car} & 15.95(40)$^{B}$ & \multirow{2}{*}{0.005} & 2.77(32) & 5.72(35) & 35.9(3.1) & 16.88(39)$^{A}$ & \multirow{2}{*}{0.005} & 2.50(20) & 6.41(25) & 38.0(2.4) & 15.80(30)$^{B}$ & 0.040(-20) & 3.06(28) & 6.11(40) & 38.7(3.3) & 16.62(45)$^{A}$ & 0.040 & 2.89(23) & 6.58(30) & 40.0(2.5)\\
  & 12.55(17)$^{B}$ & & 1.67(29) & 4.29(10) & 34.2(1.3) & 12.89(30)$^{A}$ & & 1.50(15) & 4.49(15) & 34.8(2.1) & 12.49(17)$^{A}$ & 0.040(-40) & 1.90(35) & 4.37(25) & 35.0(2.5) & \multicolumn{5}{c}{-----}\vspace{3mm}\\
  
\multirow{2}{*}{V1034\,Sco} & 17.13(10)$^{B}$ & \multirow{2}{*}{0.005} & 6.37(18) & 5.33(12) & 31.1(9) & 19.90(1.0)$^{A}$ & \multirow{2}{*}{0.005} & 5.00(50) & 7.00(60) & 35.2(5.0) & 17.15(13)$^{B}$ & 0.040(-6) & 7.00(15) & 6.42(27) & 37.4(1.9) & 18.97(77)$^{A}$ & 0.040 & 6.06(37) & 7.44(50) & 39.2(4.4)\\
  & 9.66(7)$^{B}$ & & 6.42(60) & 2.73(5) & 28.3(6) & 10.64(24)$^{A}$ & & 5.42(49) & 3.12(16) & 29.3(2.2) & 9.65(7)$^{B}$ & 0.040(-6) & 7.76(65) & 2.88(5) & 29.8(8) & 10.47(24)$^{A}$ & 0.040 & 6.51(58) & 3.29(15) & 31.4(2.2) \vspace{3mm}\\

\multirow{2}{*}{V380\,Cyg} & 11.43(19)$^{B,1}$ & \multirow{2}{*}{0.005} & 17.3(1.0) & 0.38(38) & 3.3(3.3) & 15.00(-5)$^B$ & \multirow{2}{*}{0.005} & 12.40(1.0) & 2.41(1) & 16.1(1) & 11.43(19)$^{B,1}$ & 0.005(5) & 17.3(1.0) & 0.38(38) & 3.3(3.3) & 15.00(50)$^{A,2}$ & 0.040 & 14.1(1) & 3.86(5) & 25.7(5)\\
  & 7.07(14)$^B$ & & 19.6(1.3) & 1.61(5) & 22.8(1.2) & 8.55(75)$^A$ & & 12.50(2.50) & 2.12(30) & 24.8(5.2) & 7.06(13)$^B$ & 0.040(30) & 22.6(2.7) & 1.81(30) & 25.6(4.8) & 8.32(75)$^A$ & 0.040 & 15.7(3.3) & 2.26(30) & 27.2(6.6) \vspace{3mm}\\

\multirow{2}{*}{CW\,Cep} & 13.01(6)$^A$ & \multirow{2}{*}{0.005} & 6.70(56) & 3.84(8) & 29.5(8) & 13.49(48)$^A$ & \multirow{2}{*}{0.005} & 6.50(56) & 4.09(25) & 30.3(3.1) & 13.01(7)$^A$ & 0.038(-38) & 7.09(65) & 4.38(60) & 33.7(4.8) & \multicolumn{5}{c}{-----}\\
  & 11.99(6)$^B$ & & 5.66(57) & 3.65(6) & 30.4(7) & 12.38(16)$^A$ & & 6.39(55) & 3.69(14) & 29.8(1.5) & 11.95(9)$^A$ & 0.040(-30) & 7.62(65) & 3.86(45) & 32.3(4.0) & \multicolumn{5}{c}{-----}\vspace{3mm} \\

\multirow{2}{*}{U\,Oph} & 5.10(2)$^A$ & \multirow{2}{*}{0.005} & 51.7(5.8) & 0.93(5) & 18.2(1.1) & 5.04(20)$^A$ & \multirow{2}{*}{0.005} & 51.3(5.8) & 0.93(5) & 18.5(1.7) & 5.10(5)$^A$ & 0.013(13) & 52.9(3.0) & 1.01(10) & 19.8(2.2) & \multicolumn{5}{c}{-----}\\
  & 4.60(2)$^A$ & & 57.5(7.5) & 0.87(5) & 18.9(1.2) & 4.63(18)$^A$ & & 57.6(10.0) & 0.87(5) & 18.8(1.9) & 4.60(5)$^A$ & 0.017(12) & 57.4(3.3) & 0.93(7) & 20.2(1.8) & \multicolumn{5}{c}{-----}\vspace{3mm} \\  

V621\,Per & 9.44(46)$^{B,1}$ & 0.005 & 21.8(1) & 1.55(75) & 16.4(9.2) & 12.5(1)$^B$ & 0.005 & 14.9(4) & 2.13(5) & 17.0(6) & 9.44(46)$^{B,1}$ & 0.030(5) & 24.1(1) & 2.08(15) & 22.0(2.8) & 11.81(30)$^A$ & 0.030 & 17.5(7) & 2.80(12) & 23.7(1.7)\vspace{1mm}\\

\hline
\end{tabular}
\begin{tablenotes}
      \tiny
      \item $^{A/B}$ within/outside error box; $^1$ dynamical mass was enforced; $^2$ maximum amount of the core-boundary mixing is assumed (see Sect.~\ref{Sect:V380Cyg})
    \end{tablenotes}
\label{Table:SingleStarCaseParameters}
\end{threeparttable}
\end{table}
\end{landscape}

\begin{figure*}
   \centering
   \includegraphics[width=8.7cm]{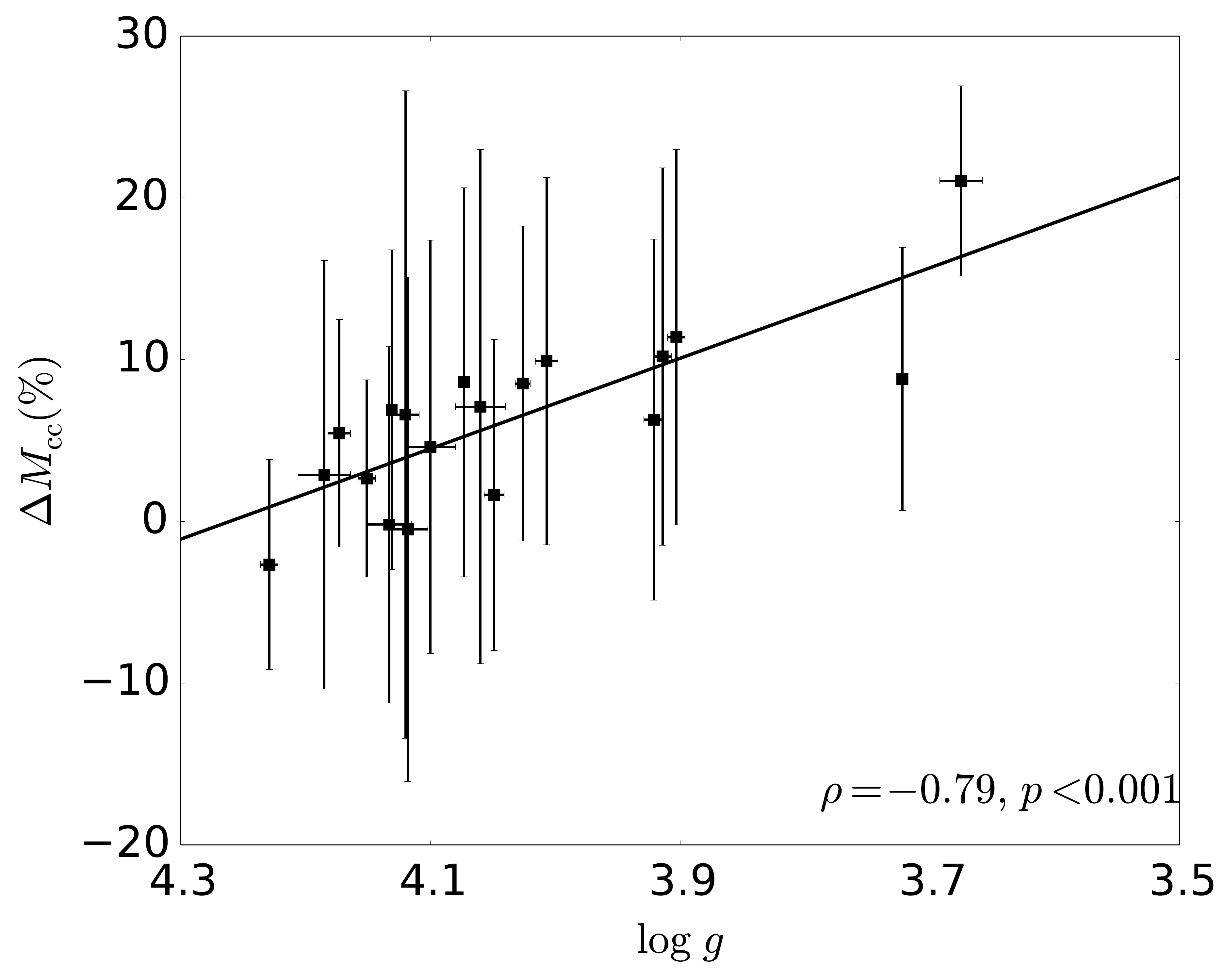}\hspace{5mm}
   \includegraphics[width=8.7cm]{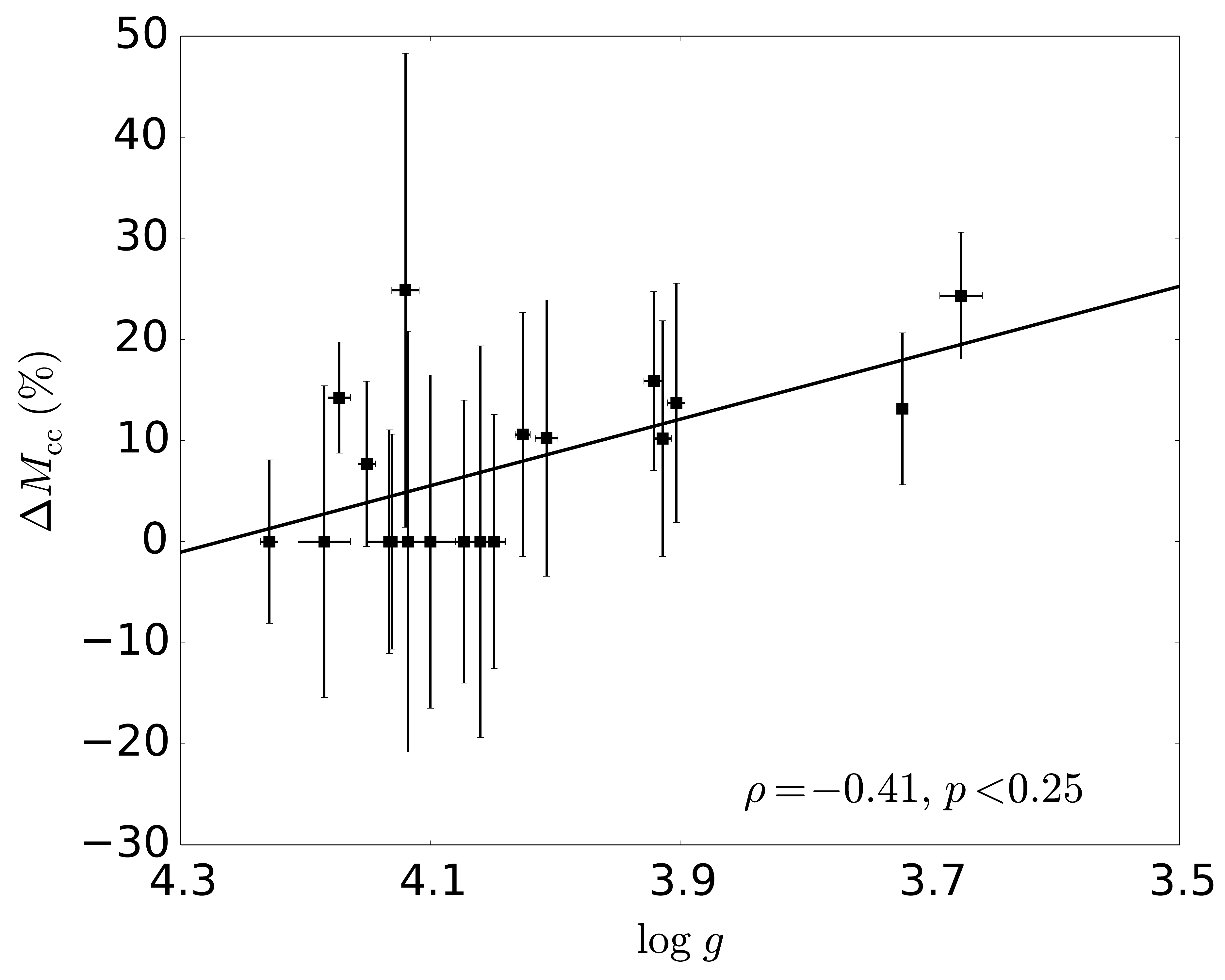}
   \includegraphics[width=8.7cm]{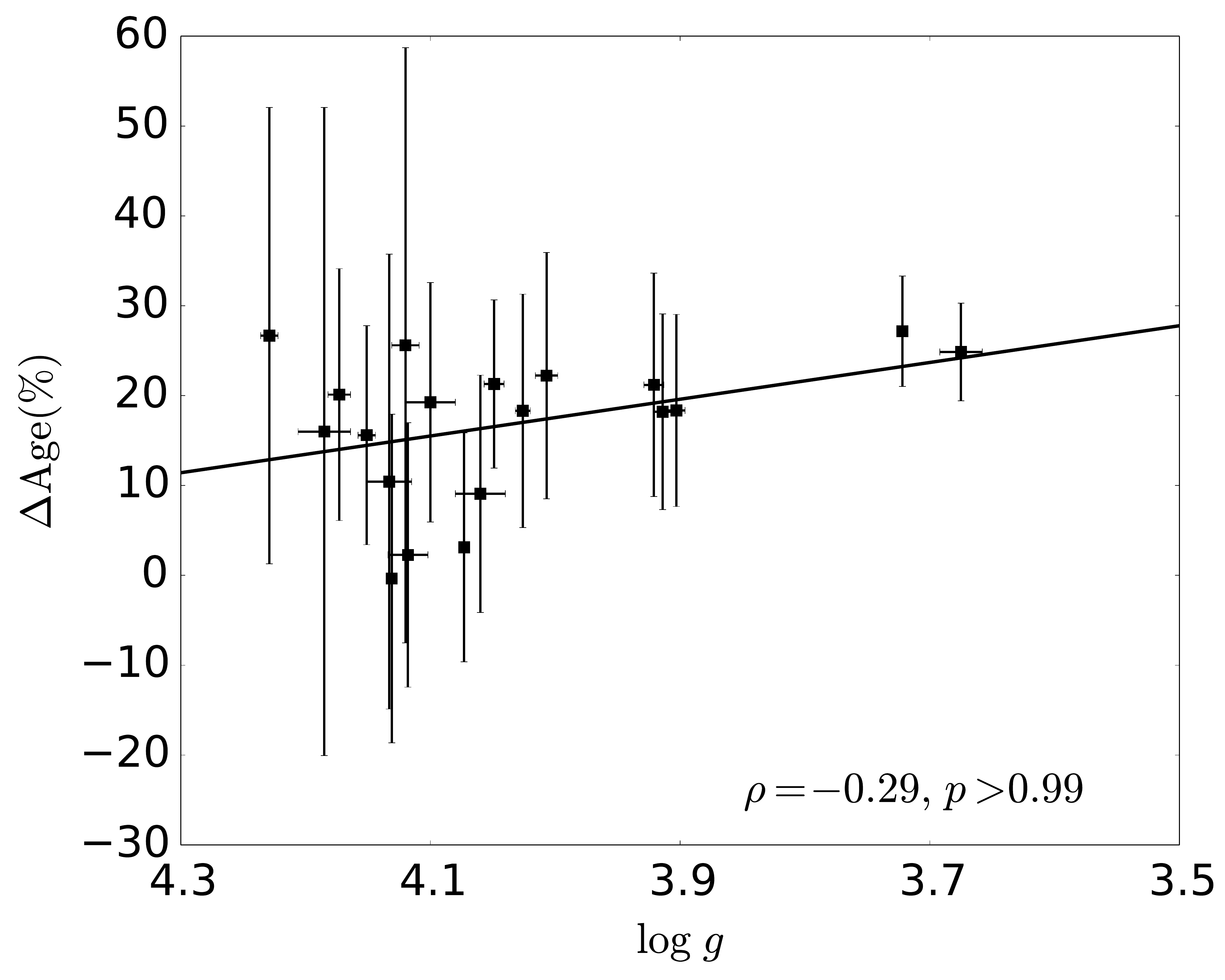}\hspace{5mm}
   \includegraphics[width=8.7cm]{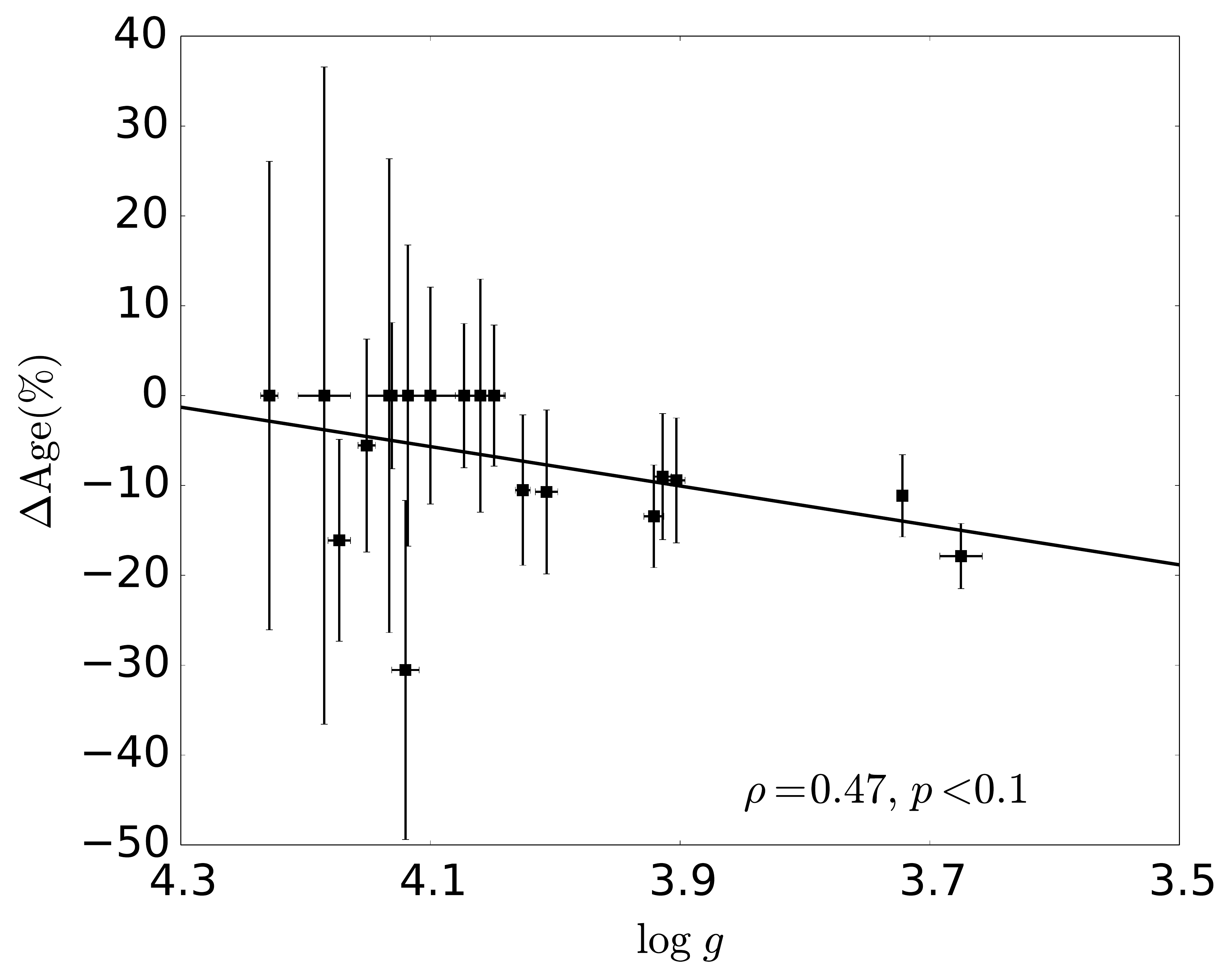}
      \caption{Same as Figure~\ref{Fig:Solution3} but for the IM-CBM solution
        (cf. Sect.~\ref{Sect: step-by-step} and
        Table~\ref{Table:SingleStarCaseParameters}, column IM-CBM
        solution). Left and right columns represent comparison against the IM
        (cf. Table~\ref{Table:SingleStarCaseParameters}, column IM solution) and
        CBM (cf. Table~\ref{Table:SingleStarCaseParameters}, column CBM
        solution) solutions, respectively.} 
         \label{Fig:Solution4}
   \end{figure*}

%\noindent
Our solution is detailed in Table~\ref{Table:SingleStarCaseParameters} (the
`IM-CBM solution' column) and is compared with the IM and CBM solutions in the
left and right columns of Fig.~\ref{Fig:Solution4}, respectively. The primary
components of V621\,Per and V380\,Cyg are not included in the comparison because
the IM solution did not reveal satisfactory fits for both stars and the mass
discrepancy could not be quantified for them. One can see that the convective
core mass has generally increased by $\sim$5-10\% (top left panel) while the
stars have generally become older (bottom left panel) compared to the IM
solution. One also notices a strong negative correlation between the increase of
the convective core mass and surface gravity of the star ($p<0.001$), while the
relationship for the age difference is not statistically significant
($p>0.99$). At the same time, the sample is generally younger than the CBM
solution (bottom right panel), yet the increase in the convective core mass is
evident (top right panel). A noteworthy feature in the right panels of
Fig.~\ref{Fig:Solution4} is the two populations of stars --- those clustering at
zero value on the ordinate axis and rest of the sample. This bimodality is
simply the result of adopting the exact parameters from the CBM solution for
stars whose atmospheric properties and dynamical masses are reproduced with
extra amount of the near-core mixing and did not require any further change of
the initial mass in stellar evolution models. In other words, these are the
systems that no longer show the mass discrepancy after introducing the
exponentially decaying CBM in a diffusive approximation with the free parameter
$f_{\rm ov}$.

\begin{figure*}
   \centering
   \includegraphics[width=8.7cm]{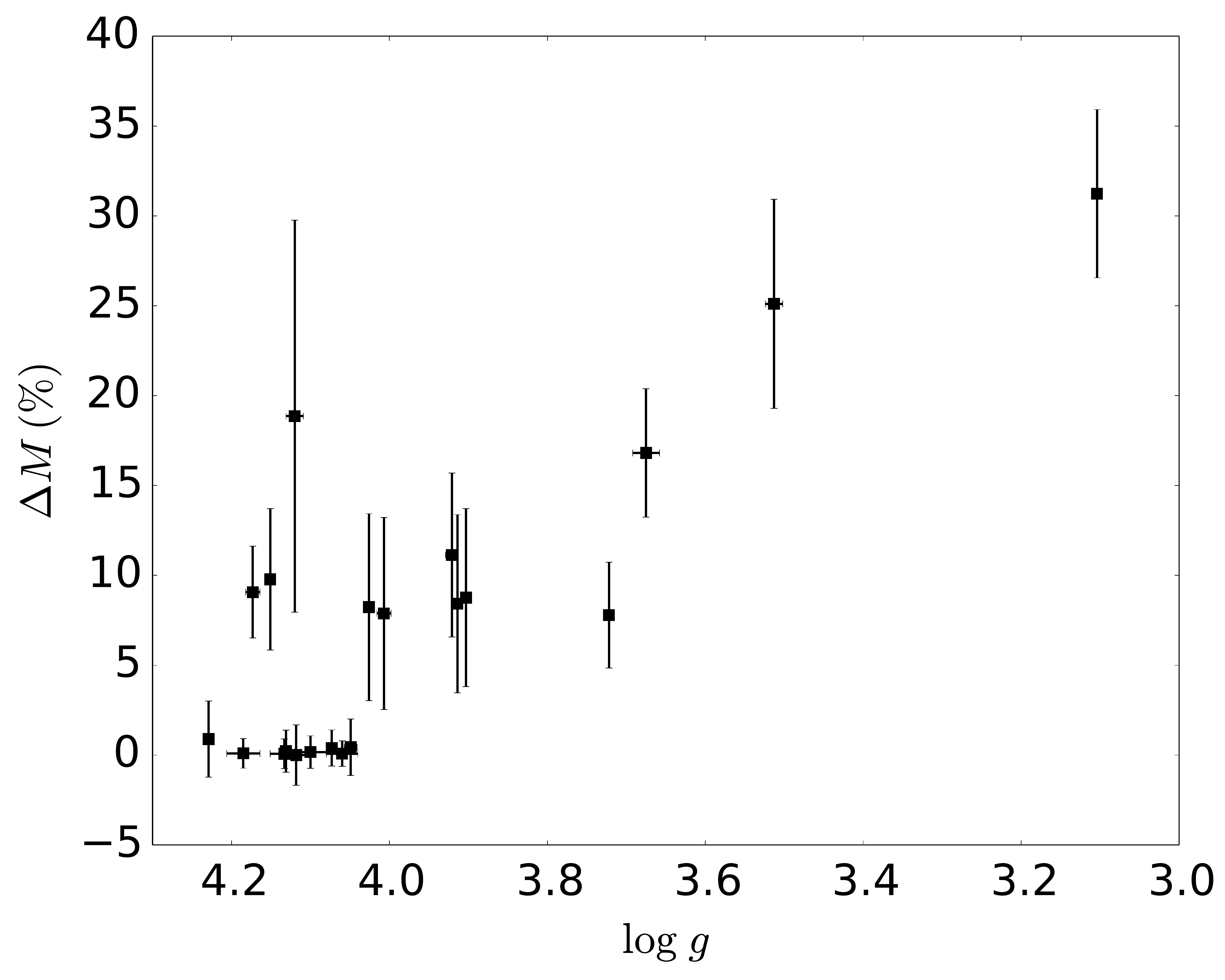}\hspace{5mm}
   \includegraphics[width=8.7cm]{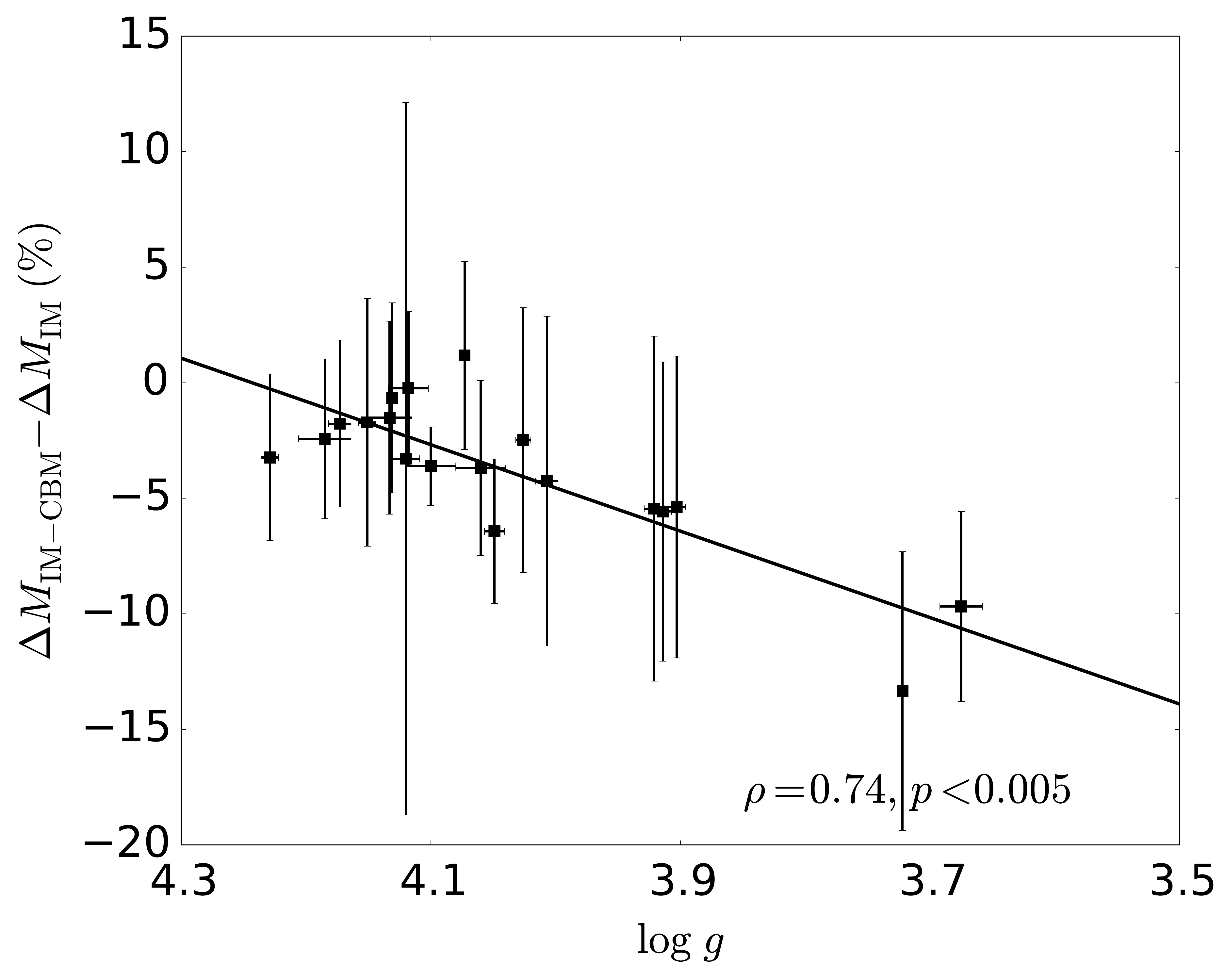}
      \caption{{\bf Left:} The remaining mass discrepancy after introducing an
        exponentially decaying CBM profile with radiative
        temperature gradient (the IM-CBM solution, cf. Sect.~\ref{Sect:
          step-by-step} and respective column in
        Table~\ref{Table:SingleStarCaseParameters}), defined as percentage of
        the dynamical mass of the star. {\bf Right:} Percentage difference
        between the mass discrepancy after (the IM-CBM solution) and before (the
        IM solution) the inclusion of CBM in the models.} 
         \label{Fig:Solution4_MassDiscrepancy}
   \end{figure*}

Finally, Fig.~\ref{Fig:Solution4_MassDiscrepancy} (left panel) shows the mass
discrepancy as a function of the surface gravity of the star, where the
discrepancy is determined with respect to the dynamical mass of the
star. Similar to the previous case, two stellar populations are evident: 1)
stars clustering at the zero mass discrepancy have positions in the Kiel diagram
that are well reproduced by adopting their dynamical masses in combination with
extra near-core mixing (cf.\ CBM solution); and 2)
the rest of the sample requires higher evolutionary masses in addition to the
CBM to reproduce the atmospheric properties of the respective
stars. For this latter sub-sample, the mass discrepancy amounts to some 10\% on
average with no statistically significant dependence on the surface gravity of
the star. The two most evolved stars in the sample are the primary components of
V621\,Per and V380\,Cyg, which are also included in
Fig.~\ref{Fig:Solution4_MassDiscrepancy} and show a mass discrepancy $\Delta M
\sim$25\% and $\sim$30\%, respectively. The mass discrepancy for the other 19
stars is reduced compared to the case of minimum amount of the near-core mixing
(cf.\ the IM solution), which is evidenced by the right panel of
Fig.~\ref{Fig:Solution4_MassDiscrepancy}. Just as the mass discrepancy itself,
the reduction is the largest for more evolved stars, i.e. those with the lowest
measured surface gravities.

\subsubsection{Step-like CBM with instantaneous mixing}
\label{Sect: ConvectivePenetration}

\begin{figure*}
   \centering
   \includegraphics[width=8.7cm]{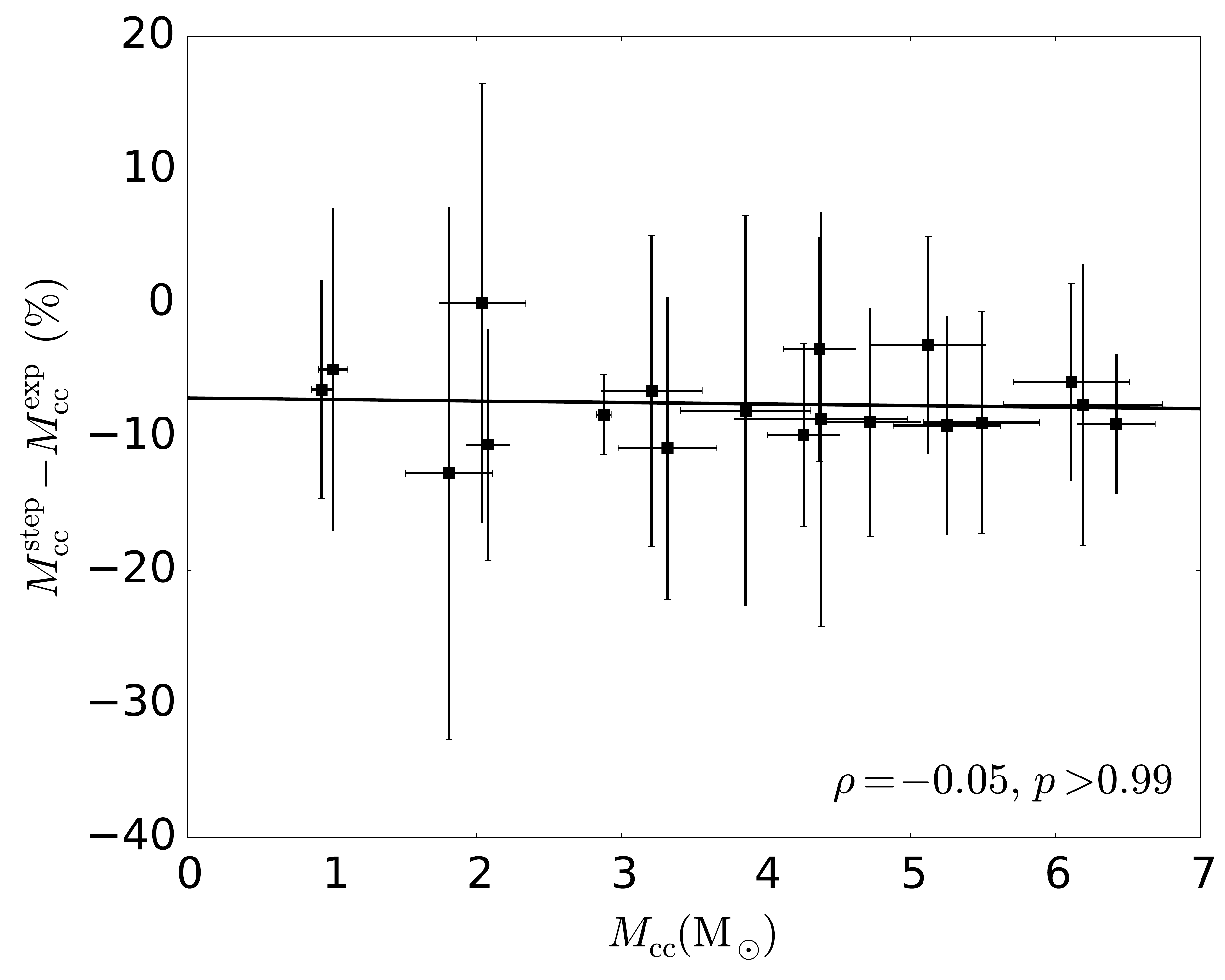}\hspace{5mm}
   \includegraphics[width=8.7cm]{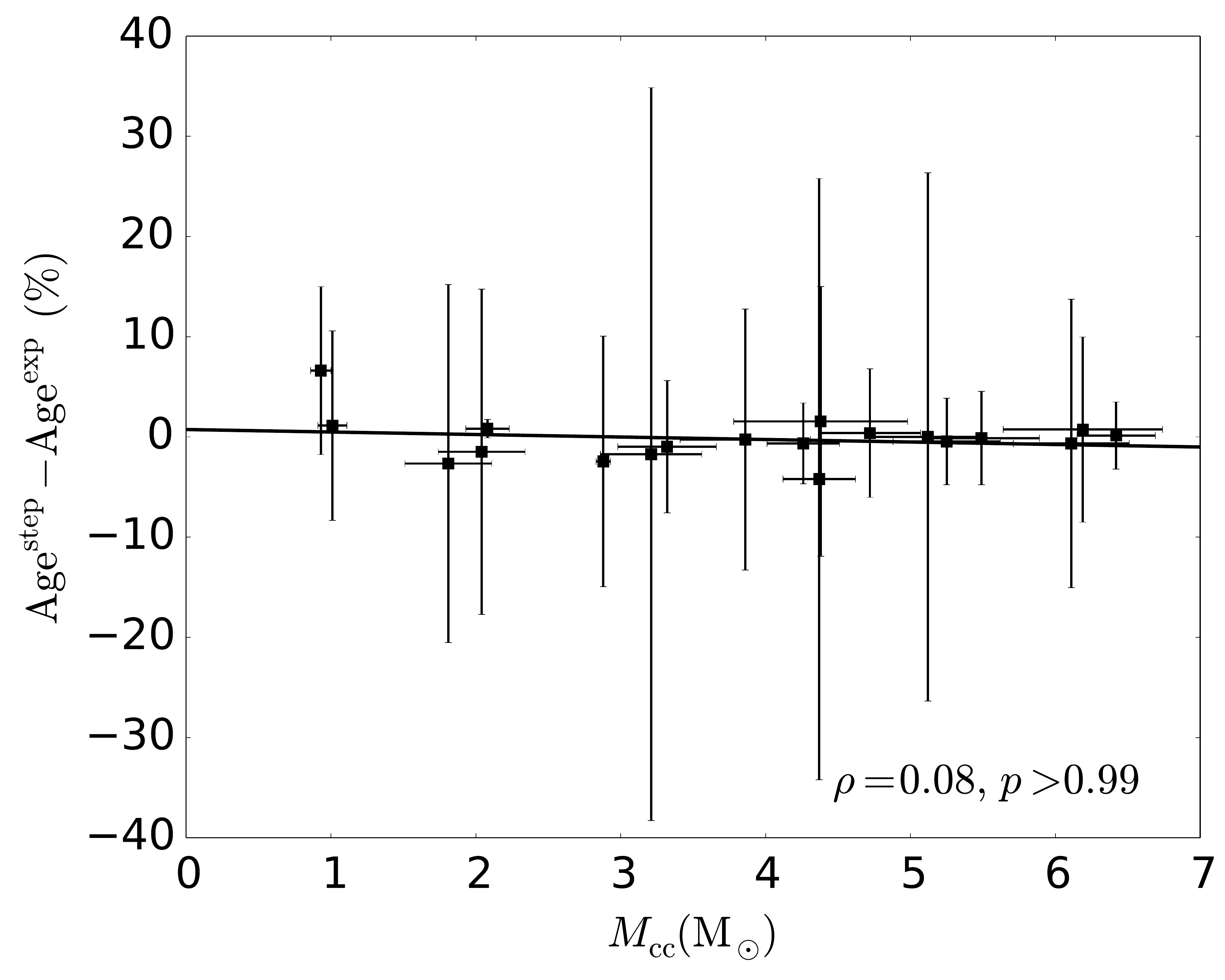}
      \caption{Relative change in the convective core mass (left) and age
        (right) of the star between solutions assuming a step-like and an
        exponentially decaying functional form of the CBM. The Y-axes
        are in percent relative to the value dictated by the solution with the
        exponentially decaying CBM.} 
         \label{Fig:StepVsExponentialOvershoot}
\end{figure*}

   Subsequently, we test a different functional form of the CBM, having a different temperature gradient in that zone. More precisely, the
   exponentially decaying formalism is replaced with a step function with the
   temperature gradient in the near-core region also changed from radiative to
   fully adiabatic \citep[i.e. {\it convective penetration};][]{Zahn1991}. The
   results corresponding to this form of CBM are summarized in
   Table~\ref{Table:DeterminedParametersSolution5}, while
   Fig.~\ref{Fig:StepVsExponentialOvershoot} shows a comparison between the two
   functional forms of the CBM. Left and right panels show the change of the
   convective core mass and age, respectively, where the parameter in question
   corresponding to the exponential functional form is subtracted from the one
   based on the step formalism. One can see that there is no statistically
   significant change in the age of the stellar sample while the mass of the
   convective core is consistently lower by about 5-10\%. As evidenced by
   $p>0.99$, neither the mass of the convective core nor the age of the star
   show any significant correlation with stellar surface gravity.

Similarly to the case of the exponentially decaying CBM, we reach the upper grid limit
for the $\alpha_{\rm ov}$ parameter in the majority of cases, as evidenced from
Table~\ref{Table:DeterminedParametersSolution5}. The two parameters, $f_{\rm
  ov}$ and $\alpha_{\rm ov}$, differ by a factor of ten for almost all stars in
the sample (the IM solution column in
Tables~\ref{Table:SingleStarCaseParameters} and
\ref{Table:DeterminedParametersSolution5}). The fact that we do not reach the
same mass of the convective core in the case of the $\alpha_{\rm ov}$ parameter
suggests a somewhat larger conversion factor between the parametrizations, when
interpreted  in terms of efficiency of the mixing as to increasing the mass of
the convective core. This is fully consistent with 
\citet{Claret2017}, who found the scaling $\alpha_{\rm
  ov}$/$f_{\rm ov}=11.36\pm0.22$ based on their study of 56 individual stellar
components of binary systems, while earlier studies by \citet{Moravveji2016} and
\citet{Valle2017} suggested $\alpha_{\rm ov}$/$f_{\rm ov}\approx$13
and 12, respectively.

\subsection{Binary equal age-case scenario}
\label{Sect:BinaryEqualAgeCase}
So far, our results have been obtained under the assumption that the individual
stellar components are single stars with their masses and surface gravities
being obtained in a model-independent way. In reality, they are members of
binary star systems which puts further strong constraints on stellar evolution
models, namely that both components of a binary system should have the same
age.

In this section, we repeat the sequence of analyses outlined in Sect.~\ref{Sect:
  step-by-step} for the single-star case scenario in
Sect~\ref{Sect:SingleStarCase}, but also forcing an equal age condition for
both stellar components of each binary system in the sample. The obtained
results are summarized in Table~\ref{Table:BinaryStarCaseParameters} in columns
designated RM, IM, CBM, and IM-CBM solutions and in
Figures~\ref{Fig:Solution2_Binary}--\ref{Fig:Solution4_Binary_MassDiscrepancy}. 
The distributions look somewhat  different from those for the single star-case
scenario due to some systems showing
considerably different ages for their individual stellar components (e.g.,
AH\,Cep in Table~\ref{Table:SingleStarCaseParameters}, the RM solution
column). Enforcing the equal-age condition for these systems propagates into
the other free parameters in the fitting, like initial mass and amount of near-core
mixing. This results in unsatisfactory fits for a total of four stellar
components
(primary/secondary component of the V380\,Cyg/V346\,Cen system and both
components of V573\,Car) in our ultimate IM-CBM
solution (cf.\ Table~\ref{Table:BinaryStarCaseParameters}) 
compared to none in
the single star-case
scenario. 
Figures~\ref{Fig:Solution2_Binary}--\ref{Fig:Solution4_Binary_MassDiscrepancy}
confirm our main conclusions from Sect.~\ref{Sect:SingleStarCase} that: 1) the
(convective core) mass discrepancy tends to increase as the surface gravity of
the star decreases; 2) the mass discrepancy can be partially accounted for by
increasing the amount of the near-core mixing and the mass of the convective
core; and 3) because of the link between the mass discrepancy and stellar age,
the near-core mixing should not be increased infinitely as it makes the star
older with increasing convective core mass. For this reason, we have limited the
$f_{\rm ov}$ and $\alpha_{\rm ov}$ 
values to those covered from asteroseismic inference of single
pulsators in the mass regime under study \citep[for both convective penetration
and diffusive exponential CBM;][]{Aerts2020}.

\begin{table}
\begin{threeparttable}
\small
\tabcolsep 0.9mm \caption{Same as the CBM solution in Table~\ref{Table:SingleStarCaseParameters} but for the step functional form of the core-boundary mixing with adiabatic temperature gradient in the mixing region.}
\label{Table:DeterminedParametersSolution5}      % is used to refer this table in the text
\centering                          % used for centering table
\begin{tabular}{l c c c c c}        % centered columns (4 columns)
\toprule
Object/ & $M$ & $\alpha_{\rm ov}$ & age & \multicolumn{2}{c}{$M_{\rm cc}$} \\
Parameter & (M$_{\odot}$) & ($H_{\rm p}$) & (Myr) & (M$_{\odot}$) & (\%)\\   
\midrule                        % inserts single horizontal line
  \multirow{2}{*}{V578\,Mon}\rule{0pt}{9pt} & 14.55(9)$^A$ & 0.40(-40) & 4.13(77) & 4.96(12) & 34.1(1.0)\\
  & 10.30(6)$^A$ & 0.40(-40) & 5.7(1.5) &  3.00(13) & 29.1(1.5)\vspace{1mm} \\

  \multirow{2}{*}{V453\,Cyg} & 14.05(15)$^B$ & 0.40(-6) & 12.30(30) & 3.84(15) & 27.3(1.4)\\
  & 11.12(15)$^B$ & 0.40(-15) & 12.20(60) & 2.96(16) & 26.6(1.8)\vspace{1mm} \\
  
  \multirow{2}{*}{V478\,Cyg} & 15.61(21)$^B$ & 0.40(-10) & 8.20(29) & 5.00(22) & 32.0(1.9)\\
  & 15.25(22)$^B$ & 0.40(-10) & 8.65(28) & 4.77(22) & 31.3(1.9)\vspace{1mm} \\

  \multirow{2}{*}{AH\,Cep} & 16.27(18)$^B$ & 0.40(-25) & 5.45(40) & 5.72(35) & 35.2(2.5)\\
  & 13.80(14)$^B$ & 0.40(-15) & 7.83(40) & 4.30(20) & 31.2(1.7)\vspace{1mm} \\
  
  \multirow{2}{*}{V346\,Cen} & 12.00(5)$^B$ & 0.05(5) & 15.80(5) & 0.00(1) & 0.0(1)\\
  & 8.41(10)$^A$ & 0.05(+35) & 13.3(1.5) & 2.04(15) & 24.3(2.1)\vspace{1mm}\\
  
  \multirow{2}{*}{V573\,Car} & 15.86(28)$^B$ & 0.40(-30) & 3.04(34) & 5.75(21) & 36.2(2.1)\\
  & 12.52(17)$^B$ & 0.20(20) & 1.82(45) & 4.22(27) & 33.7(3.3)\vspace{1mm} \\
  
  \multirow{2}{*}{V1034\,Sco} & 17.17(13)$^B$ & 0.40(-10) & 7.01(18) & 5.84(21) & 34.0(1.5)\\
  & 9.66(7)$^B$ & 0.40(-25) & 7.57(72) & 2.64(7) & 27.3(1.0)\vspace{1mm} \\
  
  \multirow{2}{*}{V380\,Cyg} & 11.43(19)$^{B,1}$ & 0.20(5) & 17.9(2) & 2.00(15) & 17.5(1.6)\\
  & 7.07(14)$^B$ & 0.40(35) & 22.0(3.0) & 1.58(20) & 22.3(3.4)\vspace{1mm} \\
  
  \multirow{2}{*}{CW\,Cep} & 13.01(7)$^A$ & 0.40(-35) & 7.20(70) & 4.00(32) & 30.7(2.7)\\
  & 11.95(8)$^A$ & 0.40(-35) & 7.60(75) & 3.55(34) & 29.7(3.1)\vspace{1mm} \\
  
  \multirow{2}{*}{U\,Oph} & 5.10(5)$^A$ & 0.25(15) & 53.5(4.0) & 0.96(7) & 18.8(1.6)\\
  & 4.60(5)$^A$ & 0.30(15) & 61.2(3.5) & 0.87(3) & 18.9(9)\vspace{1mm} \\  
  
  V621\,Per & 9.44(46)$^{B,1}$ & 0.40(-10) & 24.3(2) & 1.86(10) & 19.7(2.1)\\
\bottomrule
\end{tabular}
\begin{tablenotes}
      \tiny
      \item $^{A/B}$ within/outside error box; $^1$ dynamical mass was enforced
    \end{tablenotes}
\end{threeparttable}
\end{table}

\section{Revisiting the case of V380\,Cyg}
\label{Sect:V380Cyg}
As discussed in the previous sections, the primary components of V380\,Cyg and
V621\,Per are by far the most evolved stars in our sample. We consider the
evolved primary component of the V380\,Cyg system as representative of the class
in this section and provide a possible scenario of why the mass discrepancy
shows a strong negative correlation with the surface gravity of the star. A
notable characteristic of the primary component of V380\,Cyg is its high value
of the required microturbulent velocity $\xi=15\pm1$~km\,s$^{-1}$ as measured by
\citet{Tkachenko2014} from high-resolution optical spectra. Having high values
of the microturbulent velocity is a common phenomenon in massive evolved stars
as demonstrated by \citet{Cantiello2009} from samples of high-mass stars in the
Galaxy and in the LMC. The authors also report a strong anti-correlation between
the observed microturbulent velocities and stellar surface gravities, namely
$\xi$ tends to increase as $\log\,g$ decreases. This is related to the formation
of a sub-surface convection zone in high-mass stars that gives rise to the
microturbulent velocity field. {\it Micro}turbulence results predominantly in
radial velocities and accompanying spectral line variations
and should not be confused with {\it
  macro}turbulent broadening, which requires horizontal velocities to be
dominant over radial ones in the line-forming region. A connection between
macroturbulent line broadening and the horizontal velocities due to internal
gravity waves has been suggested \citep{Aerts2015}.

In this Section, we take a closer look at the problem of high microturbulence in
the primary component of V380\,Cyg and discuss it in the context of the mass
discrepancy problem. Standard spectrum analysis procedures are based on grids of
pre-computed atmosphere models that are fed into spectrum synthesis algorithms
to allow comparison between observed and theoretical spectra in arbitrary
wavelength intervals. Though spectral synthesis algorithms often allow for
optimization of all fundamental atmospheric parameters of stars, including the
microturbulent velocity parameter $\xi$, atmosphere model grids are typically
available for only a few values of microturbulent velocity, with
$\xi=2$~km\,s$^{-1}$ being the most commonly adopted value. A similar approach
was also used by \citet{Tkachenko2014}, where the authors adopted the grid of
$\xi=2$~km\,s$^{-1}$ {\sc atlas9} models \citep{Kurucz1993} and determined the
optimal microturbulent velocity along with other atmospheric parameters, by
varying it in the spectrum synthesis code. Despite the large inconsistency
between the derived microturbulent velocity of $\xi=15\pm1$~km\,s$^{-1}$ from
the observed spectrum and $\xi=2\pm1$~km\,s$^{-1}$ adopted in the {\sc atlas9}
atmosphere models, no further iterations have been done to maintain
self-consistency in the spectrum analysis procedure.

\begin{figure*}
   \centering
   \includegraphics[width=8.7cm]{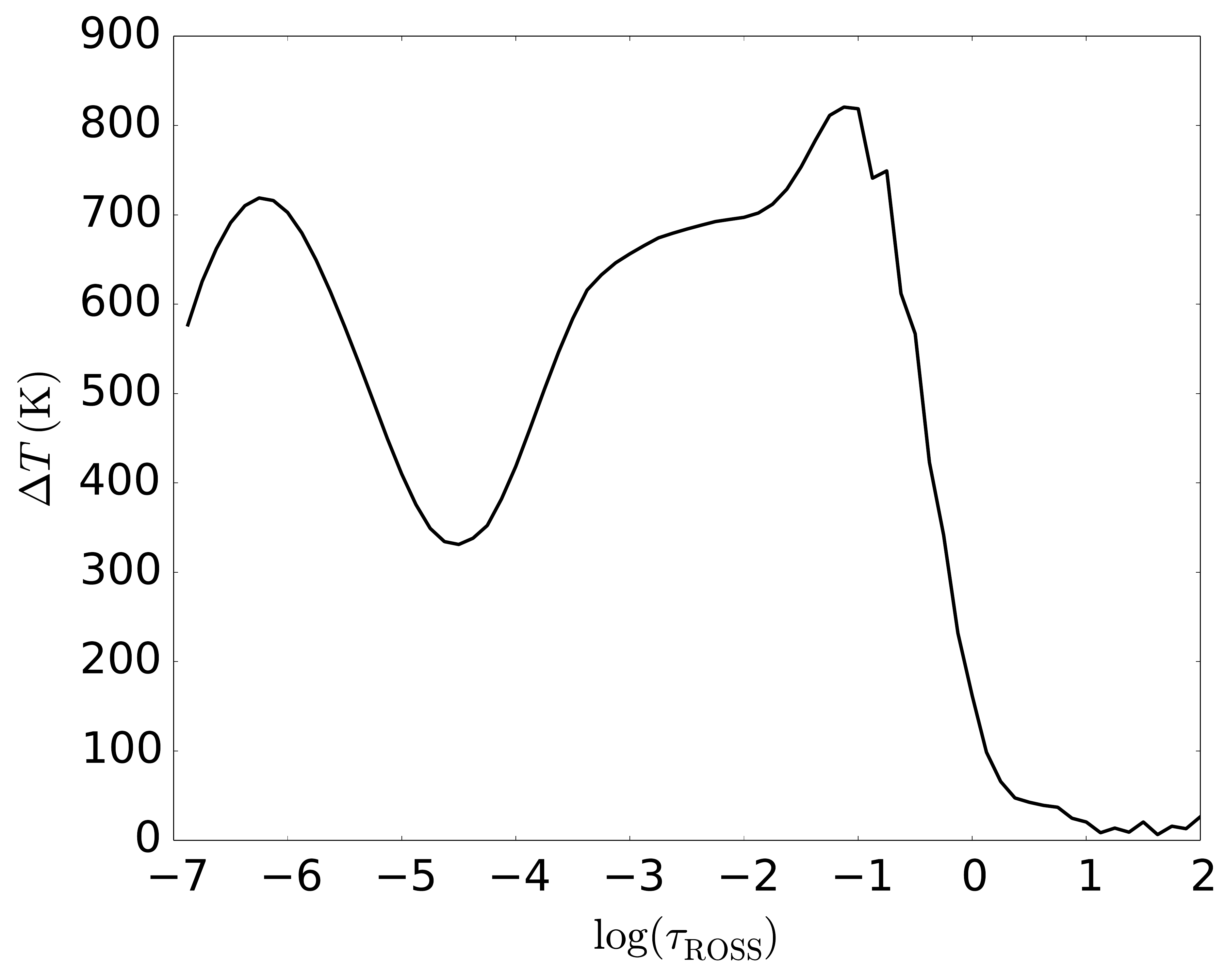}\hspace{5mm}
   \includegraphics[width=8.7cm]{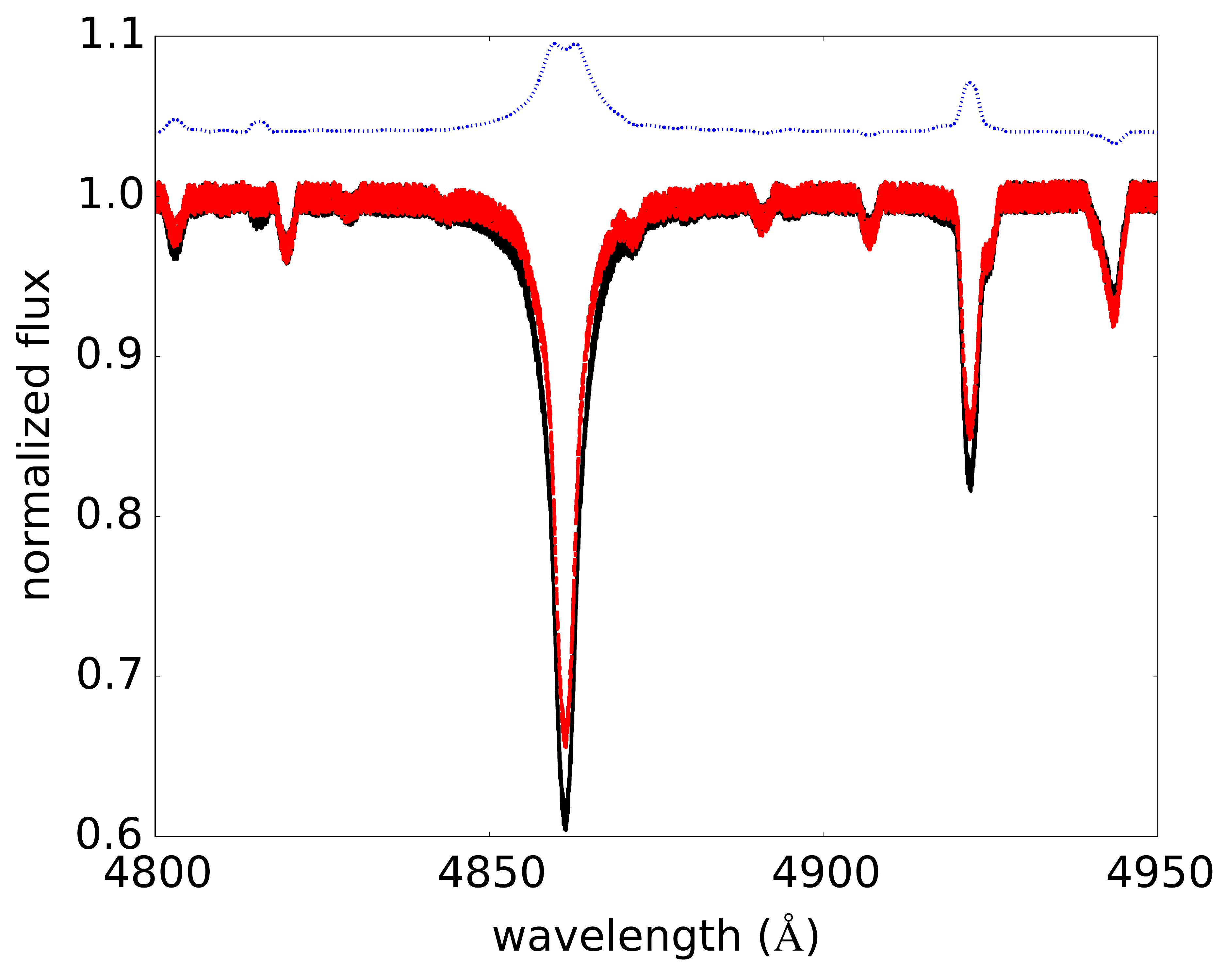}
      \caption{{\bf Left:} temperature difference between a model atmosphere
        computed with a microturbulent velocity of $\xi=15$ and 2~km\,s$^{-1}$, as
        a function of the logarithm of the Rosseland optical depth. {\bf Right:}
        Differences in spectral line profiles associated with the change of
        microturbulent velocity in the model atmosphere from $\xi=2$~km\,s$^{-1}$
        (solid black line) to $\xi=15$~km\,s$^{-1}$ (dashed red line). The
        difference spectrum (dotted blue line) was vertically shifted for
        clarity. Gaussian noise was added to both spectra to simulate a
        signal-to-noise ratio S/N$\sim$100.} 
         \label{Fig:ModelAtmosphere}
\end{figure*}

\begin{figure*}
   \centering
   \includegraphics[width=8.7cm]{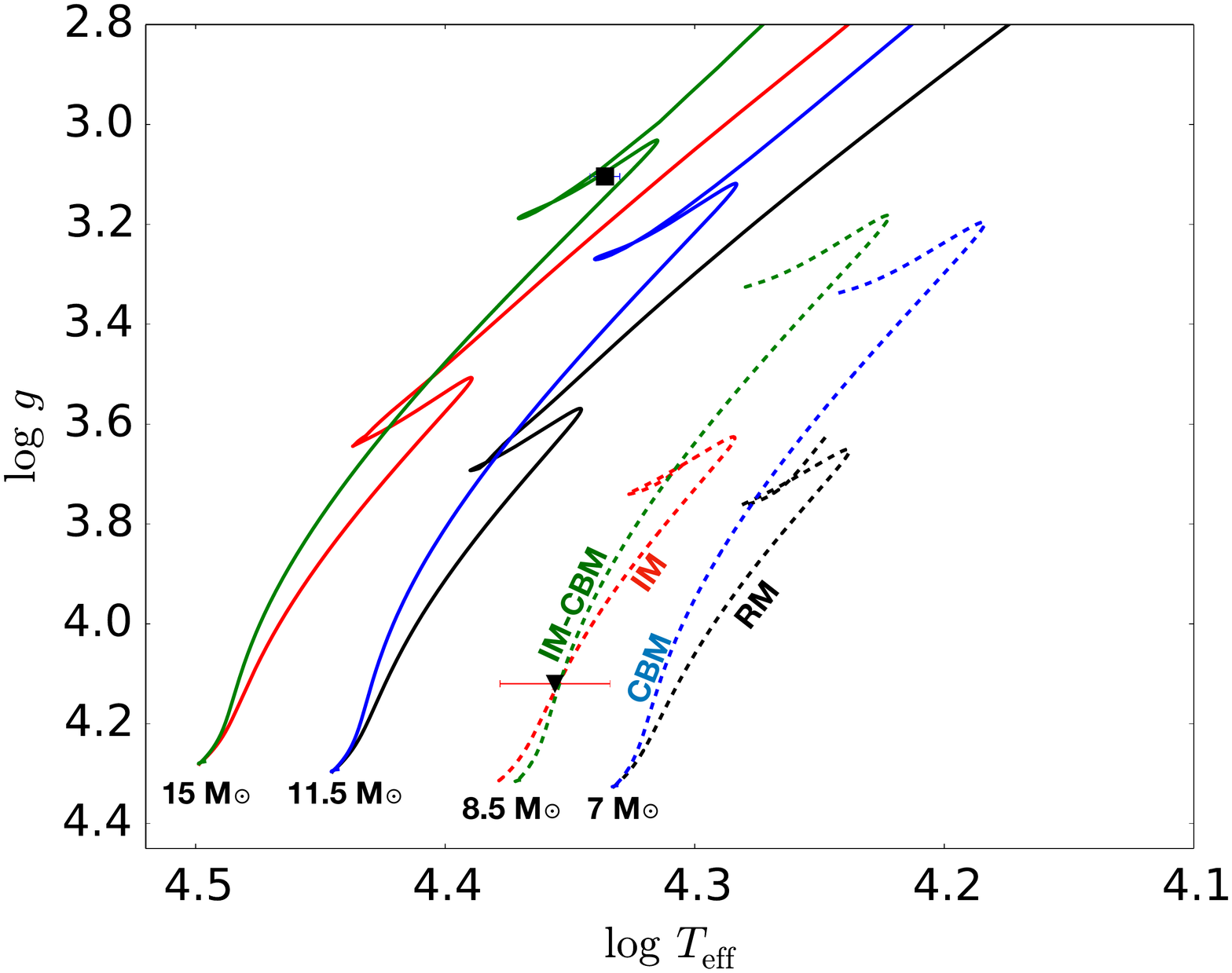}\hspace{5mm}
   \includegraphics[width=8.7cm]{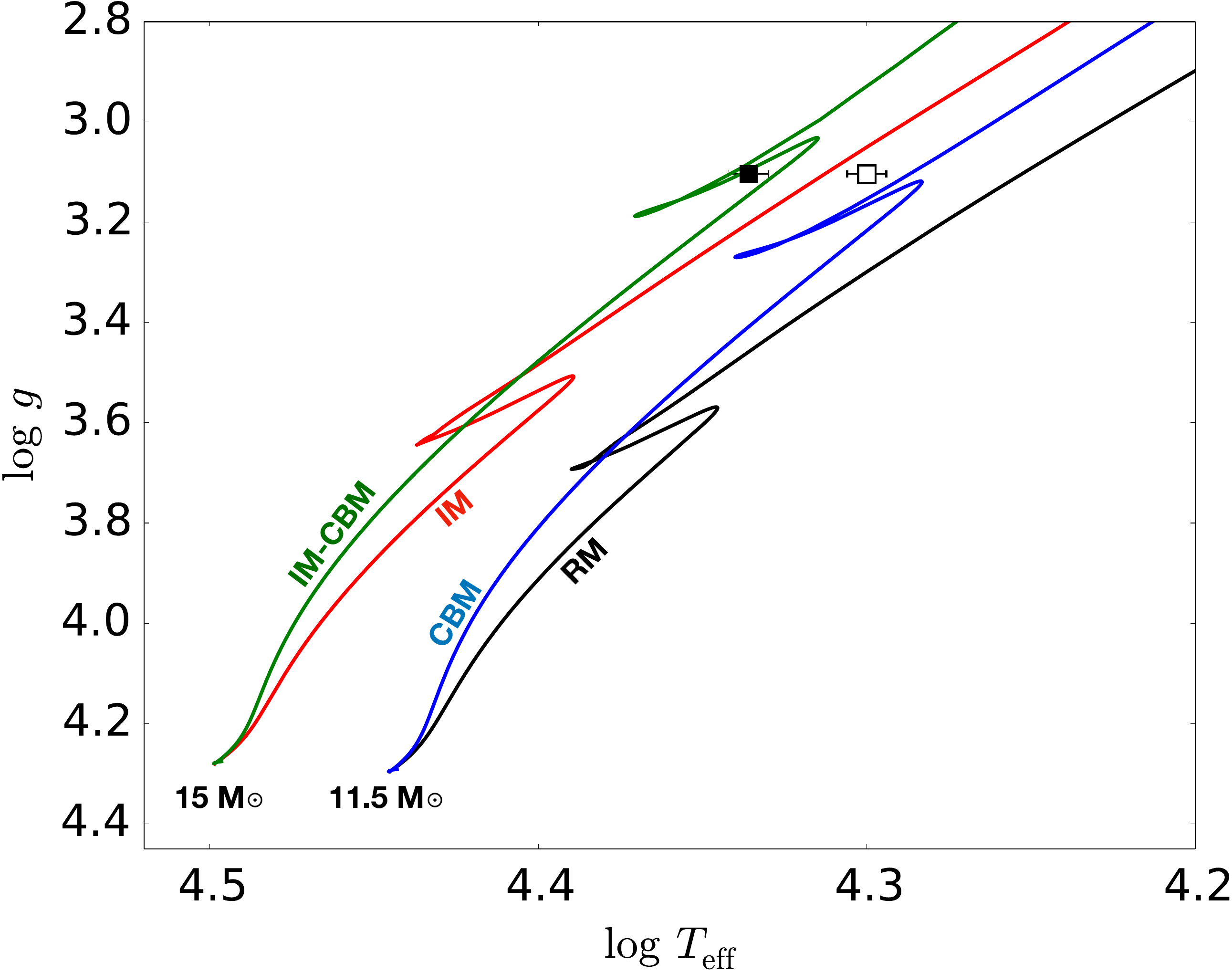}
      \caption{{\bf Left:} Four solutions obtained for the V380\,Cyg system and
        detailed in Table~\ref{Table:SingleStarCaseParameters}. The
        corresponding {\sc mesa} evolutionary tracks are shown with solid
        (dashed) lines for the primary (secondary) component. The RM, IM, CBM,
        and IM-CBM solutions are shown in black, red, blue, and green,
        respectively. The more (less) evolved primary (secondary) is shown with 
        filled square (triangle) symbols. {\bf Right:} Changing position of the
        primary component (open square) associated with a reduced $T_{\rm
          eff}$ by 8\%. The color coding is the same as in the
        left panel.} 
         \label{Fig:V380Cyg}
\end{figure*}

Including microturbulence in stellar atmosphere models leads to an increase of
the turbulent pressure, which in turn impacts the gas pressure to meet the
condition of total pressure conservation. This alters the temperature at a given
optical depth, hence influencing the entire structure of the stellar atmosphere
model. We demonstrate this effect in Fig.~\ref{Fig:ModelAtmosphere} (left panel)
where the temperature difference between atmosphere models with
$\xi=15$~km\,s$^{-1}$ and $\xi=2$~km\,s$^{-1}$ is plotted as a function of the
logarithm of the Rosseland optical depth. For these calculations
\citep[performed with the {\sc LLmodels} stellar atmosphere model code;
][]{Shulyak2004}, we assume an atmosphere model corresponding to the primary
component of V380\,Cyg, i.e. $T_{\rm eff}=21\,500$~K, $\log\,g=3.1$~dex, and
[M/H] = 0.0~dex. One can see that the temperature difference associated with an
increase of microturbulence and inclusion of turbulent pressure is significant
in the line forming regions, and unavoidably affects both shapes and
strengths of individual lines in the stellar spectrum. The latter effect is
illustrated in the right panel of Fig.~\ref{Fig:ModelAtmosphere}, where we plot
stellar spectra synthesized with the {\sc gssp} code from atmosphere models with
$\xi=15$~km\,s$^{-1}$ (dashed red line) and with $\xi=2$~km\,s$^{-1}$ (solid
black line), along with their difference (dotted blue line), where the latter is
subtracted from the former. We also added a Gaussian noise to both spectra to
simulate S/N$\sim$100 for a realistic illustration.

Analyzing the spectrum of a hot star where spectral lines are altered by a
significant contribution from turbulent pressure with a grid of
$\xi=2$~km\,s$^{-1}$ atmosphere models leads to large systematic uncertainties
in parameters such as $T_{\rm eff}$ and $\log\,g$. Given that the latter
parameter is computed from the dynamical mass and radius in the case of binary
stars, $T_{\rm eff}$ is the parameter that suffers the most. In order to
illustrate this quantitatively, we employ a grid of $\xi=2$~km\,s$^{-1}$
atmosphere models to analyze the spectrum shown with the solid black line in
Fig.~\ref{Fig:ModelAtmosphere} and having the following parameters: $T_{\rm
  eff}=21\,500$~K, $\log\,g=$3.1~dex, $\xi_{\rm model}=15$~km\,s$^{-1}$,
$\xi_{\rm spectrum}=15$~km\,s$^{-1}$, [M/H] = 0.0~dex, and
$v\,\sin\,i=$100~km\,s$^{-1}$. In this process, we keep $\log\,g$ fixed to
3.1~dex assuming it is known from the dynamical mass and radius of the star. The
obtained best fit parameters are: $T_{\rm eff}=23\,190\pm210$~K, $\xi_{\rm
  spectrum}=14.1\pm1.0$~km\,s$^{-1}$, [M/H] = -0.01$\pm0.05$~dex, and
$v\,\sin\,i=98.5\pm3.4$~km\,s$^{-1}$. One can see that, though the other
parameters agree with their input values within 1$\sigma$ uncertainties,
$T_{\rm eff}$ is overestimated by $\sim$1\,700~K, i.e., some 8\%.

The left panel of Fig.~\ref{Fig:V380Cyg} shows {\sc mesa} evolutionary tracks
corresponding to four single-star-case scenario solutions detailed in
Table~\ref{Table:SingleStarCaseParameters}. One can see that the tracks
corresponding to the dynamical masses of the stars (11.43$\pm$0.19~M$_{\odot}$
and 7.00$\pm$0.14~M$_{\odot}$ for the primary and secondary, respectively) are
significantly off from their spectroscopic positions and that increasing
near-core mixing does not solve the problem. It is essential to introduce a
mass discrepancy of $\sim$30\% ($\sim$20\%) for the primary (secondary)
component to maintain consistency between evolutionary tracks and
spectroscopically measured parameters, where a high value of the 
parameter $f_{\rm ov}\approx0.04\ H_{\rm p}$ is also essential to use for the
more evolved primary. The mass discrepancy gets substantially reduced when
$T_{\rm eff}$ of the primary is corrected by 8\% of the measured value in order
to account for the effect of the microturbulent velocity in the stellar atmosphere model
(Fig.~\ref{Fig:V380Cyg}, right panel). In practice, the mass discrepancy nearly
vanishes for the primary of V380\,Cyg when its dynamical mass and high $f_{\rm
  ov}$ are assumed. The argument of microturbulence in atmosphere models
does not apply to the secondary component of V380\,Cyg as it is an unevolved
star. However, using a wrong value for $T_{\rm eff}$ for the primary has a
significant effect on determining $T_{\rm eff}$ of the low flux contributing
secondary, both in the spectrum analysis of the disentangled spectra and in the
photometric analysis of the light curve. Indeed, \citet{Tkachenko2014} report a
significant difference of $\sim$2000~K between spectroscopic and photometric
values of $T_{\rm eff}$ for the secondary and adopt the mean value as a
compromise. It is essential to perform a full re-analysis of the system in a
self-consistent way in order to answer the question whether accounting for the
effect of microturbulence in atmosphere models and of missing convective core
mass in SSE models resolves the mass discrepancy for V380\,Cyg.

Stellar evolution dictates the formation of a sub-surface convection zone in
high-mass stars as they approach the Terminal Age Main-Sequence (TAMS), giving
rise to a higher microturbulent velocity field
\citep[e.g.,][]{Cantiello2009}. Ignoring the effect of microturbulence by fixing
it to 2~km\,s$^{-1}$ in stellar atmosphere models implies overestimation of the
effective temperature of the star by means of a spectroscopic analysis. As
demonstrated in Figs.~\ref{Fig:MesaTracks} and \ref{Fig:V380Cyg}, increasing the
initial mass of the star in stellar evolution models shifts the corresponding
track to higher temperatures (Fig.~\ref{Fig:MesaTracks}, left panel). Hence, one
{\it introduces} the mass discrepancy by overestimating the effective
temperature of the star and the effect is expected to be more pronounced for
evolved main-sequence stars. The latter correlation in terms of the dependence
of the mass discrepancy on the surface gravity of the star is exactly what we
find in our stellar sample (see, e.g., top left panels in
Figs.~\ref{Fig:Solution2} and \ref{Fig:Solution2_Binary}).

\section{Discussion and Conclusions}
\label{Sect:Conclusions}
In this paper, we presented the mass discrepancy problem as observed in
intermediate- and high-mass eclipsing SB2 binary stars. Discovered in a large
sample of single stars \citep{Herrero1992}, the problem obtained further
acknowledgement in the stellar astrophysics community when it was observed in
binary stars, largely because of the model-independent way of measuring accurate
stellar masses and radii in such systems. A common proposed solution to the mass
discrepancy problem until now was a significant increase of near-core mixing
in the form of convective core overshooting, either adopting a step formalism
with an adiabatic temperature gradient \citep[e.g.,][]{Guinan2000} or an
exponentially decaying diffusive mixing with
a radiative temperature gradient in the overshoot zone
\citep[e.g.,][]{Tkachenko2014}.

\begin{figure*}[t!]
   \centering
   \includegraphics[width=8.7cm]{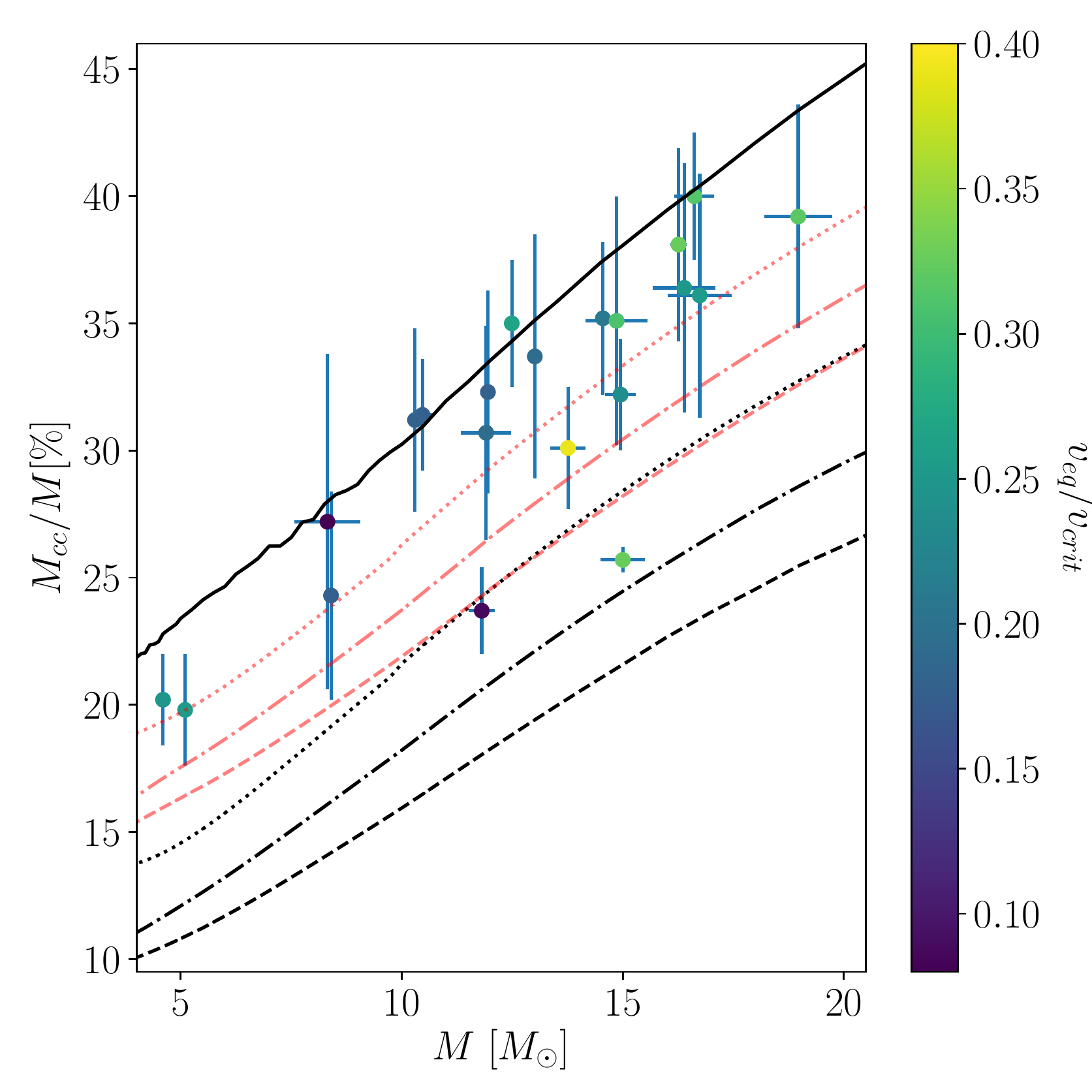}\hspace{5mm}
   \includegraphics[width=8.7cm]{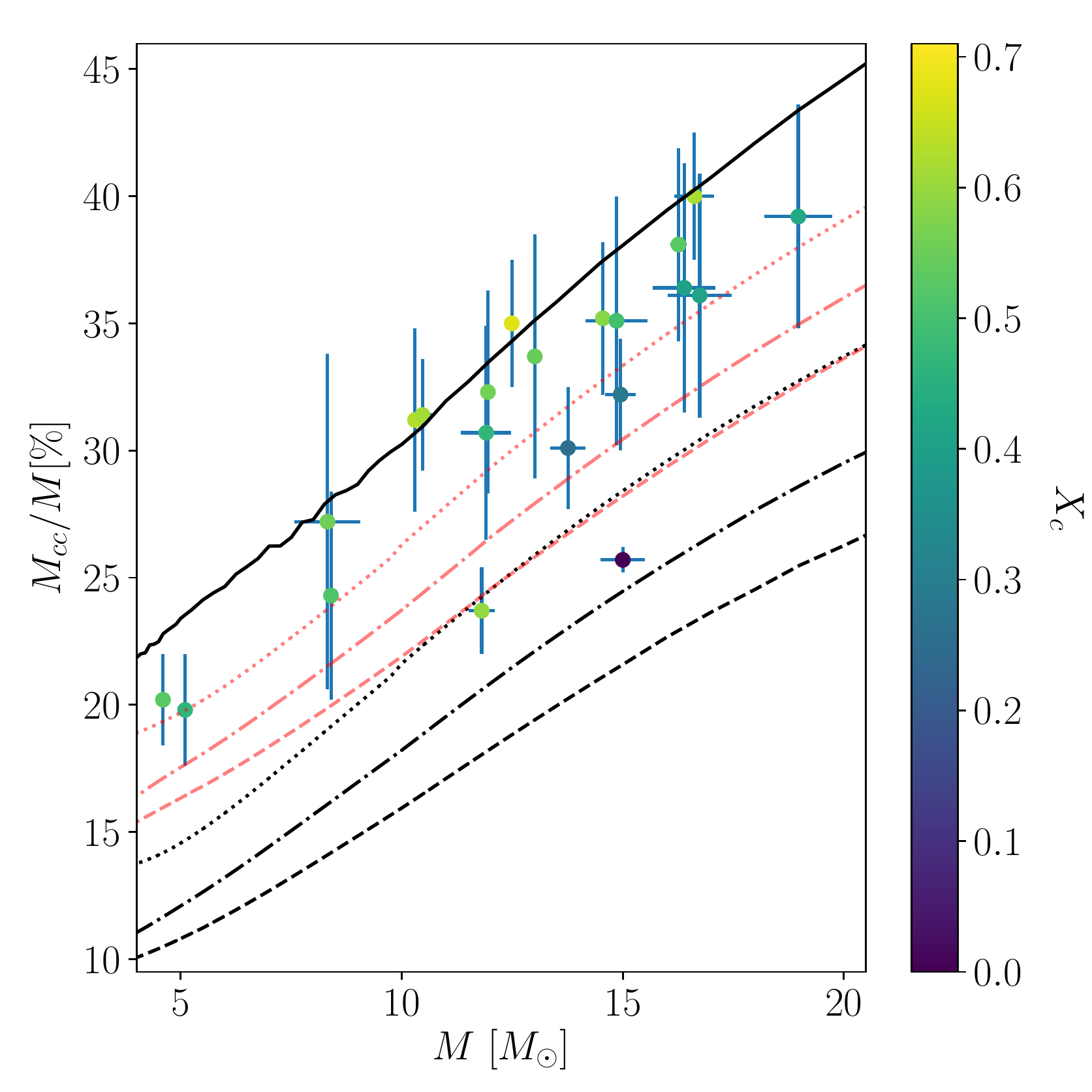}
      \caption{Convective core mass as a function of total mass for
        ZAMS models (full line) and models for the central hydrogen mass fractions of $X_c=0.35$ (red) and $X_c=0.1$ (black) when considering low (dashed), moderate (dashed-dot),
        and high (dotted) levels of CBM according to the ranges in
        Table\,\ref{Tab:MesaGrid}.  The sample stars are added with color
        coding according to their $v_{\rm eq}/v_{\rm cirt}$ (left) and their
        $X_c$ (right). } 
\label{Fig:Mcc-versus-M}
\end{figure*}

In order for the mass discrepancy problem to be properly addressed, a
homogeneously analyzed sample of binary stars whose components cover a large
range in stellar mass and evolutionary stage is needed. This avoids systematic
uncertainties propagating into the analysis of the problem, hence influencing
conclusions in a largely unpredictable way. Our stellar sample is the first that
meets the homogeneity requirement and comprises eleven eclipsing SB2 binary
systems, whose individual stellar components are slow to moderate rotators
covering the mass range from $M\approx 4.5$~M$_{\odot}$ to
$M\approx 17$~M$_{\odot}$ and evolutionary stage from the ZAMS to the TAMS. The
analysis is performed both in the single-star scenario where the binary
components are treated independent of each other, and in the binary-star
scenario where we enforce the equal-age condition for the two stellar components
of the same binary system. The results of these two analyses are qualitatively
alike and can be summarized as follows:
\begin{itemize}
    \item the mass discrepancy is clearly present in our stellar sample and
      there is strong statistical evidence ($p<0.01$) of it being
      anti-correlated with surface gravity;

    \item a strong relationship exists between the convective core mass and the
      mass discrepancy. More precisely, there is strong evidence that SSE
      models without extra near-core mixing 
underestimate the convective core mass. Statistical evidence is
      found that the problem becomes more pronounced as the star evolves during
      the main-sequence;

    \item because of its ability to supply the convective core with fresh
      hydrogen and to increase its mass, enhanced near-core boundary mixing
      can partially account for the observed mass discrepancy. The mass
      discrepancy turns out to be a convective core mass--stellar age
      correlation/problem in stellar evolution models;

    \item the mass discrepancy does not correlate with any other fundamental or
      atmospheric parameters of stars than their surface gravity (proxy for
      age/low density atmosphere);

    \item neglecting high microturbulent velocity values and turbulent pressure
      in stellar atmosphere models of hot stars results in an overestimation of
      the effective temperature of a star up to some 8\%, when its dynamical
      $\log\,g$ value is assumed. This effect is larger for more evolved stars,
      i.e., those that show lower surface gravities and are located closer to the
      TAMS. In practice, stars appear hotter than they are, which partially
      contributes to the mass discrepancy problem. 
\end{itemize}
We conclude that the mass discrepancy problem in binary stars is the combined
effect of a convective core mass-stellar age correlation in stellar evolution
models and the neglect of high microturbulent velocities and turbulent pressure
in stellar atmosphere models in spectroscopic analysis.

\citet{Claret2016,Claret2017,Claret2018,Claret2019} report an almost linear
increase of the overshooting parameter $\alpha_{\rm ov} (f_{\rm ov}$) from
0.0~$H_{\rm p}$ to $\sim$0.2 (0.02)~$H_{\rm p}$ for masses between
$M\approx 1.2$~M$_{\odot}$ and $M\approx 2.0$~M$_{\odot}$ and report
systematically lower metallicities from stellar evolution models compared to
values inferred from high-resolution spectroscopy. In their work, the authors
could not find any plausible explanation for the observed effect. In
Sect.~\ref{Sect: ModelsGrid} (cf. Fig.~\ref{Fig:MesaTracks}, right panel), we
demonstrated that lowering the initial metallicity parameter $Z$ in stellar
evolution models has a similar effect to increasing the initial mass of a
star. We estimate that for a $M=2.5$~M$_{\odot}$ star \citep[approximately at
the center of the stellar mass range studied by][]{Claret2016}, the effect of
decreasing the metallicity parameter from $Z=0.014$ to $Z=0.010$ (0.006) is
similar to the effect of increasing the initial stellar mass by 0.25
(0.35)~M$_{\odot}$, i.e., 10 (15)\% of the assumed value of
$M=2.5$~M$_{\odot}$. The low metallicity values dictated by stellar evolution
models for about half of the sample studied in
\citet{Claret2016,Claret2017,Claret2019} are hence the manifestation of the mass
discrepancy problem in binary stars born with a convective core. This moderate
mass discrepancy gets compensated by a moderate increase of the mass of the
convective core of the star that is provided by enhanced near-core boundary
mixing. We find low parameter values $f_{\rm ov}=0.013\pm0.013$ and
$f_{\rm ov}=0.017\pm0.012$ for the primary and secondary component of the U\,Oph
system, the two lowest-mass stars in our sample with $M=5.10\pm0.05$~M$_{\odot}$
and $M=4.60\pm0.05$~M$_{\odot}$, respectively.

There is a substantial overlap between targets in our sample and stars investigated by \citet{Schneider2014}, yet a quantitative comparison between the two studies proves difficult. While \citet{Schneider2014} adopted parameters from the catalog by \citet{Torres2010}, we reanalyzed all the targets in a homogeneous way to avoid systematic uncertainties, so that the parameters can in some cases differ substantially between the two studies. As a particular example, we find the primary component of AH\,Cep to be about 6\% more massive, while both components of the V478\,Cyg system are $\sim$1300~K hotter compared to the values adopted by \citet{Schneider2014}. Additionally, these authors adopted a grid of stellar evolution models computed for a lower metallicity $Z$=0.0088 (though with an increased iron abundance so that it closely resembles the solar value of $\log$(Fe/H)$ + 12 \approx 7.50$), while we adopt a metallicity of $Z=0.014$, in agreement with the spectroscopic analyses of the stars in our sample.

Finally, we point out that the two single B-type stars with asteroseismic
  estimations of their $M_{\rm cc}/M$ have values of 21\% \citep{Moravveji2015}
  and 19\% \citep{Moravveji2016}, both having a mass of $M\simeq
  3.2\,$M$_\odot$. This is very much in  line with our results for the
  binaries. To illustrate the importance of extra CBM, we plot 
$M_{\rm cc}/M$ versus $M$ for our sample stars
  in Fig.\,\ref{Fig:Mcc-versus-M}, color coded according to $v_{\rm eq}/v_{\rm
    crit}$ (left panel) and main-sequence phase according to $X_c$ (right
  panel). It can be seen that $M_{\rm cc}/M$ steadily increases as $M$
  increases, as expected from SSE theory. We also find
that there is no correlation between the rotation
  rate and $M_{\rm cc}/M$. Further, it can 
be seen that all the stars in our sample, including the most
  evolved ones, have $M_{\rm cc}/M$ far above the value predicted by SSE models
  with minimal CBM (dashed line). The results of our work on $M_{\rm cc}/M$, 
graphically
  represented in Fig.\,\ref{Fig:Mcc-versus-M}, 
  imply that stars have much larger 
helium cores near the TAMS compared to
  SSE models with no extra near-core mixing.

In the future, we plan to increase the size of our stellar sample drastically by
extending it towards both lower and higher masses. It is important to keep the
sample homogeneous in terms of data analysis to make sure systematic
uncertainties associated with the use of different algorithms do not propagate
to final conclusions. Higher-mass stars than those in our current sample tend to
have strong winds driving significant mass loss from the star. In this case,
using our suite of analysis techniques is not justified anymore. Furthermore, it
would be important to revisit some of the systems like V380\,Cyg. As we have
shown in this study, it is essential to take into account the high
microturbulent velocity and turbulent pressure in atmosphere model calculations
of such stars. In this context, it is also important to balance our sample by
including more stars at advanced stages of (main-sequence) evolution, from mid
main-sequence onwards. Last but not least, searching for eclipsing,
spectroscopic double-lined binary systems with stellar components pulsating in
gravity-mode oscillations is another promising step forward in calibrating
stellar evolution models with binary stars \citep{Aerts2020}.  
Space-missions such as TESS \citep{Ricker2015} and
PLATO \citep{Rauer2014} are excellent observational facilities to search for the
best candidate asteroseismic binary systems.

\begin{acknowledgements}
      The research leading to these results has received funding from the
      European Research Council (ERC) under the European Union's Horizon 2020
      research and innovation programme (grant agreement N$^\circ$670519:
      MAMSIE), from the KU~Leuven Research Council (grant C16/18/005:
      PARADISE), from the Research Foundation Flanders (FWO) under grant
      agreements G0H5416N (ERC Runner Up Project) and G0A2917N (BlackGEM), as
      well as from the BELgian federal Science Policy Office (BELSPO) through
      PRODEX grant PLATO. KP acknowledges financial support from the Croatian
      Science Foundation under grant IP-2014-09-8656 (STARDUST). VT acknowledges
      support by the RF Ministry of Science and Higher Education in the
      framework of the state task (project no. 3.7126.2017/8.9).
\end{acknowledgements}

\bibliographystyle{aa}
\bibliography{BinarySample}

\begin{appendix}
\section{Binary equal age-case scenario}

Here, we present Tables and Figures supplementing our results as to the equal age scenario presented in Section~\ref{Sect:BinaryEqualAgeCase}. 

\begin{figure*}
   \centering
   \includegraphics[width=8.7cm]{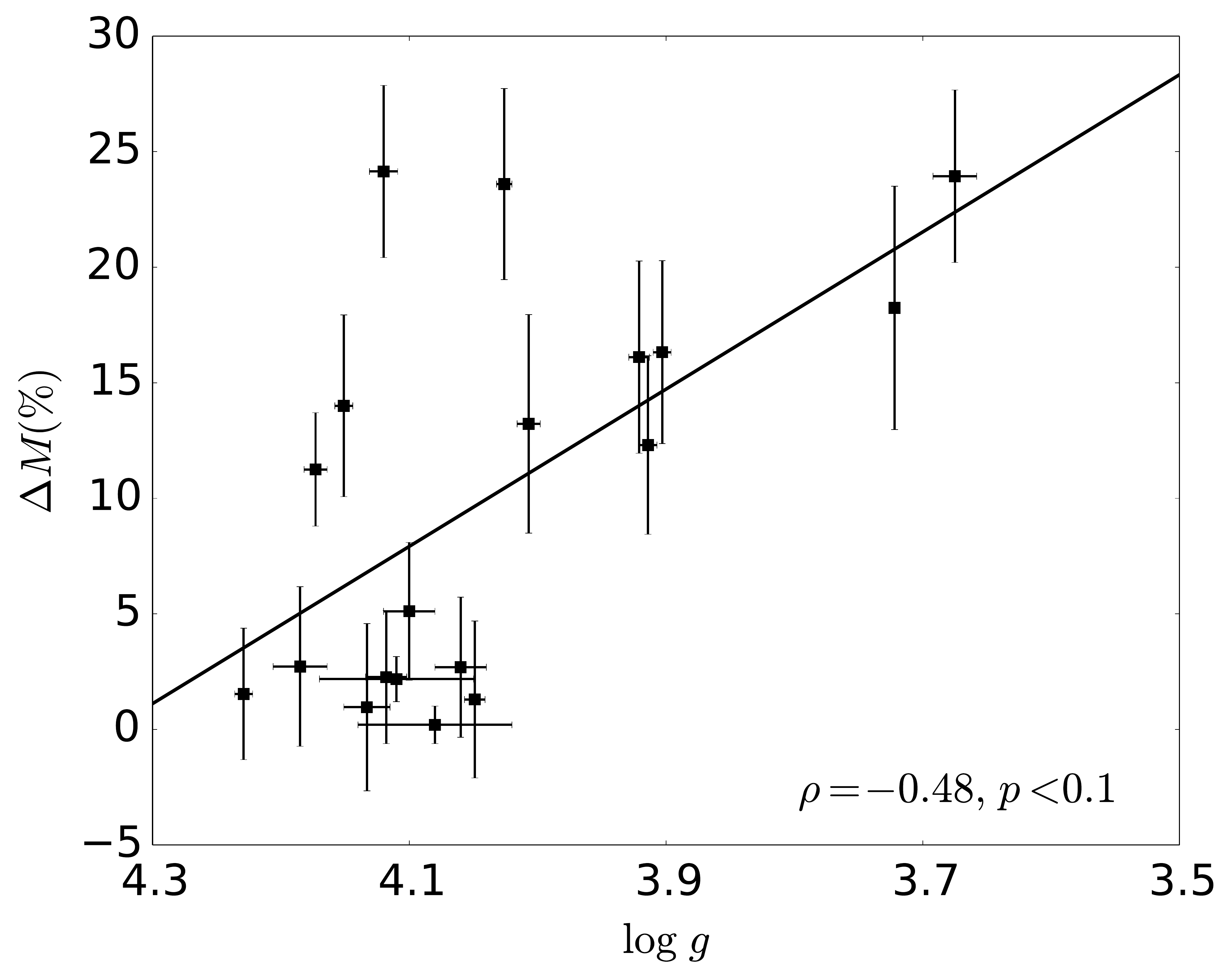}\hspace{5mm}
   \includegraphics[width=8.7cm]{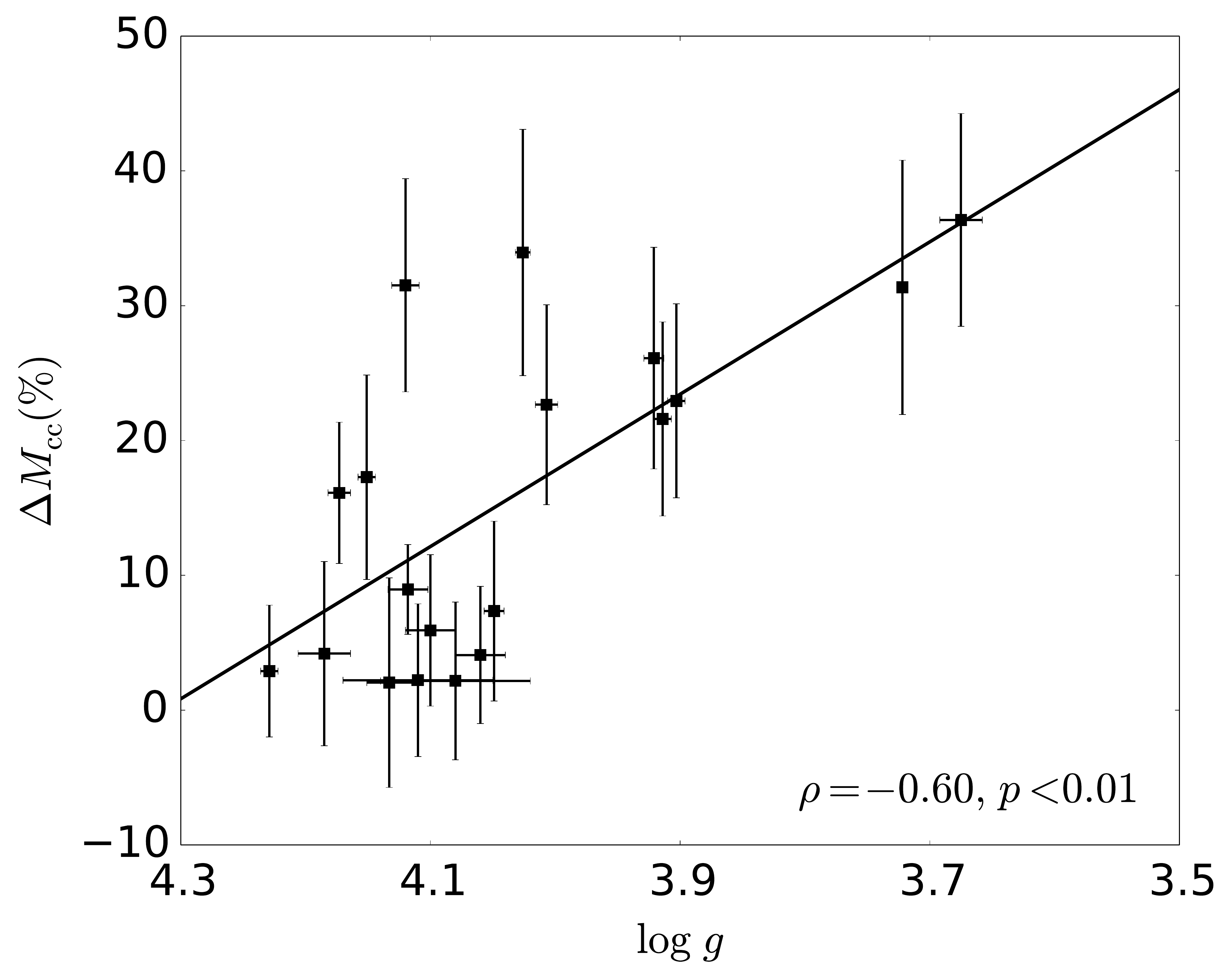}
   \includegraphics[width=8.7cm]{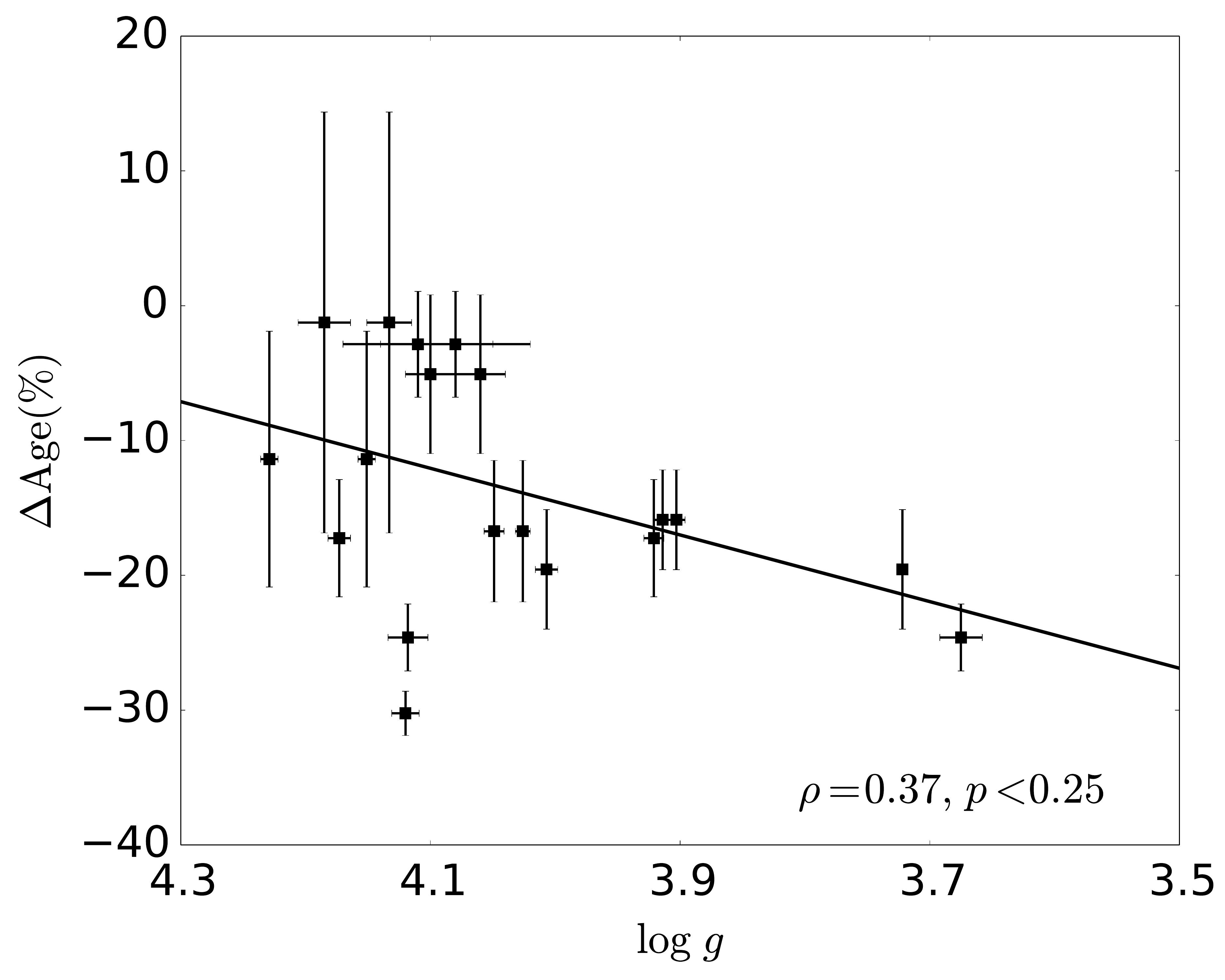}\hspace{5mm}
   \includegraphics[width=8.7cm]{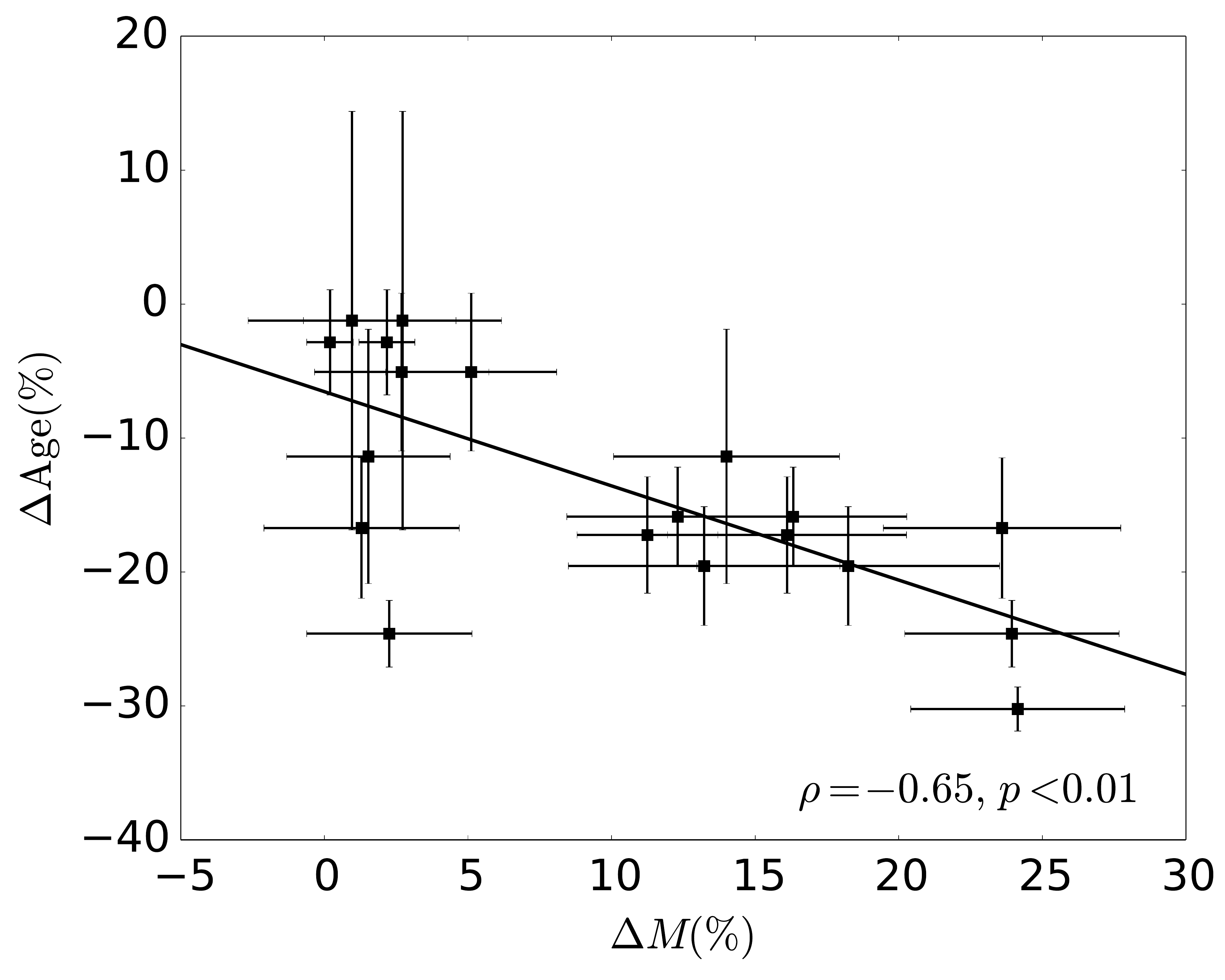}
      \caption{Same as Figure~\ref{Fig:Solution2} but for the common age scenario. The respective binary common age RM and IM solutions are detailed in Table~\ref{Table:BinaryStarCaseParameters}} 
         \label{Fig:Solution2_Binary}
\end{figure*}

\begin{figure*}
   \centering
   \includegraphics[width=8.7cm]{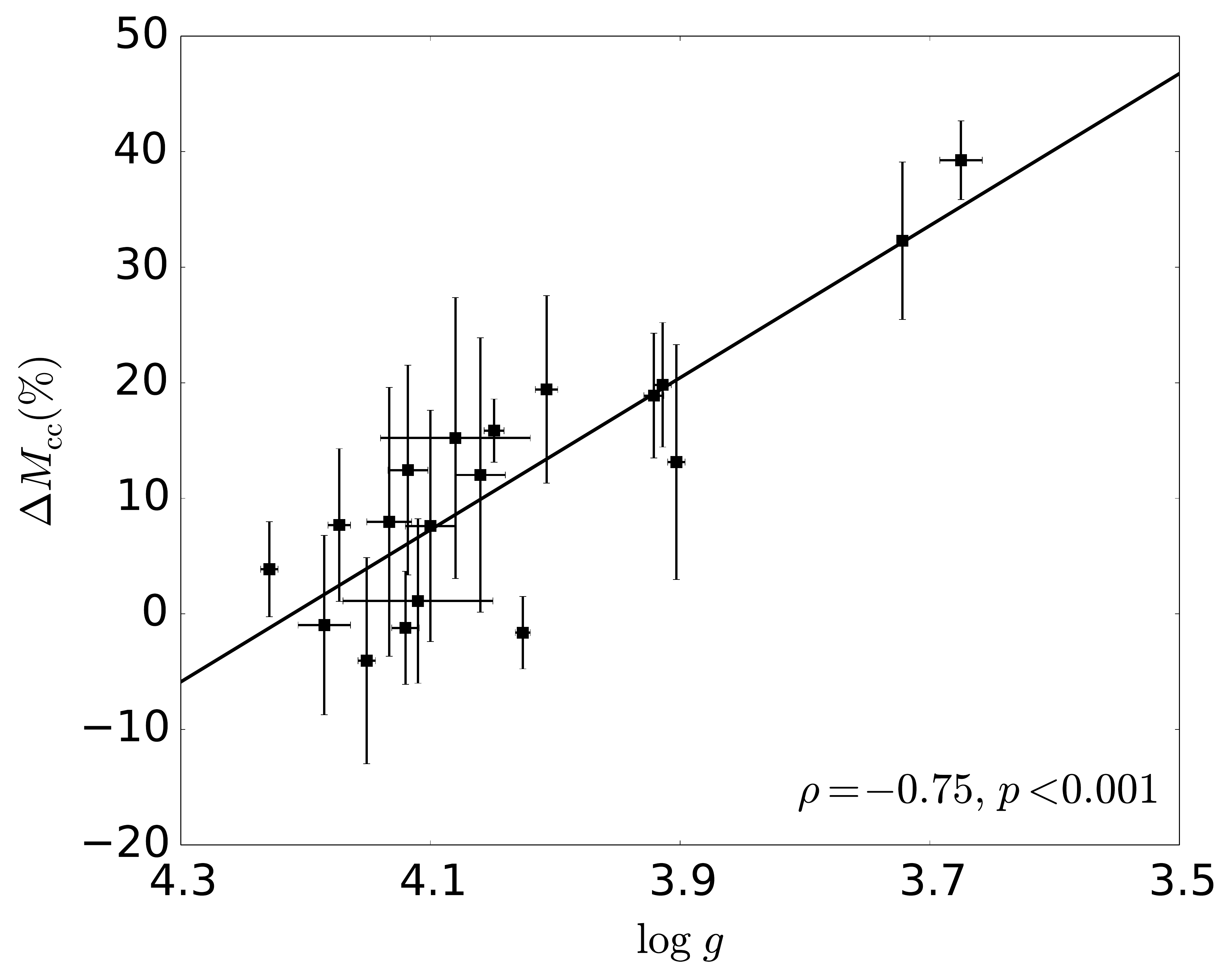}\hspace{5mm}
   \includegraphics[width=8.7cm]{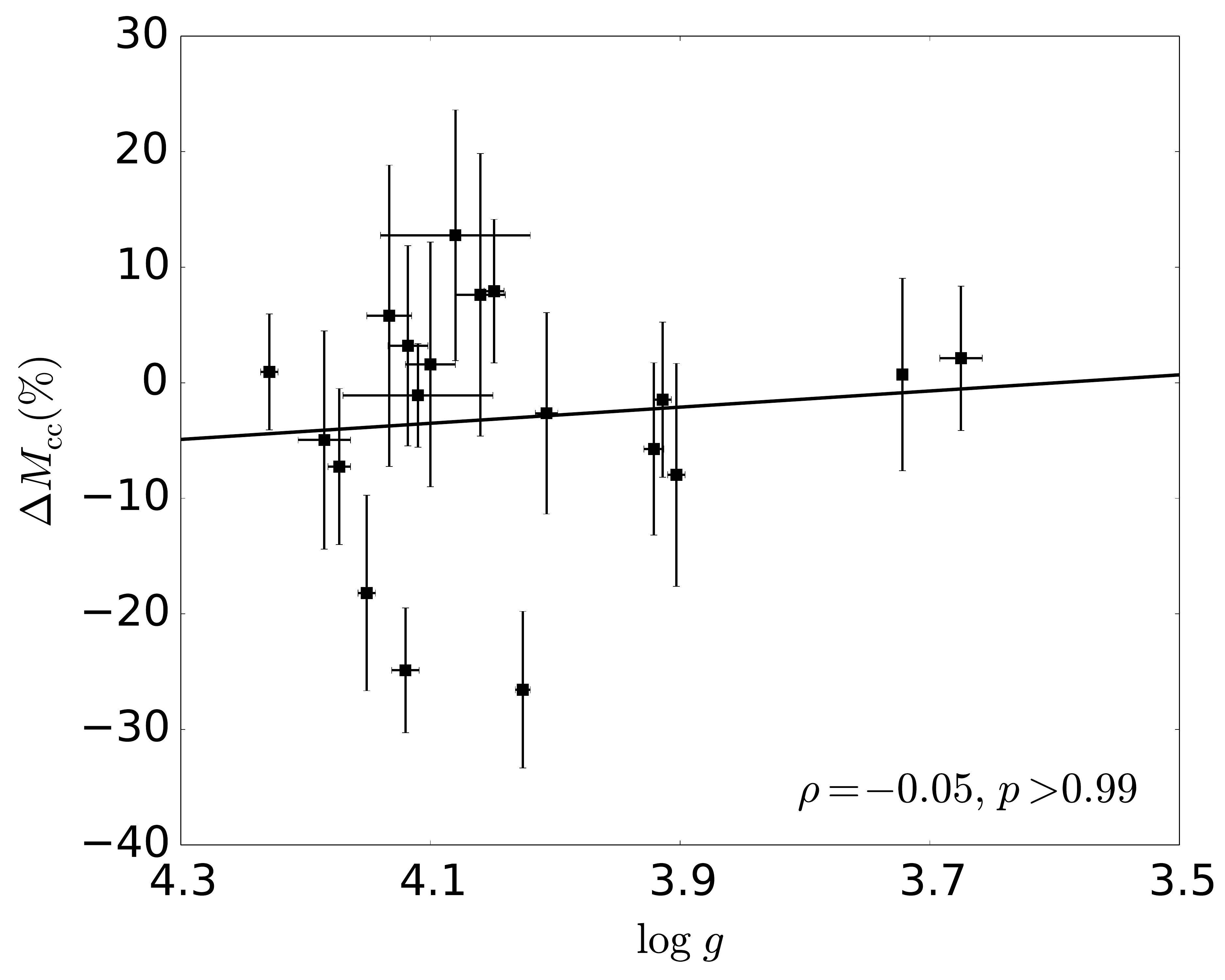}
   \includegraphics[width=8.7cm]{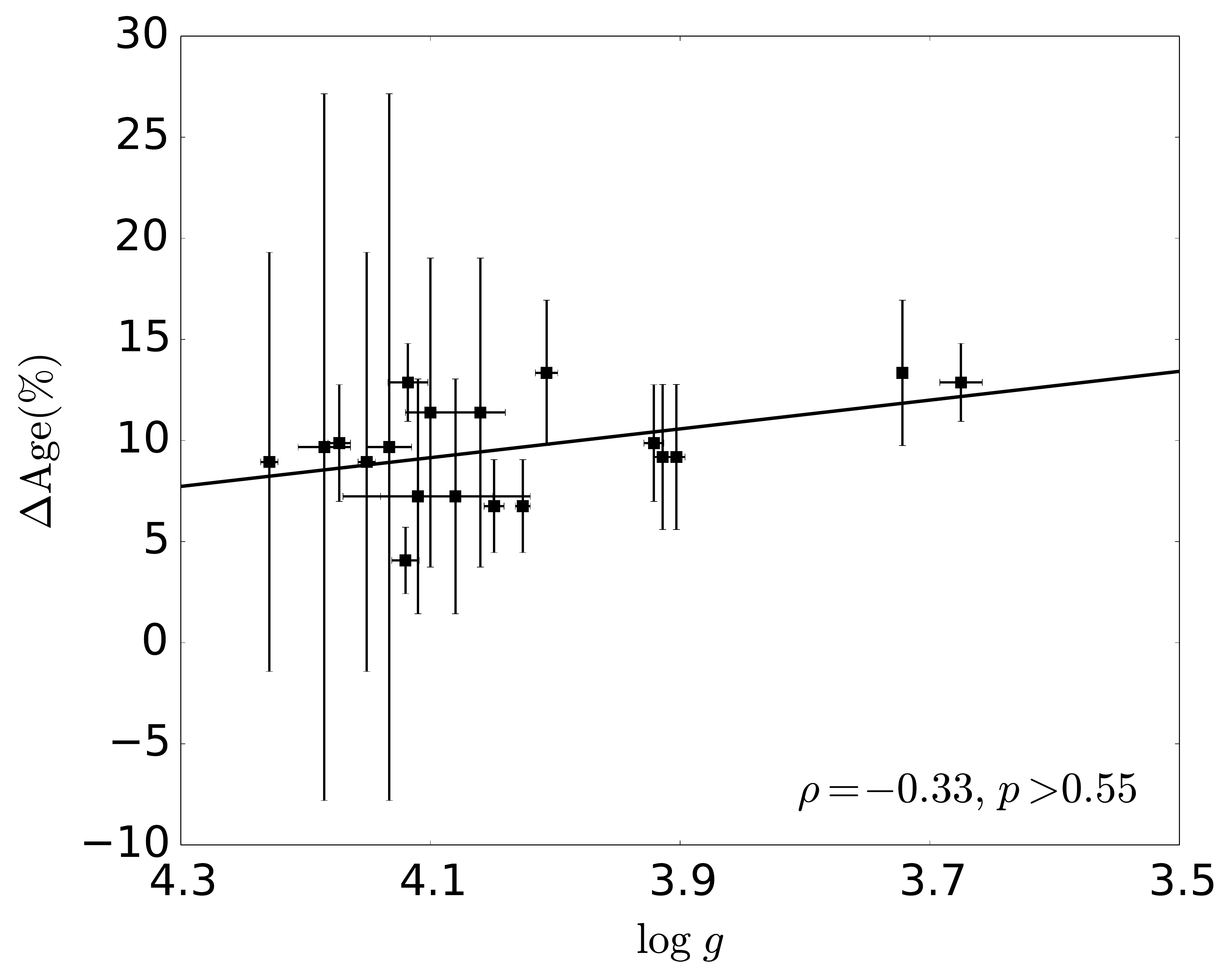}\hspace{5mm}
   \includegraphics[width=8.7cm]{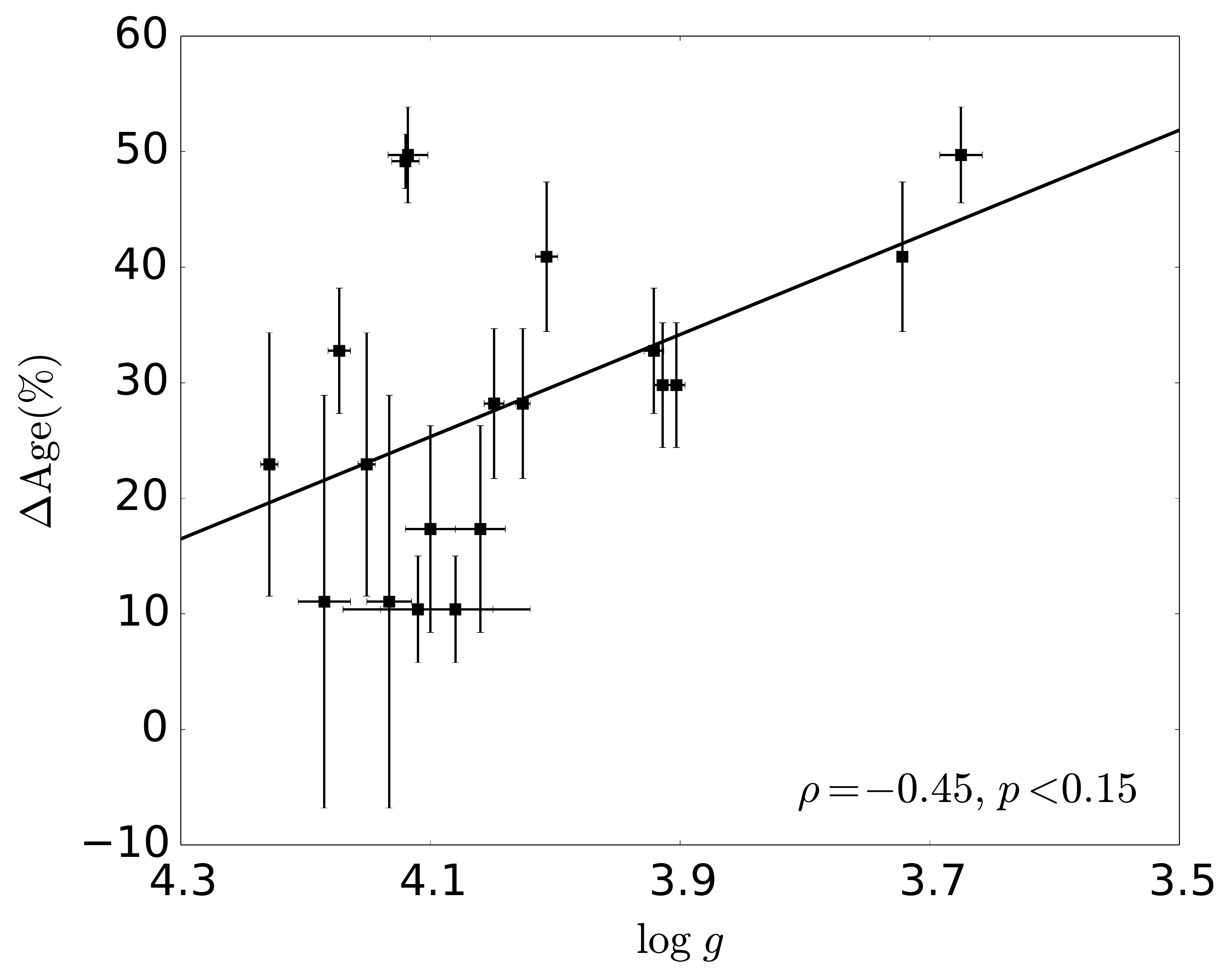}
      \caption{Same as Figure~\ref{Fig:Solution3} but for the common age scenario. The respective binary common age RM, IM, and CBM solutions are detailed in Table~\ref{Table:BinaryStarCaseParameters}} 
         \label{Fig:Solution3_Binary}
   \end{figure*}

\begin{figure*}
   \centering
   \includegraphics[width=8.7cm]{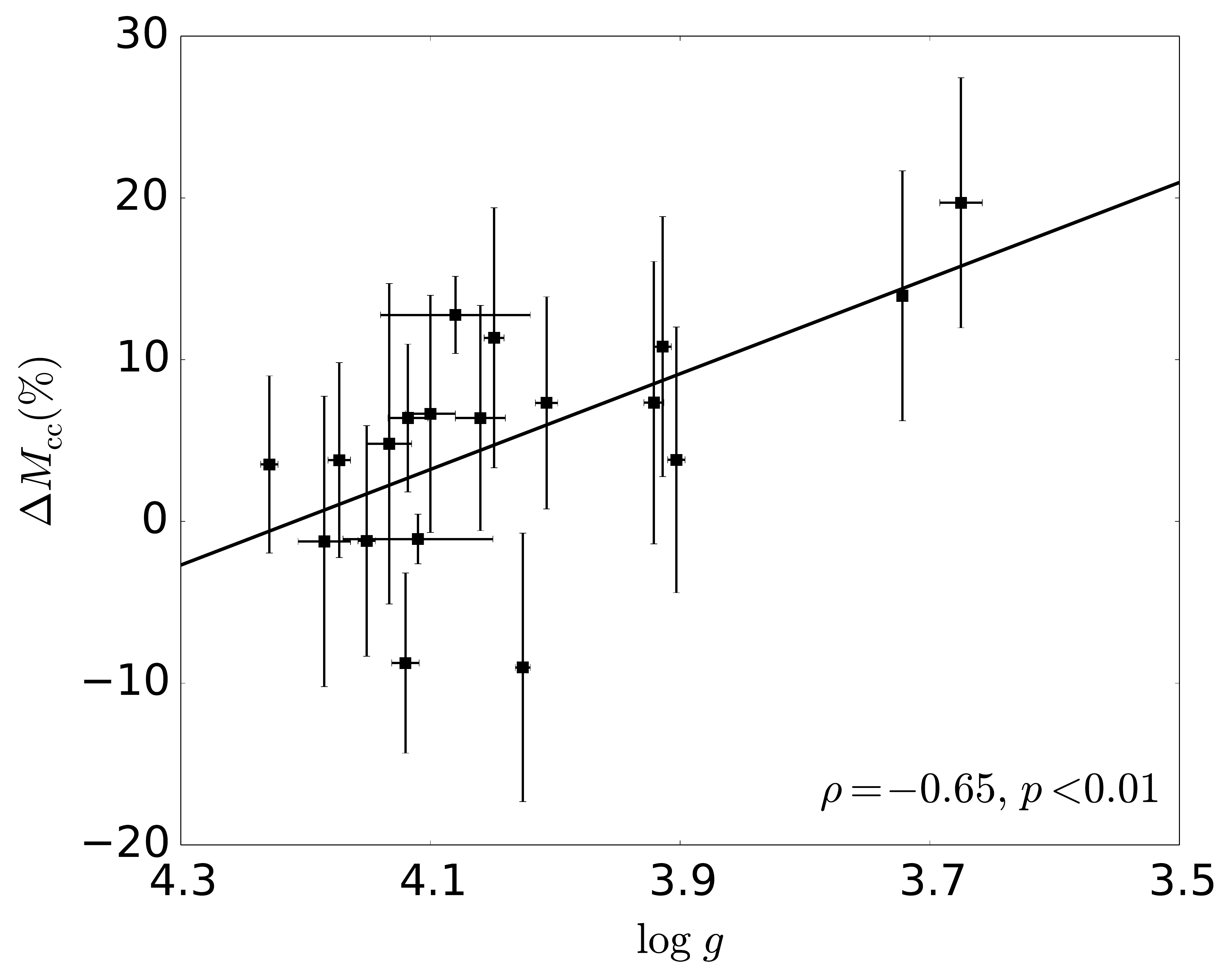}\hspace{5mm}
   \includegraphics[width=8.7cm]{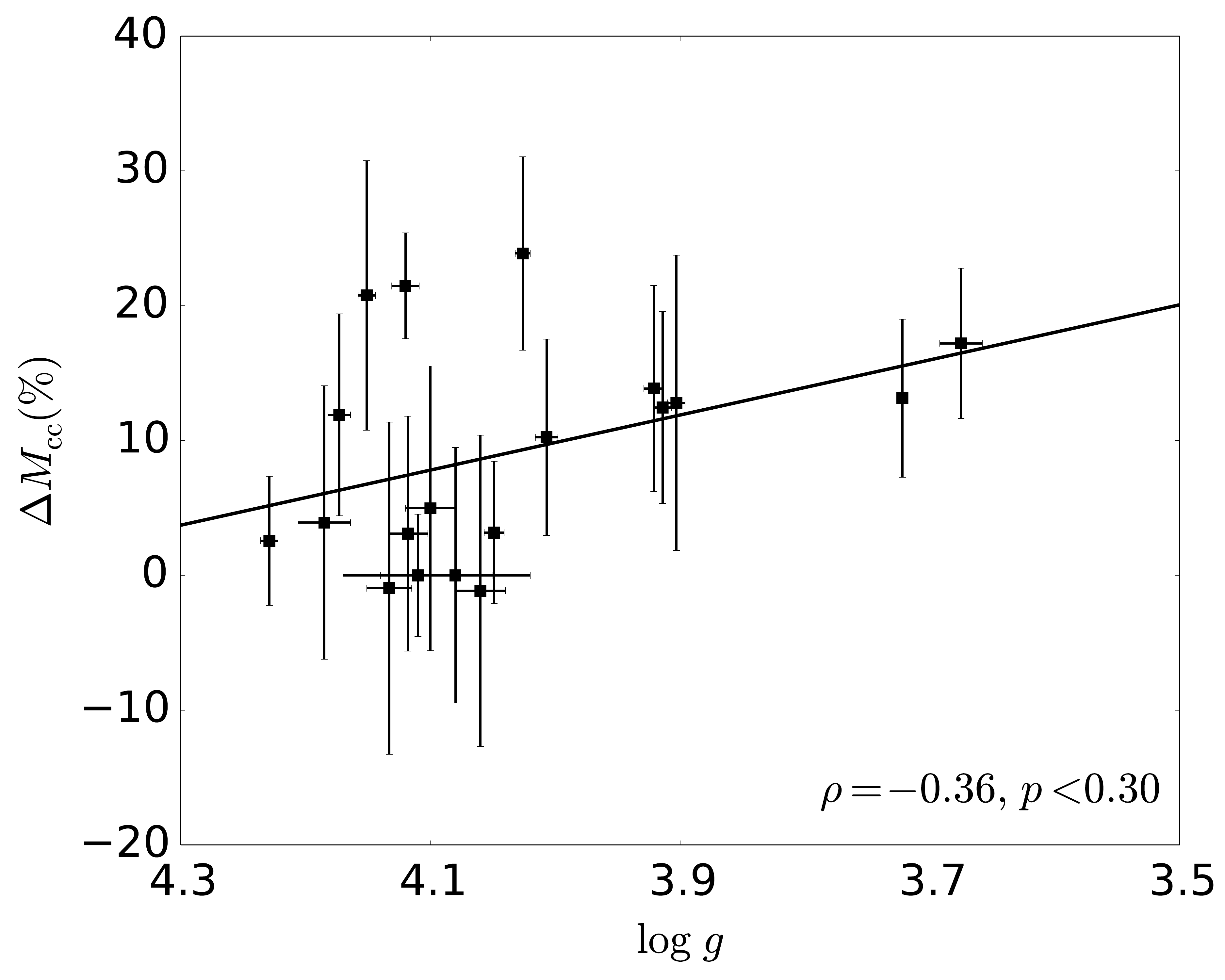}
   \includegraphics[width=8.7cm]{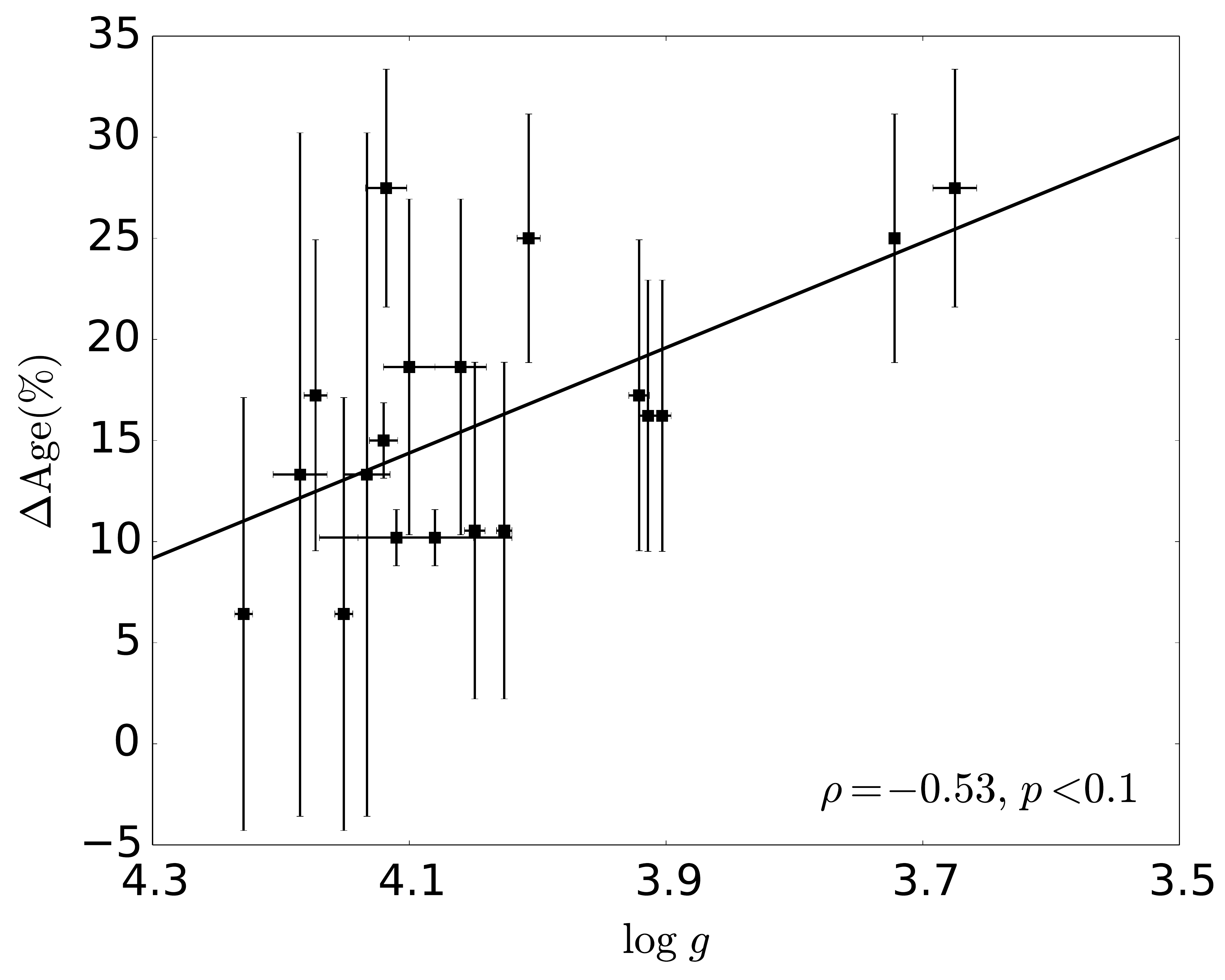}\hspace{5mm}
   \includegraphics[width=8.7cm]{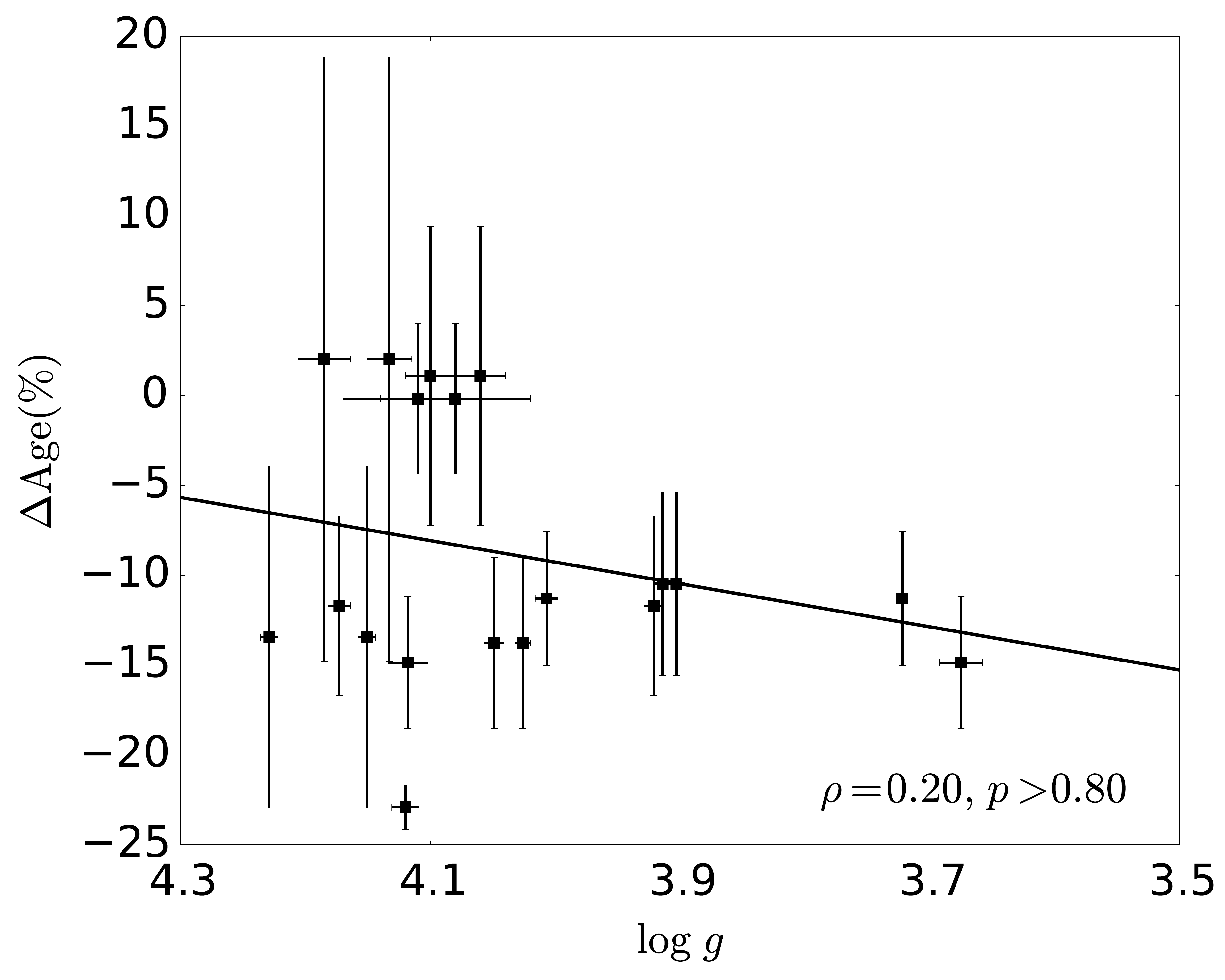}
      \caption{Same as Figure~\ref{Fig:Solution4} but with the equal age condition enforced.} 
         \label{Fig:Solution4_Binary}
   \end{figure*}

\begin{figure*}
   \centering
   \includegraphics[width=8.7cm]{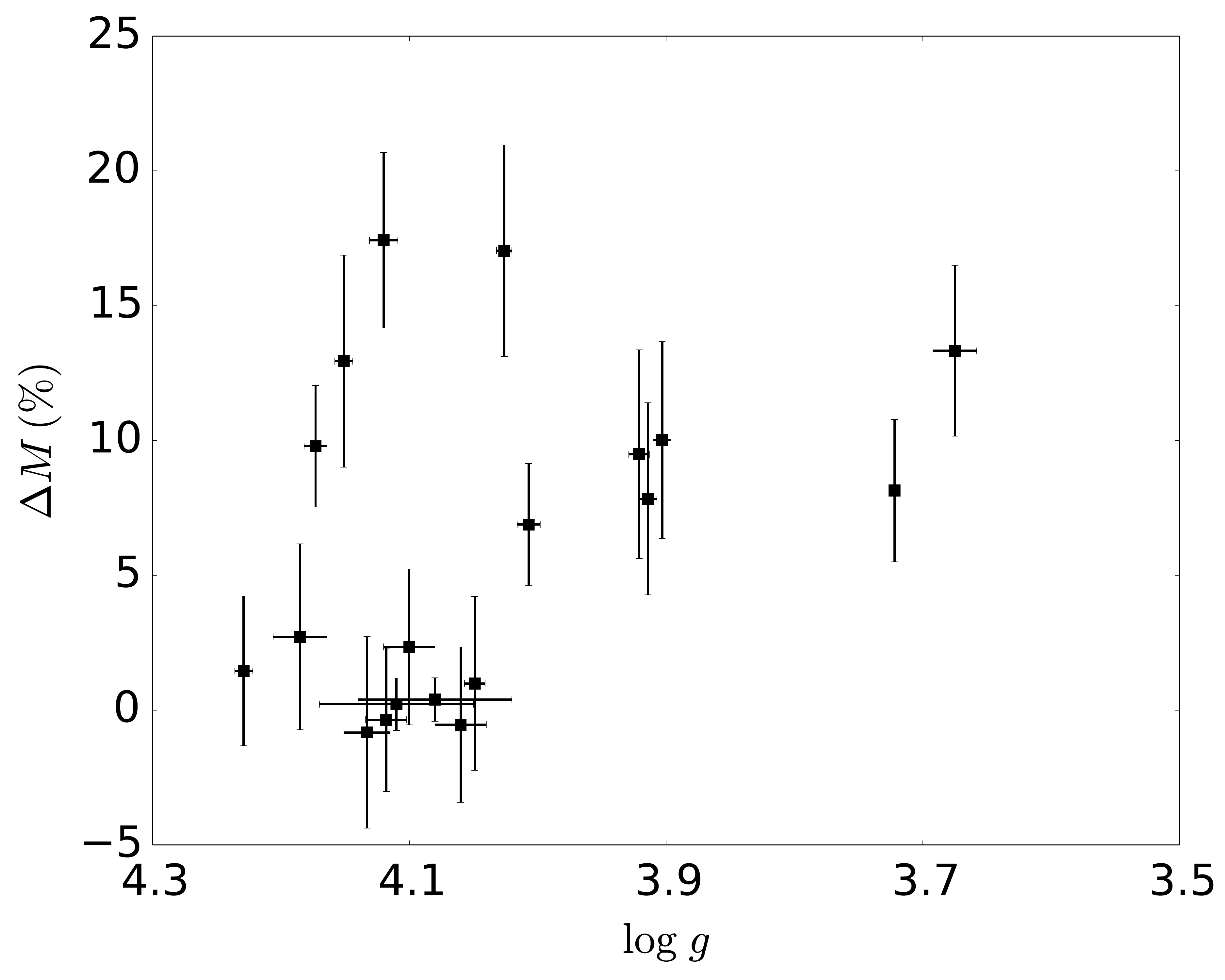}\hspace{5mm}
   \includegraphics[width=8.7cm]{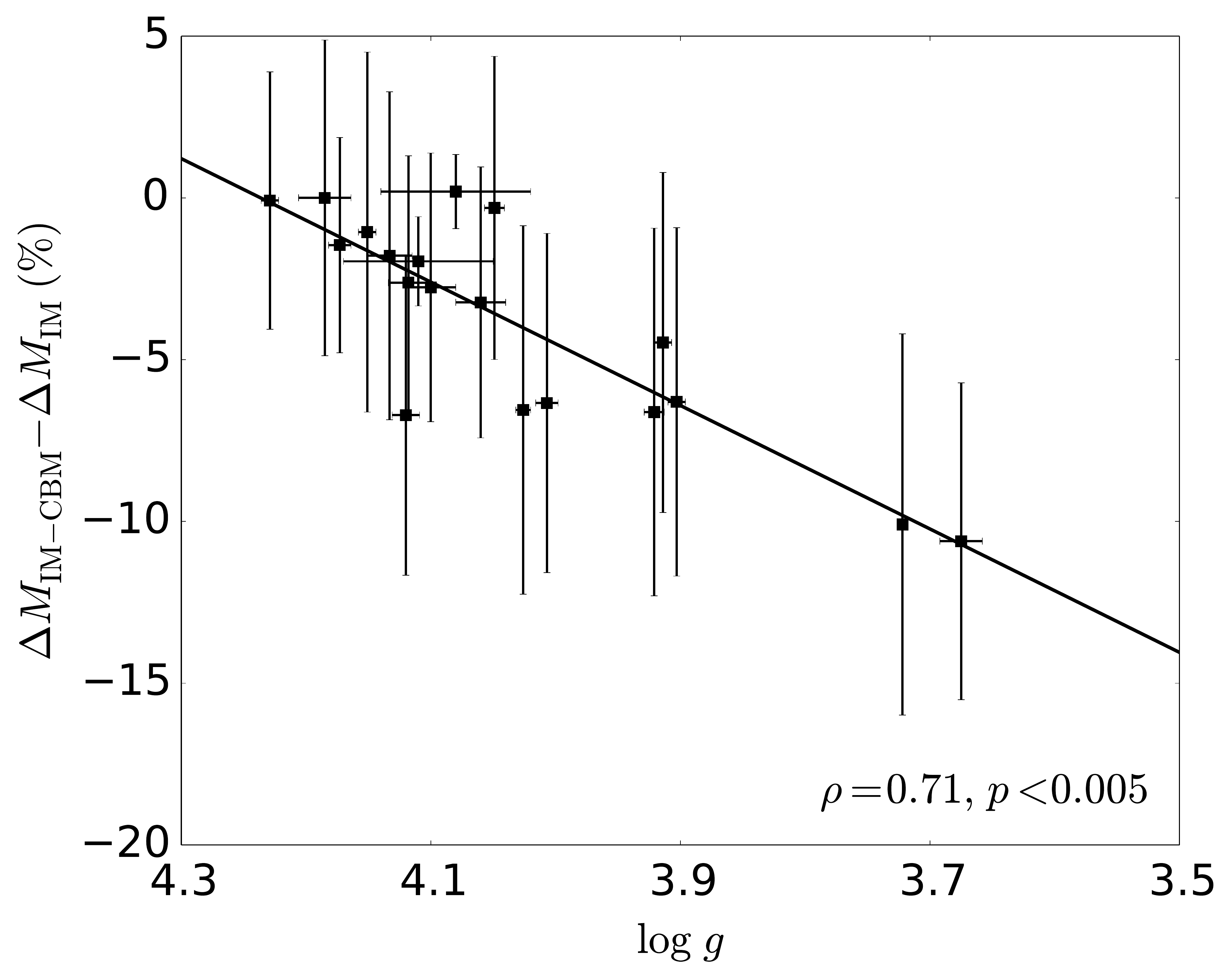}
      \caption{Same as Figure~\ref{Fig:Solution4_MassDiscrepancy} but with the equal age condition enforced.} 
         \label{Fig:Solution4_Binary_MassDiscrepancy}
   \end{figure*}

\begin{landscape}
\begin{table}
\begin{threeparttable}
\tiny
\tabcolsep 0.4mm \caption{Same as Table~\ref{Table:SingleStarCaseParameters} but with the equal age condition enforced.}
\begin{tabular}{lcccccccccccccccccccc}
\hline
 & \multicolumn{5}{c}{{\bf RM solution}}\rule{0pt}{9pt} & \multicolumn{5}{c}{{\bf IM solution}} & \multicolumn{5}{c}{{\bf CBM solution}} & \multicolumn{5}{c}{{\bf IM-CBM solution}}\\
Object/ & $M$ & $f_{\rm ov}$ & age & \multicolumn{2}{c}{$M_{\rm cc}$} & $M$ & $f_{\rm ov}$ & age & \multicolumn{2}{c}{$M_{\rm cc}$} & $M$ & $f_{\rm ov}$ & age & \multicolumn{2}{c}{$M_{\rm cc}$} & $M$ & $f_{\rm ov}$ & age & \multicolumn{2}{c}{$M_{\rm cc}$}\\
Parameter & (M$_{\odot}$) & ($H_{\rm p}$) & (Myr) & (M$_{\odot}$) & (\%) & (M$_{\odot}$) & ($H_{\rm p}$) & (Myr) & (M$_{\odot}$) & (\%) & (M$_{\odot}$) & ($H_{\rm p}$) & (Myr) & (M$_{\odot}$) & (\%) & (M$_{\odot}$) & ($H_{\rm p}$) & (Myr) & (M$_{\odot}$) & (\%)\\
\hline
\multirow{2}{*}{V578\,Mon} & 14.55(9)$^{A}$ & \multirow{2}{*}{0.005} & \multirow{2}{*}{4.03(44)} & 4.90(15) & 33.7(1.2) & 14.68(52)$^{A}$ & \multirow{2}{*}{0.005} & \multirow{2}{*}{3.98(45)} & 5.00(35) & 34.1(3.7) & 14.54(9)$^{A}$ & 0.040(-40) & \multirow{2}{*}{4.42(55)} & 5.29(55) & 36.4(4.0) & 14.42(51)$^{A}$ & 0.040 & \multirow{2}{*}{4.51(50)} & 5.24(35) & 36.3(3.9)\\
  & 10.30(6)$^{B}$ & & & 3.10(7) & 30.1(9) & 10.57(35)$^{A}$ & & & 3.23(20) & 30.6(3.0) & 10.30(6)$^{B}$ & 0.005(+35) & & 3.07(23) & 29.8(2.4) & 10.57(35)$^{A}$ & 0.005 & & 3.19(21) & 30.2(3.1)\vspace{3mm} \\

 \multirow{2}{*}{V453\,Cyg} & 13.98(23)$^{B}$ & \multirow{2}{*}{0.005} & \multirow{2}{*}{10.94(18)} & 3.22(9) & 23.0(1.1) & 16.40(70)$^{A}$ & \multirow{2}{*}{0.005} & \multirow{2}{*}{8.80(45)} & 4.23(29) & 25.8(3.0) & 13.98(14)$^{B}$ & 0.040(-5) & \multirow{2}{*}{12.40(35)} & 4.26(20) & 30.5(1.7) & 15.00(30)$^{A}$ & 0.040 & \multirow{2}{*}{11.00(30)} & 4.82(15) & 32.1(1.7)\\
  & 11.10(13)$^{B}$ & & & 2.78(5) & 25.0(8) & 12.50(50)$^{A}$ & & & 3.41(20) & 27.3(2.8) & 11.10(12)$^{B}$ & 0.040(-10) & & 3.32(22) & 30.0(2.2) & 11.80(20)$^{A}$ & 0.040 & & 3.66(10) & 31.0(1.4)\vspace{3mm} \\

\multirow{2}{*}{V478\,Cyg} & 15.46(17))$^{B}$ & \multirow{2}{*}{0.005} & \multirow{2}{*}{7.62(13)} & 4.49(12) & 29.0(1.2) & 17.34(55)$^{A}$ & \multirow{2}{*}{0.005} & \multirow{2}{*}{6.41(25)} & 5.46(30) & 31.5(2.8) & 15.48(23)$^{B}$ & 0.040(-5) & \multirow{2}{*}{8.32(24)} & 5.38(21) & 34.8(1.9) & 16.65(50)$^{A}$ & 0.040 & \multirow{2}{*}{7.45(35)} & 6.05(32) & 36.3(3.1) \\
  & 15.52(18))$^{B}$ & & & 4.49(12) & 28.9(1.2) & 17.53(55)$^{A}$ & & & 5.52(30) & 31.5(2.8) & 15.36(20)$^{B}$ & 0.030(10)& & 5.08(44) & 33.1(3.3) & 16.58(50)$^{B}$ & 0.030 & & 5.73(34) & 34.6(3.1)\vspace{3mm} \\

\multirow{2}{*}{AH\,Cep} & 16.14(26)$^{B,1}$ & \multirow{2}{*}{0.005} & \multirow{2}{*}{5.92(8)} & 5.17(10) & 32.0(1.2) & 16.40(52)$^{B}$ & \multirow{2}{*}{0.005} & \multirow{2}{*}{4.93(30)} & 5.55(33) & 33.8(3.2) & 16.14(26)$^{B,1}$ & 0.040(-2) & \multirow{2}{*}{6.32(11)} & 5.99(10) & 37.1(1.3) & 16.35(49)$^{A}$ & 0.040 & \multirow{2}{*}{5.45(28)} & 6.18(30) & 37.8(3.1)\\
  & 13.69(21)$^{B,1}$ & & & 4.30(10) & 31.4(1.2) & 16.97(55)$^{B}$ & & & 5.76(38) & 33.9(3.5) & 13.69(21)$^{B,1}$ & 0.005(+2)& & 4.23(9) & 30.9(1.1) & 16.07(52)$^{A}$ & 0.005 & & 5.24(29) & 32.6(3.0)\vspace{3mm} \\

\multirow{2}{*}{V346\,Cen} & 11.78(13)$^{B,1}$ & \multirow{2}{*}{0.005} & \multirow{2}{*}{14.14(4)} & 2.42(2) & 20.5(4) & 14.60(42)$^{A}$ & \multirow{2}{*}{0.005} & \multirow{2}{*}{10.66(35)} & 3.30(19) & 22.6(2.0) & 12.04(13)$^{B}$ & 0.040(-2) & \multirow{2}{*}{15.96(27)} & 3.37(8) & 28.0(1.0) & 13.35(35)$^{A}$ & 0.040 & \multirow{2}{*}{13.59(52)} & 3.95(17) & 29.6(2.1)\\
  & 8.40(10)$^{B,1}$ & & & 2.01(3) & 23.9(7) & 8.59(22)$^{B}$ & & & 2.19(6) & 25.5(1.4) & 8.36(10)$^{B}$ & 0.040(-20) & & 2.26(18) & 27.0(2.5) & 8.37(20)$^{B}$ & 0.040 & & 2.33(8) & 27.8(1.7)\vspace{3mm}\\
  
\multirow{2}{*}{V573\,Car} & 15.14(39)$^{B,1}$ & \multirow{2}{*}{0.005} & \multirow{2}{*}{2.46(17)} & 5.67(25) & 37.5(2.5) & 17.26(45)$^{B}$ & \multirow{2}{*}{0.005} & \multirow{2}{*}{2.18(16)} & 6.65(32) & 38.5(3.0) & 15.14(39)$^{B,1}$ & 0.005(+20) & \multirow{2}{*}{2.68(19)} & 5.44(44) & 35.9(4.1) & 17.10(45)$^{B}$ & 0.005 & \multirow{2}{*}{2.32(17)} & 6.57(32) & 38.4(3.0)\\
  & 12.42(16)$^{B}$ & & & 4.14(11) & 33.3(1.4) & 12.57(29)$^{B}$ & & & 4.26(17) & 33.9(2.2) & 12.42(16)$^{B}$ & 0.040(-5) & & 4.30(13) & 34.6(1.5) & 12.56(28)$^{B}$ & 0.040 & & 4.41(16) & 35.1(2.1)\vspace{3mm} \\

\multirow{2}{*}{V1034\,Sco} & 17.17(12)$^{B}$ & \multirow{2}{*}{0.005} & \multirow{2}{*}{6.38(12)} & 5.40(11) & 31.5(8) & 19.82(70)$^{A}$ & \multirow{2}{*}{0.005} & \multirow{2}{*}{5.28(25)} & 6.81(43) & 34.4(3.5) & 17.14(13)$^{B}$ & 0.040(-5) & \multirow{2}{*}{7.01(14)} & 6.42(27) & 37.5(1.8) & 18.69(65)$^{A}$ & 0.040 & \multirow{2}{*}{6.19(32)} & 7.31(41) & 39.1(3.7)\\
  & 9.65(7)$^{B}$ & & & 2.73(6) & 28.3(8) & 10.68(23)$^{A}$ & & & 3.17(13) & 29.7(1.9) & 9.66(7)$^{B}$ & 0.040(-15) & & 2.94(17) & 30.4(2.0) & 10.54(21)$^{A}$ & 0.040 & & 3.29(14) & 31.2(2.0) \vspace{3mm} \\
  
\multirow{2}{*}{V380\,Cyg} & 11.43(19)$^{B,1}$ & \multirow{2}{*}{0.005} & \multirow{2}{*}{17.2(2)} & 0.38(38) & 3.3(3.3) & 14.70(10)$^{B}$ & \multirow{2}{*}{0.005} & \multirow{2}{*}{12.00(20)} & 2.41(6) & 16.4(5) & 11.43(19)$^{B,1}$ & 0.012(2) & \multirow{2}{*}{17.9(2)} & 1.76(10) & 15.4(1.1) & 15.00(-5)$^{B,2}$ & 0.040 & \multirow{2}{*}{13.8(1)} & 4.07(4) & 27.1(4)\\
  & 7.00(14)$^{B,1}$ & & & 1.65(7) & 23.6(1.5) & 8.69(22)$^{A}$ & & & 2.17(11) & 25.0(1.9) & 7.00(14)$^{B,1}$ & 0.005(5) & & 1.63(4) & 23.3(1.0) & 8.22(18)$^{A}$ & 0.005 & & 1.98(5) & 24.1(1.1)\vspace{3mm} \\

\multirow{2}{*}{CW\,Cep} & 13.01(7)$^{A}$ & \multirow{2}{*}{0.005} & \multirow{2}{*}{6.50(21)} & 3.91(6) & 30.1(6) & 13.35(39)$^{A}$ & \multirow{2}{*}{0.005} & \multirow{2}{*}{6.17(32)} & 4.07(19) & 30.5(2.4) & 13.00(7)$^{A}$ & 0.040(-25) & \multirow{2}{*}{7.24(45)} & 4.38(46) & 33.7(3.6) & 12.93(37)$^{A}$ & 0.040 & \multirow{2}{*}{7.32(40)} & 4.33(21) & 33.5(2.1)\\
  & 11.97(8)$^{B}$ & & & 3.55(6) & 29.7(7) & 12.54(35)$^{A}$ & & & 3.76(19) & 30.0(2.4) & 11.96(8)$^{A}$ & 0.040(-30) & & 3.82(35) & 31.9(3.1) & 12.21(34)$^{A}$ & 0.040 & & 4.01(20) & 32.8(2.7)\vspace{3mm} \\

\multirow{2}{*}{U\,Oph} & 5.06(5)$^{A}$& \multirow{2}{*}{0.005} & \multirow{2}{*}{52.5(2.0)} & 0.92(5) & 18.2(1.2) & 5.09(4)$^{A}$ & \multirow{2}{*}{0.005} & \multirow{2}{*}{51.0(5)} & 0.94(2) & 18.5(5) & 5.10(7)$^{A}$ & 0.025(10) & \multirow{2}{*}{56.3(2.3)} & 1.06(10) & 20.8(2.3) & 5.10(4)$^{A}$ & 0.025 & \multirow{2}{*}{56.2(5)} & 1.06(1) & 20.8(4)\\
  & 4.65(7)$^{B}$ & & & 0.90(5) & 19.4(1.3) & 4.69(4)$^{B}$ & & & 0.92(1) & 19.6(4) & 4.60(4)$^{A}$ & 0.010(7) & & 0.91(4) & 19.8(1.0) & 4.60(4)$^{A}$ & 0.010 & & 0.91(1) & 19.8(4)\vspace{1mm} \\

\hline
\end{tabular}
\begin{tablenotes}
      \tiny
      \item $^{A/B}$ within/outside error box; $^1$ dynamical mass was enforced; $^2$ maximum amount of the core-boundary mixing is assumed (see Sect.~\ref{Sect:V380Cyg})
    \end{tablenotes}
\label{Table:BinaryStarCaseParameters}
\end{threeparttable}
\end{table}
\end{landscape}

\end{appendix}

\end{document}